\documentclass[twocolumn,aps,prd,floatfix,eqsecnum]{revtex4}
\usepackage{graphicx}% Include figure files
\usepackage{dcolumn}% Align table columns on decimal point
\usepackage{bm}% bold math
\usepackage{amsfonts}
\usepackage{latexsym}
\usepackage{amssymb} 
\usepackage{verbatim}
\usepackage{amsthm}
\usepackage{amsmath}

\def\fun#1#2{\lower3.6pt\vbox{\baselineskip0pt\lineskip.9pt
  \ialign{$\mathsurround=0pt#1\hfil##\hfil$\crcr#2\crcr\sim\crcr}}}
\newcommand{\MR}{{{\mathbb R}}}

\newcommand{\RF}{{{\mathbb R}}}

\newtheorem{theorem}{Theorem}[section]

\begin{document}

%\preprint{APS/123-QED}

\title{Second-order Gauge-invariant
  Cosmological Perturbation Theory: Current Status}% Force line breaks with \\  

\author{Kouji Nakamura}
%\altaffiliation[Also at ]{
%  }%Lines break automatically or can be forced with \\
% \email{kouji.nakamura@nao.ac.jp}
\affiliation{%
  Optical and Infrared Astronomy Division, National Astronomical
  Observatory of Japan, Osawa, Mitaka, Tokyo 181-8588, Japan.
}%

%\affiliation{

\date{\today}% It is always \today, today,

\begin{abstract}
  The current status of the recent developments of the
  second-order gauge-invariant cosmological perturbation theory
  is reviewed.
  To show the essence of this perturbation theory, we
  concentrate only on the universe filled with a single scalar
  field.
  Through this review, we point out the problems which should be
  clarified for the further theoretical sophistication of this
  perturbation theory.
  We also expect that this theoretical sophistication will be
  also useful to discuss the theoretical predictions of
  Non-Gaussianity in CMB and comparison with observations.
\end{abstract}

%\pacs{04.50.+h, 04.62.+v, 98.80.Jk}% PACS, the Physics and Astronomy
\maketitle

%%%%%%%%%%%%%%%%%%%%%%%%%%%%%%%%%%%%%%%%%%%%%%%%%%%%%%%%%%%%%%%%%%%%%%
\section{Introduction}
\label{sec:intro}
%%%%%%%%%%%%%%%%%%%%%%%%%%%%%%%%%%%%%%%%%%%%%%%%%%%%%%%%%%%%%%%%%%%%%%

%****************************************************************

The general relativistic cosmological {\it linear} perturbation
theory has been developed to a high degree of sophistication
during the last 30
years\cite{Bardeen-1980,Kodama-Sasaki-1984,Mukhanov-Feldman-Brandenberger-1992}.
One of the motivations of this development was to clarify the
relation between the scenarios of the early universe and
cosmological data, such as the cosmic microwave background (CMB)
anisotropies.
Recently, the first-order approximation of our universe from a
homogeneous isotropic one was revealed through the observation
of the CMB by the Wilkinson Microwave Anisotropy Probe
(WMAP)\cite{WMAP,Non-Gaussianity-observation}, the cosmological
parameters are accurately measured, we have obtained the
standard cosmological model, and the so-called ``precision 
cosmology'' has begun.
These developments in observations were also supported by the
theoretical sophistication of the linear order cosmological
perturbation theory.

%****************************************************************

The observational results of CMB also suggest that the
fluctuations of our universe are adiabatic and Gaussian at least
in the first-order approximation.
We are now on the stage to discuss the deviation from this
first-order approximation from the 
observational\cite{Non-Gaussianity-observation} and theoretical
sides\cite{Non-Gaussianity-inflation,Non-Gaussianity-in-CMB} 
through the non-Gaussianity, the non-adiabaticity, and so on.
These will be goals of future satellite missions.
With the increase of precision of the CMB data, the study of
relativistic cosmological perturbations beyond linear order is a
topical subject.
The {\it second-order} cosmological perturbation theory is one
of such perturbation theories beyond linear order.

%********************************************************************

Although the second-order perturbation theory in general
relativity is an old topic, a general framework of the
gauge-invariant formulation of the general relativistic
second-order perturbation has been
proposed\cite{kouchan-gauge-inv,kouchan-second}.
This general formulation is an extension of the works of Bruni
et al.\cite{M.Bruni-S.Matarrese-S.Mollerach-S.Soonego-CQG1997}
and has also been applied to cosmological perturbations:
The derivation of the second-order Einstein equation in a
gauge-invariant manner without any gauge
fixing\cite{kouchan-cosmo-second}; 
Applicability in more generic
situations\cite{kouchan-second-cosmo-matter};
Confirmation of the consistency between all components of the
second-order Einstein equations and equations of
motions\cite{kouchan-second-cosmo-consistency}.
We also note that the radiation case has recently been discussed
by treating the Boltzmann equation up to second
order\cite{Pitro-2007-2009} along the gauge-invariant manner of
the above series of papers by the present author.

%********************************************************************

In this review article, we summarize the current status of this
development of the second-order gauge-invariant cosmological
perturbation theory through the simple system of a sclar field.
Through this review, we point out the problems which should be
clarified and directions of the further development of the
theoretical sophistication of the general relativistic
higher-order perturbation theory, especially in cosmological
perturbations.
We expect that this sophistication will be also useful to
discuss the theoretical predictions of Non-Gaussianity in CMB
and comparison with observations.

%********************************************************************

The organization of this paper is as follows.
In
Sec.~\ref{sec:General-framework-of-GR-GI-perturbation-theory},
we review the general framework of the second-order gauge 
invariant perturbation theory developed in
Refs.~\cite{kouchan-gauge-inv,kouchan-cosmo-second,kouchan-second,kouchan-LTVII}.
This review also includes additional explanation not given in
those papers.
In Sec.~\ref{sec:Perturbation-of-the-field-equations}, we also
the derivations of the second-order perturbation of the Einstein
equation and the energy-momentum tensor from general point of
view.
For simplicity, in this paper, we only consider a single scalar
field as a matter content.
The ingredients of
Sec.~\ref{sec:General-framework-of-GR-GI-perturbation-theory}
and \ref{sec:Perturbation-of-the-field-equations} will be
applicable to perturbation theory in any theory with general 
covariance, if the decomposition formula
(\ref{eq:linear-metric-decomp}) for the linear-order metric  
perturbation is correct.
In Sec.~\ref{sec:Cosmological-Background-spacetime-equations},
we summarize the Einstein equations in the case of a background
homogeneous isotropic universe, which are used in the derivation
of the first- and second-order Einstein equations. 
In
Sec.~\ref{sec:Equations-for-the-first-order-cosmological-perturbations},
the first-order perturbation of the Einstein equations and the
Klein-Gordon equations are summarized.
The derivation of the second-order perturbations of the Einstein
equations and the Klein-Gordon equations, and their consistency
are reviewed in
Sec.~\ref{sec:Equations-for-the-second-order-cosmological-perturbations}. 
The final section, Sec.~\ref{sec:summary}, is devoted to a
summary and discussions.

%********************************************************************

%%%%%%%%%%%%%%%%%%%%%%%%%%%%%%%%%%%%%%%%%%%%%%%%%%%%%%%%%%%%%%%%%%%%%%
%%%%%%%%%%%%%%%%%%%%%%%%%%%%%%%%%%%%%%%%%%%%%%%%%%%%%%%%%%%%%%%%%%%%%%
\section{General framework of the general relativistic gauge-invariant perturbation theory}
\label{sec:General-framework-of-GR-GI-perturbation-theory}
%%%%%%%%%%%%%%%%%%%%%%%%%%%%%%%%%%%%%%%%%%%%%%%%%%%%%%%%%%%%%%%%%%%%%%
%%%%%%%%%%%%%%%%%%%%%%%%%%%%%%%%%%%%%%%%%%%%%%%%%%%%%%%%%%%%%%%%%%%%%%

%****************************************************************

In this section, we review the general framework of the gauge
invariant perturbation theory developed in 
Refs.~\cite{kouchan-gauge-inv,kouchan-second,M.Bruni-S.Matarrese-S.Mollerach-S.Soonego-CQG1997,kouchan-cosmo-second,kouchan-LTVII,R.K.Sachs-1964,J.M.Stewart-M.Walker11974,S.Sonego-M.Bruni-CMP1998,Matarrese-Mollerach-Bruni-1998,Bruni-Gualtieri-Sopuerta-2003,Sopuerta-Bruni-Gualtieri-2004}.
To develop the general relativistic gauge-invariant perturbation
theory, we first explain the general arguments of the Taylor
expansion on a manifold without introducing an explicit
coordinate system in
Sec.\ref{sec:Taylor-expansion-of-tensors-on-a-manifold}. 
Further, we also have to clarify the notion of ``gauge'' in
general relativity to develop the gauge-invariant perturbation
theory from general point of view, which is explained in
Sec.~\ref{sec:Gauge-degree-of-freedom-in-general-relativity}.
After clarifying the notion of ``gauge'' in general relativistic
perturbations, in
Sec.~\ref{sec:Formulation-of-perturbation-theory}, we explain
the formulation of the general relativistic gauge-invariant
perturbation theory from general point of view.
Although our understanding of ``gauge'' in general relativistic
perturbations essentially is different from ``degree of freedom
of coordinates'' as in many literature, ``a coordinate
transformation'' is induced by our understanding of ``gauge''.
This situation is explained in
Sec.~\ref{sec:Induced-coordiante-transformations}.
To exclude ``gauge degree of freedom'' which is unphysical
degree of freedom in perturbations, we construct ``gauge-invariant
variables'' of perturbations as reviewed in
Sec.~\ref{sec:gauge-invariant-variables}.
These ``gauge-invariant variables'' are regarded as physical
quantities.

%****************************************************************

%%%%%%%%%%%%%%%%%%%%%%%%%%%%%%%%%%%%%%%%%%%%%%%%%%%%%%%%%%%%%%%%%%%%%%
\subsection{Taylor expansion of tensors on a manifold}
\label{sec:Taylor-expansion-of-tensors-on-a-manifold}
%%%%%%%%%%%%%%%%%%%%%%%%%%%%%%%%%%%%%%%%%%%%%%%%%%%%%%%%%%%%%%%%%%%%%%

%****************************************************************

First, we briefly review the issues on the general form of the
Taylor expansion of tensors on a manifold ${\cal M}$.
The gauge issue of general relativistic perturbation theories
which we will discuss is related to the coordinate
transformation. 
Therefore, we have to discuss the general form of the Taylor
expansion without the explicit introduction of coordinate
systems.
Although we only consider the Taylor expansion of a scalar
function $f:{\cal M}\mapsto\MR$, here, the resulting formula is
extended to that for any tensor field on a manifold as in
Appendix \ref{sec:derivation-of-Taylor-expansion}.
We have to emphasize that the general formula of the Taylor
expansion shown here is the starting point of our
gauge-invariant formulation of the second-order general
relativistic perturbation theory.

%****************************************************************

The Taylor expansion of a function $f$ is an approximated form
of $f(q)$ at $q\in{\cal M}$ in terms of the variables at
$p\in{\cal M}$, where $q$ is in the neighborhood of $p$.
To derive the formula for the Taylor expansion of $f$, we have
to compare the values of $f$ at the different points on the
manifold.
To accomplish this, we introduce a one-parameter family of
diffeomorphisms $\Phi_{\lambda}:{\cal M}\mapsto{\cal M}$, where 
$\Phi_{\lambda}(p)=q$ and $\Phi_{\lambda=0}(p)=p$.
One example of a diffeomorphisms $\Phi_{\lambda}$ is
an exponential map with a generator.
However, we consider a more general class of diffeomorphisms.

%****************************************************************

The diffeomorphism $\Phi_{\lambda}$ induces the pull-back
$\Phi_{\lambda}^{*}$ of the function $f$ and this pull-back
enable us to compare the values of the function $f$ at different 
points.
Further, the Taylor expansion of the function $f(q)$ is given by 
\begin{eqnarray}
  f(q)
  &=& f(\Phi_{\lambda}(p))
  =: (\Phi^{*}_{\lambda}f)(p)
  \nonumber\\
  &=&
  f(p)
  +
  \left.\frac{\partial}{\partial\lambda}(\Phi^{*}_{\lambda}f)\right|_{p}
  \lambda 
  +
  \frac{1}{2}
  \left.\frac{\partial^{2}}{\partial\lambda^{2}}(\Phi^{*}_{\lambda}f)\right|_{p}
  \lambda^{2} 
  \nonumber\\
  && \quad\quad
  + O(\lambda^{3}).
  \label{eq:symbolic-Taylor-expansion-of-f}
\end{eqnarray}
Since this expression hold for an arbitrary smooth function
$f$, the function $f$ in
Eq.~(\ref{eq:symbolic-Taylor-expansion-of-f}) can be regarded as
a dummy. 
Therefore, we may regard the Taylor expansion
(\ref{eq:symbolic-Taylor-expansion-of-f}) to be the expansion of
the pull-back $\Phi_{\lambda}^{*}$ of the diffeomorphism 
$\Phi_{\lambda}$, rather than the expansion of the function $f$.

%****************************************************************

According to this point of view, Sonego and
Bruni\cite{S.Sonego-M.Bruni-CMP1998} showed that there exist 
vector fields $\xi_{1}^{a}$ and $\xi_{2}^{a}$ such that the
expansion (\ref{eq:symbolic-Taylor-expansion-of-f}) is given by 
\begin{eqnarray}
  f(q)
  &=& (\Phi^{*}_{\lambda}f)(p)
  \nonumber\\
  &=& f(p)
  + \left.\left({\pounds}_{\xi_{1}}f\right)\right|_{p} \lambda 
  + \frac{1}{2}
  \left.\left({\pounds}_{\xi_{2}}+{\pounds}_{\xi_{1}}^{2}\right)f\right|_{p}
  \lambda^{2} 
  \nonumber\\
  && \quad\quad
  + O(\lambda^{3}),
  \label{eq:Taylor-expansion-of-f}
\end{eqnarray}
without loss of generality (see Appendix
\ref{sec:derivation-of-Taylor-expansion}).
Equation (\ref{eq:Taylor-expansion-of-f}) is not only the
representation of the Taylor expansion of the function $f$, but
also the definitions of the generators $\xi_{1}^{a}$ and
$\xi_{2}^{a}$.
These generators of the one-parameter family of diffeomorphisms
$\Phi_{\lambda}$ represent the direction along which the Taylor
expansion is carried out.
The generator $\xi_{1}^{a}$ is the first-order approximation of
the flow of the diffeomorphism $\Phi_{\lambda}$, and the
generator $\xi_{2}^{a}$ is the second-order correction to this
flow.
We should regard the generators $\xi_{1}^{a}$ and $\xi_{2}^{a}$
to be independent.
Further, as shown in Appendix
\ref{sec:derivation-of-Taylor-expansion}, the representation of
the Taylor expansion of an arbitrary scalar function $f$ is
extended to that for an arbitrary tensor field $Q$ just through
the replacement $f\rightarrow Q$.

%****************************************************************

We must note that, in general, the representation
(\ref{eq:Taylor-expansion-of-f}) of the Taylor expansion is
different from an usual exponential map which is generated by a
vector field. 
In general,
\begin{eqnarray}
  \label{eq:Phi-is-not-one-parameter-group-of-diffeomorphism}
  \Phi_{\sigma}\circ\Phi_{\lambda}\neq\Phi_{\sigma+\lambda}, \quad
  \Phi_{\lambda}^{-1}\neq\Phi_{-\lambda}.
\end{eqnarray}
As noted in
Ref.~\cite{M.Bruni-S.Matarrese-S.Mollerach-S.Soonego-CQG1997},
if the second-order generator $\xi_{2}$ in
Eq.~(\ref{eq:Taylor-expansion-of-f}) is proportional to the
first-order generator $\xi_{1}$ in
Eq.~(\ref{eq:Taylor-expansion-of-f}), the diffeomorphism
$\Phi_{\lambda}$ is reduced to an exponential map.
Therefore, one may reasonably doubt that $\Phi_{\lambda}$ forms
a group except under very special conditions.
However, we have to note that the properties
(\ref{eq:Phi-is-not-one-parameter-group-of-diffeomorphism}) does
not directly mean that $\Phi_{\lambda}$ does not form a group.
There will be possibilities that $\Phi_{\lambda}$ form a group
in a different sense from exponential maps, in which the
properties
(\ref{eq:Phi-is-not-one-parameter-group-of-diffeomorphism}) will
be maintained.

%****************************************************************

\begin{figure} 
  \begin{center}
    \includegraphics[width=0.4\textwidth]{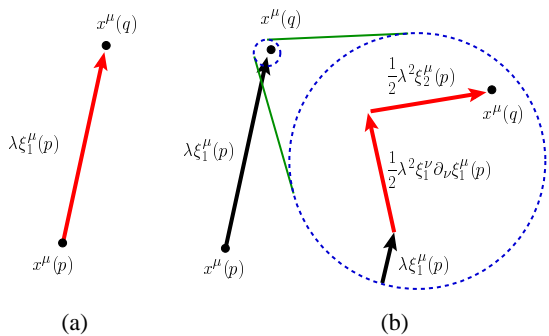}
  \end{center}
  \caption{
    (a) The second term $\lambda \xi_{1}(p)$ in
    Eq.~(\ref{eq:Taylor-expansion-of-xmu}) is the vector which 
    point from the point $x^{\mu}(p)$ to the point $x^{\mu}(q)$
    in the sense of the first-order correction. (b) If we look
    at the neighborhood of the point $x^{\mu}(q)$ in detail, the
    vector $\lambda \xi_{1}(p)$ may fail to point to
    $x^{\mu}(q)$ in the sense of the second order.
    Therefore, it is necessary to add the second-order
    correction 
    $\frac{1}{2}\lambda^{2}(\xi_{2}^{\mu}+\xi^{\nu}_{1}(p)\partial_{\nu}\xi^{\mu}_{1}(p))$.
  } 
  \label{fig:Taylor-expansion-of-coordinate-function}
\end{figure}

Now, we give an intuitive explanation of the representation
(\ref{eq:Taylor-expansion-of-f}) of the Taylor expansion through
the case where the scalar function $f$ in
Eq.~(\ref{eq:Taylor-expansion-of-f}) is a coordinate function.
When two points $p,q\in{\cal M}$ in
Eq.~(\ref{eq:Taylor-expansion-of-f}) are in the neighborhood of 
each other, we can apply a coordinate system 
${\cal M}\mapsto\RF^{n}$ ($n=\dim{\cal M}$), which denoted by
$\{x^{\mu}\}$, to an open set which includes these two points.
Then, we can measure the relative position of these two points
$p$ and $q$ in ${\cal M}$ in terms of this coordinate system in
$\RF^{n}$ through the Taylor expansion
(\ref{eq:Taylor-expansion-of-f}).
In this case, we may regard that the scalar function $f$ in
Eq.~(\ref{eq:Taylor-expansion-of-f}) is a coordinate function
$x^{\mu}$ and Eq.~(\ref{eq:Taylor-expansion-of-f}) yields
\begin{eqnarray}
  x^{\mu}(q)
  &=& (\Phi^{*}_{\lambda}x^{\mu})(p)
  \nonumber\\
  &=& x^{\mu}(p)
  + \lambda \xi_{1}(p)
  + \frac{1}{2} \lambda^{2}
  \left.\left(\xi_{2}+\xi^{\nu}_{1}\partial_{\nu}\xi^{\mu}_{1}\right)\right|_{p}
  \nonumber\\
  &&
  + O(\lambda^{3}),
  \label{eq:Taylor-expansion-of-xmu}
\end{eqnarray}
The second term $\lambda \xi_{1}(p)$ in the right hand side of
Eq.~(\ref{eq:Taylor-expansion-of-xmu}) is familiar.
This is regarded as the vector which point from the point
$x^{\mu}(p)$ to the point $x^{\mu}(q)$ in the sense of the
first-order correction as shown in
Fig.\ref{fig:Taylor-expansion-of-coordinate-function}(a).
However, in the sense of the second order, this vector 
$\lambda \xi_{1}(p)$ may fail to point to $x^{\mu}(q)$.
Therefore, it is necessary to add the second-order correction as
shown in Fig.\ref{fig:Taylor-expansion-of-coordinate-function}(b). 
As a correction of the second order, we may add the term
$\frac{1}{2} \lambda^{2} \xi^{\nu}_{1}(p)\partial_{\nu}\xi^{\mu}_{1}(p)$.
This second-order correction corresponds to that comes from the
exponential map which is generated by the vector field
$\xi_{1}^{\mu}$.
However, this correction completely determined by the vector
field $\xi_{1}^{\mu}$.
Even if we add this correction comes from the exponential map,
there is no guarantee that the corrected vector
$\lambda\xi_{1}(p)+\frac{1}{2}\lambda^{2}\xi^{\nu}_{1}(p)\partial_{\nu}\xi^{\mu}_{1}(p)$
does point to $x^{\mu}(q)$ in the sense of the second order
Thus, we have to add the new correction
$\frac{1}{2}\lambda^{2}\xi^{\nu}_{2}(p)$ of the second order, in
general.

%****************************************************************

%****************************************************************

Of course, without this correction
$\frac{1}{2}\lambda^{2}\xi^{\nu}_{2}(p)$, the vector
which comes only from the exponential map generated by the 
vector field $\xi_{1}$ might point to the point $x^{\mu}(q)$.
Actually, this is possible if we carefully choose the vector
field $\xi^{\mu}_{1}$ taking into account of the deviations at
the second order.
However, this means that we have to take care of the
second-order correction when we determine the first-order
correction.
This contradicts to the philosophy of the Taylor expansion as a
perturbative expansion, in which we can determine everything
order by order.
Therefore, we should regard that the correction
$\frac{1}{2}\lambda^{2}\xi^{\nu}_{2}(p)$ is necessary in
general situations.

%****************************************************************

%%%%%%%%%%%%%%%%%%%%%%%%%%%%%%%%%%%%%%%%%%%%%%%%%%%%%%%%%%%%%%%%%%%%%%
\subsection{Gauge degree of freedom in general relativity}
\label{sec:Gauge-degree-of-freedom-in-general-relativity}
%%%%%%%%%%%%%%%%%%%%%%%%%%%%%%%%%%%%%%%%%%%%%%%%%%%%%%%%%%%%%%%%%%%%%%

%****************************************************************

Since we want to explain the gauge-invariant perturbation theory
in general relativity, first of all, we have to explain the
notion of ``gauge'' in general relativity\cite{kouchan-LTVII}.
General relativity is a theory with general covariance, which
intuitively states that there is no preferred coordinate system
in nature.
This general covariance also introduce the notion of ``gauge''
in the theory.
In the theory with general covariance, these ``gauge'' give rise
to the unphysical degree of freedom and we have to fix the
``gauges'' or to extract some invariant quantities to obtain
physical result.
Therefore, treatments of ``gauges'' are crucial in general
relativity and this situation becomes more delicate in general
relativistic perturbation theory as explained below.

%****************************************************************

In 1964, Sachs\cite{R.K.Sachs-1964} pointed out that there are
two kinds of ``gauges'' in general relativity.
Sachs called these two ``gauges'' as the first- and the
second-kind of gauges, respectively. 
Here, we review these concepts of ``gauge''.

%****************************************************************

%%%%%%%%%%%%%%%%%%%%%%%%%%%%%%%%%%%%%%%%%%%%%%%%%%%%%%%%%%%%%%%%%%%%%%
\subsubsection{First kind gauge}
\label{sec:first-kind-gauge}

%****************************************************************

{\it The first kind gauge} is a coordinate system on a single
manifold ${\cal M}$.
Although this first kind gauge is not important in this paper, we
explain this to emphasize the ``gauge'' discussing in this paper
is different from this first kind gauge.

%****************************************************************

In the standard text book of manifolds (for example, see
\cite{Kobayashi-Nomizu-I-1996}), the following property of a
manifold is written: 
on a manifold, we can always introduce a coordinate system as a
diffeomorphism $\psi_{\alpha}$ from an open set
$O_{\alpha}\subset{\cal M}$ to an open set
$\psi_{\alpha}(O_{\alpha})\subset\RF^{n}$ ($n=\dim{\cal M}$).
This diffeomorphism $\psi_{\alpha}$, i.e., coordinate system of
the open set $O_{\alpha}$, is called {\it gauge choice} (of the
first kind).
If we consider another open set in $O_{\beta}\subset{\cal M}$,
we have another gauge choice
$\psi_{\beta}:O_{\beta}\mapsto\psi_{\beta}(O_{\beta})\subset\RF^{n}$
for $O_{\beta}$.
If these two open sets $O_{\alpha}$ and $O_{\beta}$ have the
intersection $O_{\alpha}\cap O_{\beta}\neq\emptyset$, we can
consider the diffeomorphism $\psi_{\beta}\circ\psi_{\alpha}^{-1}$.
This diffeomorphism $\psi_{\beta}\circ\psi_{\alpha}^{-1}$ is
just a coordinate transformation: $\psi_{\alpha}(O_{\alpha}\cap
O_{\beta})\subset\RF^{n}\mapsto \psi_{\beta}(O_{\alpha}\cap
O_{\beta})\subset\RF^{n}$, which is called 
{\it gauge transformation} (of the first kind) in general
relativity.

%****************************************************************

According to the theory of a manifold, coordinate system are not
on a manifold itself but we can always introduce a coordinate 
system through a map from an open set in the manifold ${\cal M}$
to an open set of $\RF^{n}$.
For this reason, general covariance in general relativity is
automatically included in the premise that our spacetime is
regarded as a single manifold.
The first kind gauge does arise due to this general covariance.
The gauge issue of the first kind is represented by the
question, which coordinate system is convenient?
The answer to this question depends on the problem which we are
addressing, i.e., what we want to clarify.
In some case, this gauge issue of the first kind is an
important. 
However, in many case, it becomes harmless if we apply a
covariant theory on the manifold.

%****************************************************************

%%%%%%%%%%%%%%%%%%%%%%%%%%%%%%%%%%%%%%%%%%%%%%%%%%%%%%%%%%%%%%%%%%%%%%
\subsubsection{Second kind gauge}
\label{sec:second-kind-gauge}

%****************************************************************

{\it The second kind gauge} appears in perturbation theories in
a theory with general covariance.
This notion of the second kind ``gauge'' is the main issue of
this paper.
To explain this, we have to remind what we are doing in
perturbation theories.

%****************************************************************

First, in any perturbation theories, we always treat two
spacetime manifolds.
One is the physical spacetime ${\cal M}$.
This physical spacetime ${\cal M}$ is our nature itself and we
want to describe the properties of this physical spacetime
${\cal M}$ through perturbations.
The other is the background spacetime ${\cal M}_{0}$.
This background spacetime have nothing to do with our nature and  
this is a fictitious manifold which is prepared by us.
This background spacetime is just a reference to carry out
perturbative analyses.
We emphasize that these two spacetime manifolds
${\cal M}$ and ${\cal M}_{0}$ are distinct.
Let us denote the physical spacetime by $({\cal M},\bar{g}_{ab})$
and the background spacetime by $({\cal M}_{0},g_{ab})$, where 
$\bar{g}_{ab}$ is the metric on the physical spacetime manifold,
${\cal M}$, and $g_{ab}$ is the metric on the background
spacetime manifold, ${\cal M}_{0}$.
Further, we formally denote the spacetime metric and the other
physical tensor fields on ${\cal M}$ by $Q$ and its background
value on ${\cal M}_{0}$ by $Q_{0}$.

%****************************************************************

Second, in any perturbation theories, we always write equations
for the perturbation of the physical variable $Q$ in the form
\begin{equation}
  \label{eq:variable-symbolic-perturbation}
  Q(``p\mbox{''}) = Q_{0}(p) + \delta Q(p).
\end{equation}
Usually, this equation is simply regarded as a relation
between the physical variable $Q$ and its background value
$Q_{0}$, or as the definition of the deviation $\delta Q$ of
the physical variable $Q$ from its background value $Q_{0}$.
However, Eq.~(\ref{eq:variable-symbolic-perturbation}) has
deeper implications.
Keeping in our mind that we always treat two different
spacetimes, ${\cal M}$ and ${\cal M}_{0}$, in perturbation
theory, Eq.~(\ref{eq:variable-symbolic-perturbation}) is a
rather curious equation in the following sense: 
The variable on the left-hand side of
Eq.~(\ref{eq:variable-symbolic-perturbation}) is a variable on
${\cal M}$, while the variables on the right-hand side of
Eq.~(\ref{eq:variable-symbolic-perturbation}) are variables on
${\cal M}_{0}$.
Hence, Eq.~(\ref{eq:variable-symbolic-perturbation}) gives a
relation between variables on two different manifolds.

%****************************************************************

Further, through Eq.~(\ref{eq:variable-symbolic-perturbation}),
we have implicitly identified points in these two different
manifolds.
More specifically, $Q(``p\mbox{''})$ on the left-hand side of
Eq.~(\ref{eq:variable-symbolic-perturbation}) is a field on
${\cal M}$, and $``p\mbox{''}\in{\cal M}$.
Similarly, we should regard the background value
$Q_{0}(p)$ of $Q(``p\mbox{''})$ and its deviation $\delta Q(p)$
of $Q(``p\mbox{''})$ from $Q_{0}(p)$, which are on the
right-hand side of
Eq.~(\ref{eq:variable-symbolic-perturbation}), as fields on
${\cal M}_{0}$, and $p\in{\cal M}_{0}$.
Because Eq.~(\ref{eq:variable-symbolic-perturbation}) is
regarded as an equation for field variables, it implicitly
states that the points $``p\mbox{''}\in{\cal M}$ and 
$p\in{\cal M}_{0}$ are same.
This represents the implicit assumption of the existence of a map 
${\cal M}_{0}\rightarrow{\cal M}$ $:$ $p\in{\cal M}_{0}\mapsto
``p\mbox{''}\in{\cal M}$, which is usually called a 
{\it gauge choice} (of the second kind) in perturbation
theory\cite{J.M.Stewart-M.Walker11974}.

%****************************************************************

It is important to note that the second kind gauge choice
between points on ${\cal M}_{0}$ and ${\cal M}$, which is
established by such a relation as
Eq.~(\ref{eq:variable-symbolic-perturbation}), is not unique to
the theory with general covariance.
Rather, Eq.~(\ref{eq:variable-symbolic-perturbation}) involves
the degree of freedom corresponding to the choice of the map
${\cal X}$ $:$ ${\cal M}_{0}\mapsto{\cal M}$.
This is called the {\it gauge degree of freedom} (of the second
kind).
Such a degree of freedom always exists in perturbations of a
theory with general covariance.
General covariance intuitively means that there is no preferred
coordinate system in the theory as mentioned above.
If general covariance is not imposed on the theory, there is a
preferred coordinate system in the theory, and we naturally
introduce this preferred coordinate system onto both 
${\cal M}_{0}$ and ${\cal M}$.
Then, we can choose the identification map ${\cal X}$ using this 
preferred coordinate system.
However, there is no such coordinate system in general
relativity due to the general covariance, and we have no
guiding principle to choose the identification map ${\cal X}$.
Indeed, we may identify $``p\mbox{''}\in{\cal M}$ with 
$q\in{\cal M}_{0}$ ($q\neq p$) instead of $p\in{\cal M}_{0}$.
In the above understanding of the concept of ``gauge'' (of the
second kind) in general relativistic perturbation theory, a
gauge transformation is simply a change of the map ${\cal X}$.

%****************************************************************

\begin{figure} 
  \begin{center}
    \includegraphics[width=0.35\textwidth]{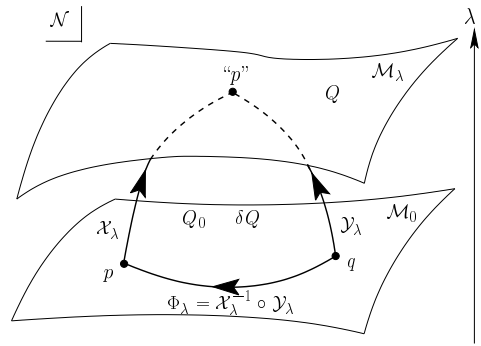}
  \end{center}
  \caption{
    The second kind gauge is a point-identification between the
    physical spacetime ${\cal M}_{\lambda}$ and the background
    spacetime ${\cal M}_{0}$ on the extended manifold ${\cal N}$.
    Through Eq.~(\ref{eq:variable-symbolic-perturbation}), we
    implicitly assume the existence of a point-identification
    map between ${\cal M}_{\lambda}$ and ${\cal M}_{0}$.
    However, this point-identification is not unique by virtue
    of the general covariance in the theory.
    We may chose the gauge of the second kind so that 
    $p\in{\cal M}_{0}$ and ``$p$''$\in{\cal M}_{\lambda}$ is
    same (${\cal X}_{\lambda}$).
    We may also choose the gauge so that $q\in{\cal M}_{0}$ and
    ``$p$''$\in{\cal M}_{\lambda}$ is same (${\cal Y}_{\lambda}$).
    These are different gauge choices.
    The gauge transformation 
    ${\cal X}_{\lambda}\rightarrow{\cal Y}_{\lambda}$ is given
    by the diffeomorphism 
    $\Phi={\cal X}_{\lambda}^{-1}\circ{\cal Y}_{\lambda}$.
  } 
  \label{fig:gauge-choice-is-point-identification}
\end{figure}

%****************************************************************

These are the basic ideas of gauge degree of freedom (of the
second kind) in the general relativistic perturbation theory
which are pointed out by Sacks\cite{R.K.Sachs-1964} and
mathematically clarified by Stewart and
Walker\cite{J.M.Stewart-M.Walker11974}.
Based on these ideas, higher-order perturbation theory has been
developed in Refs.~\cite{kouchan-gauge-inv,kouchan-second,M.Bruni-S.Matarrese-S.Mollerach-S.Soonego-CQG1997,kouchan-cosmo-second,kouchan-second-cosmo-matter,kouchan-second-cosmo-consistency,kouchan-LTVII,Bruni-Gualtieri-Sopuerta-2003,Sopuerta-Bruni-Gualtieri-2004,M.Bruni-S.Sonego-CQG1999}.

%****************************************************************

%%%%%%%%%%%%%%%%%%%%%%%%%%%%%%%%%%%%%%%%%%%%%%%%%%%%%%%%%%%%%%%%%%%%%%
\subsection{Formulation of perturbation theory}
\label{sec:Formulation-of-perturbation-theory}
%%%%%%%%%%%%%%%%%%%%%%%%%%%%%%%%%%%%%%%%%%%%%%%%%%%%%%%%%%%%%%%%%%%%%%

%****************************************************************

To formulate the above understanding in more detail, we
introduce an infinitesimal parameter $\lambda$ for the
perturbation.
Further, we consider the $4+1$-dimensional manifold 
${\cal N}={\cal M}\times\MR$, where $4=\dim{\cal M}$ and
$\lambda\in\MR$.
The background spacetime 
${\cal M}_{0}=\left.{\cal N}\right|_{\lambda=0}$ and the
physical spacetime 
${\cal M}={\cal M}_{\lambda}=\left.{\cal N}\right|_{\MR=\lambda}$ are
also submanifolds embedded in the extended manifold ${\cal N}$.
Each point on ${\cal N}$ is identified by a pair $(p,\lambda)$,
where $p\in{\cal M}_{\lambda}$, and each point in 
${\cal M}_{0}\subset{\cal N}$ is identified by $\lambda=0$.

%****************************************************************

Through this construction, the manifold ${\cal N}$ is foliated by
four-dimensional submanifolds ${\cal M}_{\lambda}$ of each
$\lambda$, and these are diffeomorphic to ${\cal M}$ and 
${\cal M}_{0}$.
The manifold ${\cal N}$ has a natural differentiable structure
consisting of the direct product of ${\cal M}$ and $\MR$.
Further, the perturbed spacetimes ${\cal M}_{\lambda}$ for each
$\lambda$ must have the same differential structure with this
construction.
In other words, we require that perturbations be continuous in
the sense that ${\cal M}$ and ${\cal M}_{0}$ are connected by a
continuous curve within the extended manifold ${\cal N}$.
Hence, the changes of the differential structure resulting from
the perturbation, for example the formation of singularities and
singular perturbations in the sense of fluid mechanics, are
excluded from consideration.

%****************************************************************

Let us consider the set of field equations 
\begin{equation}
  \label{eq:field-eq-for-Q}
  {\cal E}[Q_{\lambda}] = 0
\end{equation}
on the physical spacetime ${\cal M}_{\lambda}$ for the physical
variables $Q_{\lambda}$ on ${\cal M}_{\lambda}$.
The field equation (\ref{eq:field-eq-for-Q}) formally represents
the Einstein equation for the metric on ${\cal M}_{\lambda}$ and
the equations for matter fields on ${\cal M}_{\lambda}$.
If a tensor field $Q_{\lambda}$ is given on each 
${\cal M}_{\lambda}$, $Q_{\lambda}$ is automatically extended to a
tensor field on ${\cal N}$ by $Q(p,\lambda):=Q_{\lambda}(p)$,
where $p\in{\cal M}_{\lambda}$.
In this extension, the field equation (\ref{eq:field-eq-for-Q})
is regarded as an equation on the extended manifold ${\cal N}$.
Thus, we have extended an arbitrary tensor field and the field
equations (\ref{eq:field-eq-for-Q}) on each ${\cal M}_{\lambda}$
to those on the extended manifold ${\cal N}$.

%****************************************************************

Tensor fields on ${\cal N}$ obtained through the above construction
are necessarily ``tangent'' to each ${\cal M}_{\lambda}$.
To consider the basis of the tangent space of ${\cal N}$, we
introduce the normal form and its dual, which are normal to each
${\cal M}_{\lambda}$ in ${\cal N}$.
These are denoted by $(d\lambda)_{a}$ and
$(\partial/\partial\lambda)^{a}$, respectively, and they satisfy
$(d\lambda)_{a}(\partial/\partial\lambda)^{a}=1$.
The form $(d\lambda)_{a}$ and its dual,
$(\partial/\partial\lambda)^{a}$, are normal to any tensor field
extended from the tangent space on each ${\cal M}_{\lambda}$
through the above construction. 
The set consisting of $(d\lambda)_{a}$,
$(\partial/\partial\lambda)^{a}$ and the basis of the tangent
space on each ${\cal M}_{\lambda}$ is regarded as the basis of
the tangent space of ${\cal N}$.

%****************************************************************

Now, we define the perturbation of an arbitrary tensor field
$Q$.
We compare $Q$ on ${\cal M}_{\lambda}$ with $Q_{0}$ on 
${\cal M}_{0}$, and it is necessary to identify the points of
${\cal M}_{\lambda}$ with those of ${\cal M}_{0}$ as mentioned
above.
This point identification map is the gauge choice of the second
kind as mentioned above.
The gauge choice is made by assigning a diffeomorphism
${\cal X}_{\lambda}$ $:$ ${\cal N}$ $\rightarrow$ ${\cal N}$
such that ${\cal X}_{\lambda}$ $:$ ${\cal M}_{0}$ $\rightarrow$
${\cal M}_{\lambda}$.
Following the paper of Bruni et
al.\cite{M.Bruni-S.Matarrese-S.Mollerach-S.Soonego-CQG1997}, we
introduce a gauge choice ${\cal X}_{\lambda}$ as an
one-parameter groups of diffeomorphisms, i.e., an exponential 
map, for simplicity.
We denote the generator of this exponential map by 
${}_{{\cal X}}\!\eta^{a}$.
This generator ${}_{{\cal X}}\!\eta^{a}$ is decomposed by the 
basis on ${\cal N}$ which are constructed above.
Although the generator ${}_{{\cal X}}\!\eta^{a}$ should satisfy
some appropriate properties\cite{kouchan-gauge-inv}, the
arbitrariness of the gauge choice ${\cal X}_{\lambda}$ is
represented by the tangential component of the generator
${}_{{\cal X}}\!\eta^{a}$ to ${\cal M}_{\lambda}$.

%****************************************************************

The pull-back ${\cal X}_{\lambda}^{*}Q$, which is induced by the
exponential map ${\cal X}_{\lambda}$, maps a tensor field $Q$
on the physical manifold ${\cal M}_{\lambda}$ to a tensor field
${\cal X}_{\lambda}^{*}Q$ on the background spacetime.
In terms of this generator ${}_{{\cal X}}\!\eta^{a}$, the
pull-back ${\cal X}_{\lambda}^{*}Q$ is represented by the Taylor
expansion
\begin{eqnarray}
  Q(r)
  &=&
  Q({\cal X}_{\lambda}(p))
  =
  {\cal X}_{\lambda}^{*}Q(p)
  \nonumber\\
  &=&
  Q(p)
  + \lambda \left.{\pounds}_{{}_{{\cal X}}\!\eta}Q \right|_{p}
  + \frac{1}{2} \lambda^{2} 
  \left.{\pounds}_{{}_{{\cal X}}\!\eta}^{2}Q\right|_{p}
  \nonumber\\
  && \quad
  + O(\lambda^{3}),
  \label{eq:Taylor-expansion-of-calX-org}
\end{eqnarray}
where $r={\cal X}_{\lambda}(p)\in{\cal M}_{\lambda}$.
Because $p\in{\cal M}_{0}$, we may regard the equation
\begin{eqnarray}
  {\cal X}_{\lambda}^{*}Q(p)
  &=&
  Q_{0}(p)
  + \lambda \left.{\pounds}_{{}_{{\cal X}}\!\eta}Q\right|_{{\cal M}_{0}}(p)
  + \frac{1}{2} \lambda^{2} 
  \left.{\pounds}_{{}_{{\cal X}}\!\eta}^{2}Q\right|_{{\cal M}_{0}}(p)
  \nonumber\\
  && \quad
  + O(\lambda^{3})
  \label{eq:Taylor-expansion-of-calX}
\end{eqnarray}
as an equation on the background spacetime ${\cal M}_{0}$,
where $Q_{0}=\left.Q\right|_{{\cal M}_{0}}$ is the background
value of the physical variable of $Q$.
Once the definition of the pull-back of the gauge choice 
${\cal X}_{\lambda}$ is given, the first- and the second-order
perturbations ${}^{(1)}_{\;\cal X}\!Q$ and ${}^{(2)}_{\;\cal X}\!Q$
of a tensor field $Q$ under the gauge choice 
${\cal X}_{\lambda}$ are simply given by the expansion 
\begin{equation}
  \label{eq:Bruni-35}
  \left.{\cal X}^{*}_{\lambda}Q_{\lambda}\right|_{{\cal M}_{0}}
  =
  Q_{0}
  + \lambda {}^{(1)}_{\;\cal X}\!Q
  + \frac{1}{2} \lambda^{2} {}^{(2)}_{\;\cal X}\!Q
  + O(\lambda^{3})
\end{equation}
with respect to the infinitesimal parameter $\lambda$.
Comparing Eqs.~(\ref{eq:Taylor-expansion-of-calX}) and
(\ref{eq:Bruni-35}), we define the first- and the second-order
perturbations of a physical variable $Q_{\lambda}$ under the
gauge choice ${\cal X}_{\lambda}$ by 
\begin{eqnarray}
  {}^{(1)}_{\;\cal X}\!Q := 
  \left.{\pounds}_{{}_{\cal X}\!\eta} Q\right|_{{\cal M}_{0}},
  \quad
  {}^{(2)}_{\;\cal X}\!Q := 
  \left.{\pounds}_{{}_{\cal X}\!\eta}^{2} Q\right|_{{\cal M}_{0}}.
  \label{eq:representation-of-each-order-perturbation}
\end{eqnarray}
We note that all variables in Eq.~(\ref{eq:Bruni-35}) are
defined on ${\cal M}_{0}$.

%****************************************************************

Now, we consider two {\it different gauge choices} based on
the above understanding of the second kind gauge choice.
Suppose that ${\cal X}_{\lambda}$ and ${\cal Y}_{\lambda}$ are
two exponential maps with the generators ${}_{\cal X}\!\eta^{a}$
and ${}_{\cal Y}\!\eta^{a}$ on ${\cal N}$, respectively.
In other words, ${\cal X}_{\lambda}$ and ${\cal Y}_{\lambda}$
are two gauge choices (see
Fig.~\ref{fig:gauge-choice-is-point-identification}).
Then, the integral curves of each ${}_{\cal X}\!\eta^{a}$ and
${}_{\cal Y}\!\eta^{a}$ in ${\cal N}$ are the orbits of the
actions of the gauge choices ${\cal X}_{\lambda}$ and 
${\cal Y}_{\lambda}$, respectively.
Since we choose the generators ${}_{\cal X}\!\eta^{a}$ and
${}_{\cal Y}\!\eta^{a}$ so that these are transverse to each
${\cal M}_{\lambda}$ everywhere on ${\cal N}$, the integral
curves of these vector fields intersect with each 
${\cal M}_{\lambda}$.
Therefore, points lying on the same integral curve of either of
the two are to be regarded as {\it the same point} within the
respective gauges.
When these curves are not identical, i.e., the tangential
components to each ${\cal M}_{\lambda}$ of 
${}_{\cal X}\!\eta^{a}$ and ${}_{\cal Y}\!\eta^{a}$ are
different, these point identification maps ${\cal X}_{\lambda}$
and ${\cal Y}_{\lambda}$ are regarded as
{\it two different gauge choices}.

%****************************************************************

We next introduce the concept of {\it gauge invariance}.
In particular, in this paper, we consider the concept of
{\it order by order gauge invariance}\cite{kouchan-second-cosmo-matter}.
Suppose that ${\cal X}_{\lambda}$ and ${\cal Y}_{\lambda}$ are
two different gauge choices which are generated by the vector
fields ${}_{\cal X}\!\eta^{a}$ and ${}_{\cal Y}\!\eta^{a}$,
respectively.
These gauge choices also pull back a generic tensor field $Q$ on
${\cal N}$ to two other tensor fields, ${\cal X}_{\lambda}^{*}Q$
and ${\cal Y}_{\lambda}^{*}Q$, for any given value of $\lambda$.
In particular, on ${\cal M}_{0}$, we now have three tensor
fields associated with a tensor field $Q$; one is the background
value $Q_{0}$ of $Q$, and the other two are the pulled-back
variables of $Q$ from ${\cal M}_{\lambda}$ to ${\cal M}_{0}$ by
the two different gauge choices,
\begin{eqnarray}
  {}_{\cal X}\!Q_{\lambda} &:=&
  \left.{\cal X}^{*}_{\lambda}Q\right|_{{\cal M}_{0}}
  \nonumber\\
  &=& 
  Q_{0}
  + \lambda {}^{(1)}_{\;{\cal X}}\!Q
  + \frac{1}{2} \lambda^{2} {}^{(2)}_{\;{\cal X}}\!Q
  + O(\lambda^{3})
  \label{eq:Bruni-39-one}
  \\
  {}_{\cal Y}\!Q_{\lambda} &:=&
  \left.{\cal Y}^{*}_{\lambda}Q\right|_{{\cal M}_{0}}
  \nonumber\\
  &=& 
  Q_{0}
  + \lambda {}^{(1)}_{\;{\cal Y}}\!Q
  + \frac{1}{2} \lambda^{2} {}^{(2)}_{\;{\cal Y}}\!Q
  + O(\lambda^{3})
  \label{eq:Bruni-40-one}
\end{eqnarray}
Here, we have used Eq.~(\ref{eq:Bruni-35}).
Because ${\cal X}_{\lambda}$ and ${\cal Y}_{\lambda}$ are gauge
choices which map from ${\cal M}_{0}$ to ${\cal M}_{\lambda}$,
${}_{\cal X}Q_{\lambda}$ and ${}_{\cal Y}Q_{\lambda}$ are the
different representations on ${\cal M}_{0}$ in the two different
gauges of the same perturbed tensor field $Q$ on ${\cal M}_{\lambda}$.
The quantities ${}^{(k)}_{\;\cal X}\!Q$ and 
${}^{(k)}_{\;\cal Y}\!Q$ in Eqs.~(\ref{eq:Bruni-39-one}) and
(\ref{eq:Bruni-40-one}) are the perturbations of $O(k)$ in the
gauges ${\cal X}_{\lambda}$ and ${\cal Y}_{\lambda}$, respectively.
We say that the $k$th-order perturbation 
${}^{(k)}_{\;\cal X}\!Q$ of $Q$ is 
{\it order by order gauge invariant} iff for any two gauges
${\cal X}_{\lambda}$ and ${\cal Y}_{\lambda}$ 
the following holds:
\begin{equation}
  {}^{(k)}_{\;\cal X}\!Q = {}^{(k)}_{\;\cal Y}\!Q.
\end{equation}

%****************************************************************

Now, we consider the {\it gauge transformation rules}
between different gauge choices.
In general, the representation ${}^{\cal X}Q_{\lambda}$ on
${\cal M}_{0}$ of the perturbed variable $Q$ on 
${\cal M}_{\lambda}$ depends on the gauge choice 
${\cal X}_{\lambda}$.
If we employ a different gauge choice, the representation of
$Q_{\lambda}$ on ${\cal M}_{0}$ may change.
Suppose that ${\cal X}_{\lambda}$ and ${\cal Y}_{\lambda}$ are
different gauge choices, which are the point identification maps
from ${\cal M}_{0}$ to ${\cal M}_{\lambda}$, and the generators
of these gauge choices are given by ${}_{\cal X}\!\eta^{a}$ and
${}_{\cal Y}\!\eta^{a}$, respectively.
Then, the change of the gauge choice from ${\cal X}_{\lambda}$
to ${\cal Y}_{\lambda}$ is represented by the diffeomorphism
\begin{equation}
  \label{eq:diffeo-def-from-Xinv-Y}
  \Phi_{\lambda} :=
  ({\cal X}_{\lambda})^{-1}\circ{\cal Y}_{\lambda}.
\end{equation}
This diffeomorphism $\Phi_{\lambda}$ is the map $\Phi_{\lambda}$
$:$ ${\cal M}_{0}$ $\rightarrow$ ${\cal M}_{0}$ for each value
of $\lambda\in\MR$.
The diffeomorphism $\Phi_{\lambda}$ does change the point
identification, as expected from the understanding of the gauge
choice discussed above.
Therefore, the diffeomorphism $\Phi_{\lambda}$ is regarded as
the gauge transformation $\Phi_{\lambda}$ $:$
${\cal X}_{\lambda}$ $\rightarrow$ ${\cal Y}_{\lambda}$.

%****************************************************************

The gauge transformation $\Phi_{\lambda}$ induces a pull-back
from the representation ${}_{\cal X}\!Q_{\lambda}$ of the
perturbed tensor field $Q$ in the gauge choice 
${\cal X}_{\lambda}$ to the representation 
${}_{\cal Y}\!Q_{\lambda}$ in the gauge choice 
${\cal Y}_{\lambda}$.
Actually, the tensor fields ${}_{\cal X}\!Q_{\lambda}$ and
${}_{\cal Y}\!Q_{\lambda}$, which are defined on ${\cal M}_{0}$,
are connected by the linear map $\Phi^{*}_{\lambda}$ as
\begin{eqnarray}
  {}_{\cal Y}\!Q_{\lambda}
  &=&
  \left.{\cal Y}^{*}_{\lambda}Q\right|_{{\cal M}_{0}}
  =
  \left.\left(
      {\cal Y}^{*}_{\lambda}
      \left({\cal X}_{\lambda}
      {\cal X}_{\lambda}^{-1}\right)^{*}Q\right)
  \right|_{{\cal M}_{0}}
  \nonumber\\
  &=&
  \left.
    \left(
      {\cal X}^{-1}_{\lambda}
      {\cal Y}_{\lambda}
    \right)^{*}
    \left(
      {\cal X}^{*}_{\lambda}Q
    \right)
  \right|_{{\cal M}_{0}}
  =  \Phi^{*}_{\lambda} {}_{\cal X}\!Q_{\lambda}.
  \label{eq:Bruni-45-one}
\end{eqnarray}
According to generic arguments concerning the Taylor expansion
of the pull-back of a tensor field on the same manifold, given
in \S\ref{sec:Taylor-expansion-of-tensors-on-a-manifold}, it
should be possible to express the gauge transformation
$\Phi^{*}_{\lambda} {}_{\cal X}\!Q_{\lambda}$ in the form
\begin{eqnarray}
  \Phi^{*}_{\lambda} {}_{\cal X}\!Q = {}_{\cal X}\!Q
  + \lambda {\pounds}_{\xi_{1}} {}_{\cal X}\!Q
  + \frac{\lambda^{2}}{2} \left\{
    {\pounds}_{\xi_{2}} + {\pounds}_{\xi_{1}}^{2}
  \right\} {}_{\cal X}\!Q
  \nonumber\\
  + O(\lambda^{3}),
  \label{eq:Bruni-46-one} 
\end{eqnarray}
where the vector fields $\xi_{1}^{a}$ and $\xi_{2}^{a}$ are the
generators of the gauge transformation $\Phi_{\lambda}$ (see
Eq.~(\ref{eq:Taylor-expansion-of-f})).

%****************************************************************

Comparing the representation (\ref{eq:Bruni-46-one}) of the
Taylor expansion in terms of the generators $\xi_{1}^{a}$ and
$\xi_{2}^{a}$ of the pull-back 
$\Phi_{\lambda}^{*}{}_{\cal X}\!Q$ and that in terms of the
generators ${}_{\cal X}\!\eta^{a}$ and ${}_{\cal Y}\!\eta^{a}$
of the pull-back 
${\cal Y}^{*}_{\lambda}\circ\left({\cal X}_{\lambda}^{-1}\right)^{*}\;{}_{{\cal X}}\!Q$ 
($=\Phi_{\lambda}^{*}{}_{\cal X}\!Q$), we readily obtain explicit
expressions for the generators $\xi_{1}^{a}$ and $\xi_{2}^{a}$
of the gauge transformation 
$\Phi={\cal X}^{-1}_{\lambda}\circ{\cal Y}_{\lambda}$ in terms
of the generators ${}_{\cal X}\!\eta^{a}$ and 
${}_{\cal Y}\!\eta^{a}$ of each gauge choices as follows:
\begin{eqnarray}
  \xi_{1}^{a}
  =
  {}_{\cal Y}\!\eta^{a}
  -
  {}_{\cal X}\!\eta^{a},
  \quad
  \xi_{2}^{a}
  = 
  \left[
    {}_{\cal Y}\!\eta
    ,
    {}_{\cal X}\!\eta
  \right]^{a}.
  \label{eq:relation-between-xi-eta}
\end{eqnarray}
Further, because the gauge transformation $\Phi_{\lambda}$ is a
map within the background spacetime ${\cal M}_{0}$, the
generator should consist of vector fields on ${\cal M}_{0}$.
This can be satisfied by imposing some appropriate conditions on
the generators ${}_{\cal Y}\!\eta^{a}$ and ${}_{\cal X}\!\eta^{a}$.

%****************************************************************

We can now derive the relation between the perturbations in the
two different gauges.
Up to second order, these relations are derived by substituting
(\ref{eq:Bruni-39-one}) and (\ref{eq:Bruni-40-one}) into
(\ref{eq:Bruni-46-one}):
\begin{eqnarray}
  \label{eq:Bruni-47-one}
  {}^{(1)}_{\;{\cal Y}}\!Q - {}^{(1)}_{\;{\cal X}}\!Q &=& 
  {\pounds}_{\xi_{1}}Q_{0}, \\
  \label{eq:Bruni-49-one}
  {}^{(2)}_{\;\cal Y}\!Q - {}^{(2)}_{\;\cal X}\!Q &=& 
  2 {\pounds}_{\xi_{1}} {}^{(1)}_{\;\cal X}\!Q 
  +\left\{{\pounds}_{\xi_{2}}+{\pounds}_{\xi_{1}}^{2}\right\} Q_{0}.
\end{eqnarray}

%****************************************************************

Here, we should comment on the gauge choice in the above
explanation.
We have introduced an exponential map ${\cal X}_{\lambda}$ (or
${\cal Y}_{\lambda}$) as the gauge choice, for simplicity.
However, this simplified introduction of ${\cal X}_{\lambda}$ as
an exponential map is not essential to the gauge transformation
rules (\ref{eq:Bruni-47-one}) and (\ref{eq:Bruni-49-one}).
Actually, we can generalize the diffeomorphism 
${\cal X}_{\lambda}$ from an exponential map.
For example, the diffeomorphism whose pull-back is represented
by the Taylor expansion (\ref{eq:Taylor-expansion-of-f}) is a
candidate of the generalization.
If we generalize the diffeomorphism ${\cal X}_{\lambda}$, the
representation (\ref{eq:Taylor-expansion-of-calX}) of the
pulled-back variable ${\cal X}_{\lambda}^{*}Q(p)$, the
representations of the perturbations
(\ref{eq:representation-of-each-order-perturbation}), and the
relations (\ref{eq:relation-between-xi-eta}) between generators
of $\Phi_{\lambda}$, ${\cal X}_{\lambda}$, and 
${\cal Y}_{\lambda}$ will be changed.
However, the gauge transformation rules (\ref{eq:Bruni-47-one})
and (\ref{eq:Bruni-49-one}) are direct consequences of the
generic Taylor expansion (\ref{eq:Bruni-46-one}) of
$\Phi_{\lambda}$.
Generality of the representation of the Taylor expansion
(\ref{eq:Bruni-46-one}) of $\Phi_{\lambda}$ implies that the
gauge transformation rules (\ref{eq:Bruni-47-one}) and
(\ref{eq:Bruni-49-one}) will not be changed, even if we
generalize the each gauge choice ${\cal X}_{\lambda}$.
Further, the relations (\ref{eq:relation-between-xi-eta})
between generators also imply that, even if we employ simple
exponential maps as gauge choices, both of the generators
$\xi_{1}^{a}$ and $\xi_{2}^{a}$ are naturally induced by the
generators of the original gauge choices.
Hence, we conclude that the gauge transformation rules
(\ref{eq:Bruni-47-one}) and (\ref{eq:Bruni-49-one}) are quite
general and irreducible.
In this paper, we review the development of a second-order
gauge-invariant cosmological perturbation theory based on the
above understanding of the gauge degree of freedom only through
the gauge transformation rules (\ref{eq:Bruni-47-one}) and
(\ref{eq:Bruni-49-one}).
Hence, the developments of the cosmological perturbation theory
presented below will not be changed even if we generalize the
gauge choice ${\cal X}_{\lambda}$ from a simple exponential map.

%****************************************************************

We also have to emphasize the physical implication of the gauge 
transformation rules (\ref{eq:Bruni-47-one}) and
(\ref{eq:Bruni-49-one}).
According to the above construction of the perturbation theory,
gauge degree of freedom, which induces the transformation rules 
(\ref{eq:Bruni-47-one}) and (\ref{eq:Bruni-49-one}), is
unphysical degree of freedom.
As emphasized above, the physical spacetime ${\cal M}_{\lambda}$
is our nature itself, while there is no background spacetime
${\cal M}_{0}$ in our nature. 
The background spacetime ${\cal M}_{0}$ is a fictitious
spacetime and it have nothing to do with our nature.
Since the gauge choice ${\cal X}_{\lambda}$ just gives a
relation between ${\cal M}_{\lambda}$ and ${\cal M}_{0}$, the
gauge choice ${\cal X}_{\lambda}$ also have nothing to do with
our nature.
On the other hand, any observations and experiments are carried
out only on the physical spacetime ${\cal M}_{\lambda}$ through
the physical processes on the physical spacetime 
${\cal M}_{\lambda}$.
Therefore, any direct observables in any observations and
experiments should be independent of the gauge choice 
${\cal X}_{\lambda}$, i.e., should be gauge invariant. 
Keeping this fact in our mind, the gauge transformation rules
(\ref{eq:Bruni-47-one}) and (\ref{eq:Bruni-49-one}) imply that
the perturbations ${}^{(1)}_{\;{\cal X}}\!Q$ and
${}^{(2)}_{\;\cal X}\!Q$ include unphysical degree of freedom,
i.e., gauge degree of freedom, if these perturbations are
transformed as (\ref{eq:Bruni-47-one}) or
(\ref{eq:Bruni-49-one}) under the gauge transformation
${\cal X}_{\lambda}\rightarrow{\cal Y}_{\lambda}$.
If the perturbations ${}^{(1)}_{\;{\cal X}}\!Q$ and
${}^{(2)}_{\;\cal X}\!Q$ are independent of the gauge choice,
these variables are order by order gauge invariant.
Therefore, order by order gauge-invariant variables does not
include unphysical degree of freedom and should be related to
the physics on the physical spacetime ${\cal M}_{\lambda}$.

%****************************************************************

%%%%%%%%%%%%%%%%%%%%%%%%%%%%%%%%%%%%%%%%%%%%%%%%%%%%%%%%%%%%%%%%%%%%%%
\subsection{Coordinate transformations induced by the second
  kind gauge transformation}
\label{sec:Induced-coordiante-transformations}
%%%%%%%%%%%%%%%%%%%%%%%%%%%%%%%%%%%%%%%%%%%%%%%%%%%%%%%%%%%%%%%%%%%%%%

%****************************************************************

In many literature, gauge degree of freedom is regarded as the
degree of freedom of the coordinate transformation.
In the linear-order perturbation theory, these two degree of
freedom are equivalent with each other.
However, in the higher order perturbations, we should regard
that these two degree of freedom are different.
Although the essential understanding of the gauge degree of
freedom (of the second kind) is as that explained above, the
gauge transformation (of the second kind) also induces the
infinitesimal coordinate transformation on the physical
spacetime ${\cal M}_{\lambda}$ as a result.
In many case, the understanding of ``gauges'' in perturbations
based on coordinate transformations leads mistakes.
Therefore, we did not use any ingredient of this subsection in
our series of
papers\cite{kouchan-gauge-inv,kouchan-second,kouchan-cosmo-second,kouchan-second-cosmo-matter,kouchan-second-cosmo-consistency}  
concerning about higher-order general relativistic
gauge-invariant perturbation theory.
However, we comment on the relations between the coordinate
transformation, briefly.
Details can be seen in Refs.~\cite{Matarrese-Mollerach-Bruni-1998,Bruni-Gualtieri-Sopuerta-2003,kouchan-gauge-inv}.

%****************************************************************

To see that the gauge transformation of the second kind
induces the coordinate transformation, we introduce the
coordinate system $\{O_{\alpha},\psi_{\alpha}\}$ on the
``background spacetime'' ${\cal M}_{0}$, where $O_{\alpha}$ are
open sets on the background spacetime and $\psi_{\alpha}$ are
diffeomorphisms from $O_{\alpha}$ to $\RF^{4}$ 
($4=\dim{{\cal M}_{0}}$).
The coordinate system $\{O_{\alpha},\psi_{\alpha}\}$ is the set
of the collection of the pair of open sets $O_{\alpha}$ and
diffeomorphism $O_{\alpha}\mapsto\RF^{4}$.
If we employ a gauge choice ${\cal X}_{\lambda}$, we have the
correspondence of ${\cal M}_{\lambda}$ and ${\cal M}_{0}$.
Together with the coordinate system $\psi_{\alpha}$ on 
${\cal M}_{0}$, this correspondence between 
${\cal M}_{\lambda}$ and ${\cal M}_{0}$ induces the coordinate
system on ${\cal M}_{\lambda}$.
Actually, $X_{\lambda}(O_{\alpha})$ for each $\alpha$ is an open
set of ${\cal M}_{\lambda}$.
Then, $\psi_{\alpha}\circ{\cal X}_{\lambda}^{-1}$ becomes a
diffeomorphism from an open set
$X_{\lambda}(O_{\alpha})\subset{\cal M}_{\lambda}$ to $\RF^{4}$.
This diffeomorphism $\psi_{\alpha}\circ{\cal X}_{\lambda}^{-1}$
induces a coordinate system of an open set on 
${\cal M}_{\lambda}$.

%****************************************************************

When we have two different gauge choices ${\cal X}_{\lambda}$
and ${\cal Y}_{\lambda}$, 
$\psi_{\alpha}\circ{\cal X}_{\lambda}^{-1}$ and 
$\psi_{\alpha}\circ{\cal Y}_{\lambda}^{-1}$ become different
coordinate systems on ${\cal M}_{\lambda}$.
We can also consider the coordinate transformation from the
coordinate system $\psi_{\alpha}\circ{\cal X}_{\lambda}^{-1}$ to
another coordinate system $\psi_{\alpha}\circ{\cal Y}_{\lambda}^{-1}$.
Since the gauge transformation 
${\cal X}_{\lambda}$ $\rightarrow$ ${\cal Y}_{\lambda}$ is
induced by the diffeomorphism $\Phi_{\lambda}$ defined by
Eq.~(\ref{eq:diffeo-def-from-Xinv-Y}), the induced coordinate
transformation is given by 
\begin{eqnarray}
  y^{\mu}(q) := x^{\mu}(p) = \left(\left(\Phi^{-1}\right)^{*}x^{\mu}\right)(q)
\end{eqnarray}
in the {\it passive} point of
view\cite{Matarrese-Mollerach-Bruni-1998,Bruni-Gualtieri-Sopuerta-2003,kouchan-gauge-inv}.
If we represent this coordinate transformation in terms of the
Taylor expansion in
Sec.~\ref{sec:Taylor-expansion-of-tensors-on-a-manifold}, up to
third order, we have the coordinate transformation 
\begin{eqnarray}
  y^{\mu}(q) &=& x^{\mu}(q) - \lambda \xi^{\mu}_{1}(q) 
%  \nonumber\\
%  &&
  + \frac{\lambda^{2}}{2} \left\{
    - \xi^{\mu}_{2}(q)
    + \xi^{\nu}_{1}(q)\partial_{\nu}\xi^{\mu}_{1}(q)
  \right\}
  \nonumber\\
  &&
  + O(\lambda^{3})
  .
\end{eqnarray}

%****************************************************************

%%%%%%%%%%%%%%%%%%%%%%%%%%%%%%%%%%%%%%%%%%%%%%%%%%%%%%%%%%%%%%%%%%%%%%
\subsection{Gauge-invariant variables}
\label{sec:gauge-invariant-variables}
%%%%%%%%%%%%%%%%%%%%%%%%%%%%%%%%%%%%%%%%%%%%%%%%%%%%%%%%%%%%%%%%%%%%%%

%****************************************************************

Here, inspecting the gauge transformation rules
(\ref{eq:Bruni-47-one}) and (\ref{eq:Bruni-49-one}), we define
the gauge invariant variables for a metric perturbation and for 
arbitrary matter fields (tensor fields).
Employing the idea of order by order gauge invariance for 
perturbations\cite{kouchan-second-cosmo-matter}, we proposed a
procedure to construct gauge invariant variables of higher-order
perturbations\cite{kouchan-gauge-inv}.
This proposal is as follows.
First, we decompose a linear-order metric perturbation into its
gauge invariant and variant parts. 
The procedure for decomposing linear-order metric perturbations
is extended to second-order metric perturbations, and we can 
decompose the second-order metric perturbation into gauge
invariant and variant parts.
Then, we can define the gauge invariant variables for the first-
and second-order perturbations of an arbitrary field other
than the metric by using the gauge variant parts of the first-
and second-order metric perturbations.
Although the procedure for finding gauge invariant variables for 
linear-order metric perturbations is highly non-trivial, once we
know this procedure, we can easily define the gauge invariant
variables of a higher-order perturbation through a simple
extension of the procedure for the linear-order perturbations.

%****************************************************************

Now, we review the above strategy to construct gauge-invariant
variables.
To consider a metric perturbation, we expand the metric on the
physical spacetime ${\cal M}_{\lambda}$, which is pulled back to
the background spacetime ${\cal M}_{0}$ using a gauge choice in
the form given in (\ref{eq:Bruni-35}):
\begin{eqnarray}
  {\cal X}^{*}_{\lambda}\bar{g}_{ab}
  &=&
  g_{ab} + \lambda {}_{{\cal X}}\!h_{ab} 
  + \frac{\lambda^{2}}{2} {}_{{\cal X}}\!l_{ab}
  + O^{3}(\lambda),
  \label{eq:metric-expansion}
\end{eqnarray}
where $g_{ab}$ is the metric on ${\cal M}_{0}$.
Of course, the expansion (\ref{eq:metric-expansion}) of the
metric depends entirely on the gauge choice 
${\cal X}_{\lambda}$.
Nevertheless, henceforth, we do not explicitly express the index
of the gauge choice ${\cal X}_{\lambda}$ in an expression if
there is no possibility of confusion.

%****************************************************************

Our starting point to construct gauge invariant variables is the
assumption that 
{\it we already know the procedure for finding gauge invariant
  variables for the linear metric perturbations.}
Then, a linear metric perturbation $h_{ab}$ is decomposed as
\begin{eqnarray}
  h_{ab} =: {\cal H}_{ab} + {\pounds}_{X}g_{ab},
  \label{eq:linear-metric-decomp}
\end{eqnarray}
where ${\cal H}_{ab}$ and $X^{a}$ are the gauge invariant and
variant parts of the linear-order metric
perturbations, i.e., under the gauge transformation
(\ref{eq:Bruni-47-one}), these are transformed as 
\begin{equation}
  {}_{{\cal Y}}\!{\cal H}_{ab} - {}_{{\cal X}}\!{\cal H}_{ab} =  0, 
  \quad
  {}_{\quad{\cal Y}}\!X^{a} - {}_{{\cal X}}\!X^{a} = \xi^{a}_{1}. 
  \label{eq:linear-metric-decomp-gauge-trans}
\end{equation}
The first-order metric perturbation
(\ref{eq:linear-metric-decomp}) together with the gauge
transformation rules (\ref{eq:linear-metric-decomp-gauge-trans})
does satisfy the gauge transformation rule
(\ref{eq:Bruni-47-one}) for the first-order metric perturbation,
i.e., 
\begin{eqnarray}
  \label{eq:first-order-gauge-trans-of-metric}
  {}^{(1)}_{\;{\cal Y}}\!h_{ab} - {}^{(1)}_{\;{\cal X}}\!h_{ab} &=& 
  {\pounds}_{\xi_{1}}g_{ab}.
\end{eqnarray}

%****************************************************************

As emphasized in our series of papers
\cite{kouchan-gauge-inv,kouchan-second,kouchan-cosmo-second,kouchan-second-cosmo-matter,kouchan-second-cosmo-consistency},
the above assumption is quite non-trivial and it is not simple
to carry out the systematic decomposition
(\ref{eq:linear-metric-decomp}) on an arbitrary background
spacetime, since this procedure depends completely on the
background spacetime $({\cal M}_{0},g_{ab})$. 
However, as we will show below, this procedure exists at least
in the case of cosmological perturbations of a homogeneous and
isotropic universe in Sec.~\ref{sec:Gauge-invariant-metric-perturbations}.

%****************************************************************

Once we accept this assumption for linear-order metric
perturbations, we can always find gauge invariant variables for 
higher-order perturbations\cite{kouchan-gauge-inv}.
According to the gauge transformation rule
(\ref{eq:Bruni-49-one}), the second-order metric perturbation
$l_{ab}$ is transformed as
\begin{eqnarray}
  \label{eq:second-order-gauge-trans-of-metric}
  {}^{(2)}_{\;\cal Y}\!l_{ab} - {}^{(2)}_{\;\cal X}\!l_{ab}
  = 
  2 {\pounds}_{\xi_{1}} {}_{\;\cal X}\!h_{ab} 
%  \nonumber\\
%  &&
  +\left\{{\pounds}_{\xi_{2}}+{\pounds}_{\xi_{1}}^{2}\right\} g_{ab}
\end{eqnarray}
under the gauge transformation $\Phi_{\lambda}=({\cal X}_{\lambda})^{-1}\circ{\cal Y}_{\lambda}:{\cal X}_{\lambda}\rightarrow{\cal Y}_{\lambda}$. 
Although this gauge transformation rule is slightly complicated,
inspecting this gauge transformation rule, we first
introduce the variable $\hat{L}_{ab}$ defined by
\begin{equation}
  \label{eq:Lhatab-def}
  \hat{L}_{ab}
  :=
  l_{ab}
  - 2 {\pounds}_{X} h_{ab}
  + {\pounds}_{X}^{2} g_{ab}.
\end{equation}
Under the gauge transformation 
$\Phi_{\lambda}=({\cal X}_{\lambda})^{-1}\circ{\cal Y}_{\lambda}:{\cal X}_{\lambda}\rightarrow{\cal Y}_{\lambda}$,
the variable $\hat{L}_{ab}$ is transformed as 
\begin{eqnarray}
  {}_{\;\cal Y}\!\hat{L}_{ab} - {}_{\;\cal X}\!\hat{L}_{ab}
  &=& 
  {\pounds}_{\sigma} g_{ab},
  \label{eq:kouchan-4.67}
  \\
  \sigma^{a} &:=& \xi_{2}^{a} + [\xi_{1},X]^{a}.
  \label{eq:sigma-def}
\end{eqnarray}
The gauge transformation rule (\ref{eq:kouchan-4.67}) is
identical to that for a linear metric perturbation. 
Therefore, we may apply the above procedure to decompose
$h_{ab}$ into ${\cal H}_{ab}$ and $X_{a}$ when we decompose of
the components of the variable $\hat{L}_{ab}$.
Then, $\hat{L}_{ab}$ can be decomposed as
\begin{eqnarray}
  \label{eq:Lhatab-decomposition}
  \hat{L}_{ab} = {\cal L}_{ab} + {\pounds}_{Y}g_{ab},
\end{eqnarray}
where ${\cal L}_{ab}$ is the gauge invariant part of the
variable $\hat{L}_{ab}$, or equivalently, of the second-order
metric perturbation $l_{ab}$, and $Y^{a}$ is the gauge variant
part of $\hat{L}_{ab}$, i.e., the gauge variant part of $l_{ab}$. 
Under the gauge transformation 
$\Phi_{\lambda}=({\cal X}_{\lambda})^{-1}\circ{\cal Y}_{\lambda}$,
the variables ${\cal L}_{ab}$ and $Y^{a}$ are transformed as 
\begin{equation}
  {}_{\;\cal Y}\!{\cal L}_{ab} - {}_{\;\cal X}\!{\cal L}_{ab} = 0, 
  \quad
  {}_{\;\cal Y}\!Y_{a} - {}_{\;\cal Y}\!Y_{a} = \sigma_{a},
\end{equation}
respectively.
Thus, once we accept the assumption
(\ref{eq:linear-metric-decomp}), the second-order metric
perturbations are decomposed as
\begin{eqnarray}
  \label{eq:H-ab-in-gauge-X-def-second-1}
  l_{ab}
  &=:&
  {\cal L}_{ab} + 2 {\pounds}_{X} h_{ab}
  + \left(
      {\pounds}_{Y}
    - {\pounds}_{X}^{2} 
  \right)
  g_{ab},
\end{eqnarray}
where ${\cal L}_{ab}$ and $Y^{a}$ are the gauge invariant and
variant parts of the second order metric perturbations, i.e.,
\begin{eqnarray}
  {}_{{\cal Y}}\!{\cal L}_{ab} - {}_{{\cal X}}\!{\cal L}_{ab} = 0,
  \quad
  {}_{{\cal Y}}\!Y^{a} - {}_{{\cal X}}\!Y^{a}
  = \xi_{2}^{a} + [\xi_{1},X]^{a}.
\end{eqnarray}

%****************************************************************

Furthermore, as shown in Ref.~\cite{kouchan-gauge-inv}, using
the first- and second-order gauge variant parts, $X^{a}$ and
$Y^{a}$, of the metric perturbations, the gauge invariant
variables for an arbitrary field $Q$ other than the metric are
given by 
\begin{eqnarray}
  \label{eq:matter-gauge-inv-def-1.0}
  {}^{(1)}\!{\cal Q} &:=& {}^{(1)}\!Q - {\pounds}_{X}Q_{0}
  , \\ 
  \label{eq:matter-gauge-inv-def-2.0}
  {}^{(2)}\!{\cal Q} &:=& {}^{(2)}\!Q - 2 {\pounds}_{X} {}^{(1)}Q 
  - \left\{ {\pounds}_{Y} - {\pounds}_{X}^{2} \right\} Q_{0}
  .
\end{eqnarray}
It is straightforward to confirm that the variables
${}^{(p)}\!{\cal Q}$ defined by
(\ref{eq:matter-gauge-inv-def-1.0}) and
(\ref{eq:matter-gauge-inv-def-2.0}) are gauge invariant under
the gauge transformation rules (\ref{eq:Bruni-47-one}) and
(\ref{eq:Bruni-49-one}), respectively.

%****************************************************************

Equations (\ref{eq:matter-gauge-inv-def-1.0}) and
(\ref{eq:matter-gauge-inv-def-2.0}) have very important implications.
To see this, we represent these equations as
\begin{eqnarray}
  \label{eq:matter-gauge-inv-decomp-1.0}
  {}^{(1)}\!Q &=& {}^{(1)}\!{\cal Q} + {\pounds}_{X}Q_{0}
  , \\ 
  \label{eq:matter-gauge-inv-decomp-2.0}
  {}^{(2)}\!Q  &=& {}^{(2)}\!{\cal Q} + 2 {\pounds}_{X} {}^{(1)}Q 
  + \left\{ {\pounds}_{Y} - {\pounds}_{X}^{2} \right\} Q_{0}
  .
\end{eqnarray}
These equations imply that any perturbation of first- and
second-order can always be decomposed into gauge-invariant and
gauge-variant parts as
Eqs.~(\ref{eq:matter-gauge-inv-decomp-1.0}) and
(\ref{eq:matter-gauge-inv-decomp-2.0}), respectively. 
These decomposition formulae
(\ref{eq:matter-gauge-inv-decomp-1.0}) and
(\ref{eq:matter-gauge-inv-decomp-2.0}) are important ingredients
in the general framework of the second-order general
relativistic gauge-invariant perturbation theory.

%****************************************************************

%%%%%%%%%%%%%%%%%%%%%%%%%%%%%%%%%%%%%%%%%%%%%%%%%%%%%%%%%%%%%%%%%%%%%%
%%%%%%%%%%%%%%%%%%%%%%%%%%%%%%%%%%%%%%%%%%%%%%%%%%%%%%%%%%%%%%%%%%%%%%
\section{Perturbations of the field equations}
\label{sec:Perturbation-of-the-field-equations}
%%%%%%%%%%%%%%%%%%%%%%%%%%%%%%%%%%%%%%%%%%%%%%%%%%%%%%%%%%%%%%%%%%%%%%
%%%%%%%%%%%%%%%%%%%%%%%%%%%%%%%%%%%%%%%%%%%%%%%%%%%%%%%%%%%%%%%%%%%%%%

%****************************************************************

In terms of the gauge invariant variables defined last section,
we derive the field equations, i.e., Einstein equations and
the equation for a matter field.
To derive the perturbation of the Einstein equations and the
equation for a matter field (Klein-Gordon equation), first of
all, we have to derive the perturbative expressions of the
Einstein tensor\cite{kouchan-second}.
This is reviewed in
Sec.~\ref{sec:Perturbation-of-the-Einstein-tensor}. 
We also derive the first- and the second-order perturbations of
the energy momentum tensor for a scalar field and the
Klein-Gordon equation\cite{kouchan-second-cosmo-matter} in 
Sec.~\ref{sec:Perturbation-of-the-energy-momentum-tensor-and-KG-eq}.
Finally, we consider the first- and the second-order
the Einstein equations in
Sec.~\ref{sec:Perturbation-of-the-Einstein-equation}.

%****************************************************************

%%%%%%%%%%%%%%%%%%%%%%%%%%%%%%%%%%%%%%%%%%%%%%%%%%%%%%%%%%%%%%%%%%%%%%
\subsection{Perturbations of the Einstein curvature}
\label{sec:Perturbation-of-the-Einstein-tensor}
%%%%%%%%%%%%%%%%%%%%%%%%%%%%%%%%%%%%%%%%%%%%%%%%%%%%%%%%%%%%%%%%%%%%%%

%****************************************************************

The relation between the curvatures associated with the metrics
on the physical spacetime ${\cal M}_{\lambda}$ and the
background spacetime ${\cal M}_{0}$ is given by the relation
between the pulled-back operator
${\cal X}_{\lambda}^{*}\bar{\nabla}_{a}\left({\cal X}^{-1}_{\lambda}\right)^{*}$
of the covariant derivative $\bar{\nabla}_{a}$ associated with
the metric $\bar{g}_{ab}$ on ${\cal M}_{\lambda}$ and the
covariant derivative $\nabla_{a}$ associated with the metric
$g_{ab}$ on ${\cal M}_{0}$.
The pulled-back covariant derivative
${\cal X}_{\lambda}^{*}\bar{\nabla}_{a}\left({\cal X}^{-1}_{\lambda}\right)^{*}$
depends on the gauge choice ${\cal X}_{\lambda}$.
The property of the derivative operator
${\cal X}^{*}_{\lambda}\bar{\nabla}_{a}\left({\cal X}^{-1}_{\lambda}\right)^{*}$
as the covariant derivative on ${\cal M}_{\lambda}$ is given by
\begin{equation}
  {\cal X}^{*}_{\lambda}\bar{\nabla}_{a}
  \left(
    \left(
      {\cal X}^{-1}_{\lambda}\right)^{*}{\cal X}^{*}_{\lambda}\bar{g}_{ab}
  \right) = 0,
  \label{eq:property-as-covariant-derivative-on-phys-sp}
\end{equation}
where ${\cal X}^{*}_{\lambda}\bar{g}_{ab}$ is the pull-back of
the metric on ${\cal M}_{\lambda}$, which is expanded as
Eq.~(\ref{eq:metric-expansion}).
In spite of the gauge dependence of the operator 
${\cal X}^{*}_{\lambda}\bar{\nabla}_{a}\left({\cal X}^{-1}_{\lambda}\right)^{*}$,
we simply denote this operator by $\bar{\nabla}_{a}$, because
our calculations are carried out only on ${\cal M}_{0}$ in the
same gauge choice ${\cal X}_{\lambda}$.
Further, we denote the pulled-back metric 
${\cal X}^{*}_{\lambda}\bar{g}_{ab}$ on ${\cal M}_{\lambda}$ by
$\bar{g}_{ab}$, as mentioned above.

%****************************************************************

Since the derivative operator $\bar{\nabla}_{a}$ 
($={\cal X}^{*}\bar{\nabla}_{a}\left({\cal X}^{-1}\right)^{*}$)
may be regarded as a derivative operator on ${\cal M}_{0}$ that
satisfies the property 
(\ref{eq:property-as-covariant-derivative-on-phys-sp}), there
exists a tensor field $C^{c}_{\;\;ab}$ on ${\cal M}_{0}$ such that 
\begin{equation}
  \bar{\nabla}_{a}\omega_{b}
  = \nabla_{a}\omega_{b} - C^{c}_{\;\;ab} \omega_{c},
\end{equation}
where $\omega_{a}$ is an arbitrary one-form on ${\cal M}_{0}$.
From the property
(\ref{eq:property-as-covariant-derivative-on-phys-sp}) of the
covariant derivative operator $\bar{\nabla}_{a}$ on 
${\cal M}_{\lambda}$, the tensor field $C^{c}_{\;\;ab}$ on
${\cal M}_{0}$ is given by 
\begin{equation}
  C^{c}_{\;\;ab} = \frac{1}{2} \bar{g}^{cd}
  \left(
      \nabla_{a}\bar{g}_{db}
    + \nabla_{b}\bar{g}_{da}
    - \nabla_{d}\bar{g}_{ab}
  \right),
  \label{eq:c-connection}
\end{equation}
where $\bar{g}^{ab}$ is the inverse of $\bar{g}_{ab}$
(see Appendix \ref{sec:derivation-of-pert-Einstein-tensors}).
We note that the gauge dependence of the covariant derivative
$\bar{\nabla}_{a}$ appears only through $C^{c}_{\;\;ab}$.
The Riemann curvature $\bar{R}_{abc}^{\;\;\;\;\;\;d}$ on 
${\cal M}_{\lambda}$, which is also pulled back to 
${\cal M}_{0}$, is given by \cite{Wald-book}:
\begin{equation}
  \bar{R}_{abc}^{\;\;\;\;\;\;d} = R_{abc}^{\;\;\;\;\;\;d}
  - 2 \nabla_{[a}^{} C^{d}_{\;\;b]c}
  + 2 C^{e}_{\;\;c[a} C^{d}_{\;\;b]e},
  \label{eq:phys-riemann-back-riemann-rel}
\end{equation}
where $R_{abc}^{\;\;\;\;\;\;d}$ is the Riemann curvature on ${\cal M}_{0}$.
The perturbative expression for the curvatures are obtained from
the expansion of Eq.~(\ref{eq:phys-riemann-back-riemann-rel})
through the expansion of $C^{c}_{\;\;ab}$.

%*********************************************************************

The first- and the-second order perturbations of the Riemann, the
Ricci, the scalar, the Weyl curvatures, and the Einstein tensors
on the general background spacetime are summarized in
Ref.~\cite{kouchan-second}.  
We also derived the perturbative form of the divergence of an
arbitrary tensor field of second rank to check the
perturbative Bianchi identities in Ref.~\cite{kouchan-second}.
In this paper, we only present the perturbative expression for
the Einstein tensor, and its derivations in Appendix
\ref{sec:derivation-of-pert-Einstein-tensors}.

%*********************************************************************

We expand the Einstein tensor
$\bar{G}_{a}^{\;\;b}:=\bar{R}_{a}^{\;\;b}-\frac{1}{2}\delta_{a}^{\;\;b}\bar{R}$
on ${\cal M}_{\lambda}$ as
\begin{equation}
  \bar{G}_{a}^{\;\;b}
  =
  G_{a}^{\;\;b}
  + \lambda {}^{(1)}\!G_{a}^{\;\;b} 
  + \frac{1}{2} \lambda^{2} {}^{(2)}\!G_{a}^{\;\;b} 
  + O(\lambda^{3}).
\end{equation}
As shown in Appendix
\ref{sec:derivation-of-pert-Einstein-tensors}, each order
perturbation of the Einstein tensor is given by
\begin{eqnarray}
  \label{eq:linear-Einstein}
  {}^{(1)}\!G_{a}^{\;\;b}
  &=&
  {}^{(1)}{\cal G}_{a}^{\;\;b}\left[{\cal H}\right]
  + {\pounds}_{X} G_{a}^{\;\;b}
  ,\\
  \label{eq:second-Einstein-2,0-0,2}
  {}^{(2)}\!G_{a}^{\;\;b}
  &=& 
  {}^{(1)}{\cal G}_{a}^{\;\;b}\left[{\cal L}\right]
  + {}^{(2)}{\cal G}_{a}^{\;\;b} \left[{\cal H}, {\cal H}\right]
  \nonumber\\
  &&
  + 2 {\pounds}_{X} {}^{(1)}\!\bar{G}_{a}^{\;\;b}
  + \left\{ {\pounds}_{Y} - {\pounds}_{X}^{2} \right\} G_{a}^{\;\;b},
\end{eqnarray}
where
\begin{widetext}
\begin{eqnarray}
  \label{eq:cal-G-def-linear}
  {}^{(1)}{\cal G}_{a}^{\;\;b}\left[A\right]
  &:=&
  {}^{(1)}\Sigma_{a}^{\;\;b}\left[A\right]
  - \frac{1}{2} \delta_{a}^{\;\;b} {}^{(1)}\Sigma_{c}^{\;\;c}\left[A\right]
%  \label{eq:(1)Sigma-def-linear}
  , \quad
  {}^{(1)}\Sigma_{a}^{\;\;b}\left[A\right]
  :=
  - 2 \nabla_{[a}^{}H_{d]}^{\;\;\;bd}\left[A\right]
  - A^{cb} R_{ac}
  , \\
  \label{eq:cal-G-def-second}
  {}^{(2)}{\cal G}_{a}^{\;\;b}\left[A, B\right]
  &:=&
  {}^{(2)}\Sigma_{a}^{\;\;b}\left[A, B\right]
  - \frac{1}{2} \delta_{a}^{\;\;b} {}^{(2)}\Sigma_{c}^{\;\;c}\left[A, B\right]
  , \\
  {}^{(2)}\Sigma_{a}^{\;\;b}\left[A, B\right]
  &:=& 
    2 R_{ad} B_{c}^{\;\;(b}A^{d)c}
  + 2 H_{[a}^{\;\;\;de}\left[A\right] H_{d]\;\;e}^{\;\;\;b}\left[B\right]
  + 2 H_{[a}^{\;\;\;de}\left[B\right] H_{d]\;\;e}^{\;\;\;b}\left[A\right]
  \nonumber\\
  &&
  + 2 A_{e}^{\;\;d} \nabla_{[a}H_{d]}^{\;\;\;be}\left[B\right]
  + 2 B_{e}^{\;\;d} \nabla_{[a}H_{d]}^{\;\;\;be}\left[A\right]
  + 2 A_{c}^{\;\;b} \nabla_{[a}H_{d]}^{\;\;\;cd}\left[B\right]
  + 2 B_{c}^{\;\;b} \nabla_{[a}H_{d]}^{\;\;\;cd}\left[A\right]
  \label{eq:(2)Sigma-def-second}
  ,
\end{eqnarray}
and 
\begin{eqnarray}
  H_{ab}^{\;\;\;\;c}\left[A\right]
  &:=&
  \nabla_{(a}^{}A_{b)}^{\;\;\;c}
  - \frac{1}{2} \nabla^{c}_{}A_{ab}
  \label{eq:Habc-def-1}
  , \\
  H_{abc}\left[A\right] 
  &:=&
  g_{cd} H_{ab}^{\;\;\;\;d}\left[A\right]
  ,
  \quad
  H_{a}^{\;\;bc}\left[A\right] 
  := 
  g^{bd} H_{ad}^{\;\;\;\;c}\left[A\right]
  ,
  \quad
%  \nonumber\\
  H_{a\;\;c}^{\;\;b}\left[A\right] 
%  &:=& 
  := 
  g_{cd} H_{a}^{\;\;bd}\left[A\right].
  \label{eq:Habc-def-2}
\end{eqnarray}
We note that ${}^{(1)}{\cal G}_{a}^{\;\;b}\left[*\right]$ and
${}^{(2)}{\cal G}_{a}^{\;\;b}\left[*,*\right]$ in
Eqs.~(\ref{eq:linear-Einstein}) and
(\ref{eq:second-Einstein-2,0-0,2}) are the gauge invariant parts
of the perturbative Einstein tensors, and
Eqs.~(\ref{eq:linear-Einstein}) and
(\ref{eq:second-Einstein-2,0-0,2}) have the same forms as 
Eqs.~(\ref{eq:matter-gauge-inv-def-1.0}) and
(\ref{eq:matter-gauge-inv-decomp-2.0}), respectively.
The expression of 
${}^{(2)}{\cal G}_{a}^{\;\;b}\left[A, B\right]$ in
Eq.~(\ref{eq:cal-G-def-second}) with
Eq.~(\ref{eq:(2)Sigma-def-second}) is derived by the
consideration of the general relativistic gauge-invariant
perturbation theory with two infinitesimal parameters in
Refs.~\cite{kouchan-gauge-inv,kouchan-second}.

%*********************************************************************

We also note that ${}^{(1)}{\cal G}_{a}^{\;\;b}\left[*\right]$
and ${}^{(2)}{\cal G}_{a}^{\;\;b}\left[*,*\right]$ defined by
Eqs.~(\ref{eq:cal-G-def-linear})--(\ref{eq:(2)Sigma-def-second})
satisfy the identities
\begin{eqnarray}
  \nabla_{a}
  {}^{(1)}{\cal G}_{b}^{\;\;a}\left[A\right]
  &=& 
  - H_{ca}^{\;\;\;\;a}\left[A\right] G_{b}^{\;\;c}
  + H_{ba}^{\;\;\;\;c}\left[A\right] G_{c}^{\;\;a}
  \label{eq:linear-order-divergence-of-calGab}
  , \\
  \nabla_{a}{}^{(2)}{\cal G}_{b}^{\;\;a}\left[A, B\right]
  &=& 
  - H_{ca}^{\;\;\;\;a}\left[A\right]
    {}^{(1)}\!{\cal G}_{b}^{\;\;c}\left[B\right]
  - H_{ca}^{\;\;\;\;a}\left[B\right]
    {}^{(1)}\!{\cal G}_{b}^{\;\;c}\left[A\right]
%  \nonumber\\
%  &&
  + H_{ba}^{\;\;\;\;e}\left[A\right]
    {}^{(1)}\!{\cal G}_{e}^{\;\;a}\left[B\right]
  + H_{ba}^{\;\;\;\;e}\left[B\right]
    {}^{(1)}\!{\cal G}_{e}^{\;\;a}\left[A\right]
  \nonumber\\
  &&
  - \left(
    H_{bad}\left[B\right] A^{dc} + H_{bad}\left[A\right] B^{dc}
  \right)
  G_{c}^{\;\;a}
%  \nonumber\\
%  &&
  + \left(
    H_{cad}\left[B\right] A^{ad} + H_{cad}\left[A\right] B^{ad}
  \right)
  G_{b}^{\;\;c},
  \label{eq:second-div-of-calGab-1,1}
\end{eqnarray}
\end{widetext}
for arbitrary tensor fields $A_{ab}$ and $B_{ab}$, respectively.
We can directly confirm these identities without specifying
arbitrary tensors $A_{ab}$ and $B_{ab}$ of the second rank,
respectively. 
This implies that our general framework of the second-order
gauge invariant perturbation theory discussed here gives a
self-consistent formulation of the second-order perturbation
theory. 
These identities (\ref{eq:linear-order-divergence-of-calGab})
and (\ref{eq:second-div-of-calGab-1,1}) guarantee the first- and
second-order perturbations of the Bianchi identity
$\bar{\nabla}_{b}\bar{G}_{a}^{\;\;b}=0$ and are also useful when
we check whether the derived components of
Eqs.~(\ref{eq:cal-G-def-linear}) and (\ref{eq:cal-G-def-second})
are correct.

%*********************************************************************

%%%%%%%%%%%%%%%%%%%%%%%%%%%%%%%%%%%%%%%%%%%%%%%%%%%%%%%%%%%%%%%%%%%%%%
\subsection{Perturbations of the energy momentum tensor and
  Klein-Gordon equation}
\label{sec:Perturbation-of-the-energy-momentum-tensor-and-KG-eq}
%%%%%%%%%%%%%%%%%%%%%%%%%%%%%%%%%%%%%%%%%%%%%%%%%%%%%%%%%%%%%%%%%%%%%%

%*********************************************************************

Here, we consider the perturbations of the energy momentum
tensor of the equation of motion.
As a model of the matter field, we only consider the scalar
field, for simplicity.
Then, equation of motion for a scalar field is the Klein-Gordon
equation.

%*********************************************************************

The energy momentum tensor for a scalar field $\bar{\varphi}$
is given by
\begin{eqnarray}
  \bar{T}_{a}^{\;\;b} = 
  \bar{\nabla}_{a}\bar{\varphi} \bar{\nabla}^{b}\bar{\varphi} 
  - \frac{1}{2} \delta_{a}^{\;\;b}
  \left(
    \bar{\nabla}_{c}\bar{\varphi}\bar{\nabla}^{c}\bar{\varphi}
    + 2 V(\bar{\varphi})
  \right),
  \label{eq:MFB-6.2-again}
\end{eqnarray}
where $V(\bar{\varphi})$ is the potential of the scalar field
$\bar{\varphi}$.
We expand the scalar field $\bar{\varphi}$ as
\begin{eqnarray}
  \bar{\varphi}
  =
  \varphi
  + \lambda \hat{\varphi}_{1}
  + \frac{1}{2} \lambda^{2} \hat{\varphi}_{2}
  + O(\lambda^{3}),
  \label{eq:scalar-field-expansion-second-order}
\end{eqnarray}
where $\varphi$ is the background value of the scalar field
$\bar{\varphi}$.
Further, following to the decomposition formulae
(\ref{eq:matter-gauge-inv-def-1.0}) and
(\ref{eq:matter-gauge-inv-def-2.0}), each order perturbation of
the scalar field $\bar{\varphi}$ is decomposed as
\begin{eqnarray}
  \label{eq:varphi-1-def}
  \hat{\varphi}_{1} &=:& \varphi_{1} + {\pounds}_{X}\varphi, \\
  \label{eq:varphi-2-def}
  \hat{\varphi}_{2} &=:& \varphi_{2} 
  + 2 {\pounds}_{X}\hat{\varphi}_{1} 
  + \left( {\pounds}_{Y} - {\pounds}_{X}^{2} \right) \varphi, 
\end{eqnarray}
where $\varphi_{1}$ and $\varphi_{2}$ are the first- and the
second-order gauge-invariant perturbations of the scalar field,
respectively.

%*********************************************************************

Through the perturbative expansions
(\ref{eq:scalar-field-expansion-second-order}) and
(\ref{eq:inverse-metric-expansion}) of the scalar 
field $\bar{\varphi}$ and the inverse metric, the energy
momentum tensor (\ref{eq:MFB-6.2-again}) is also expanded as
\begin{eqnarray}
  \bar{T}_{a}^{\;\;b} = T_{a}^{\;\;b}
  + 
  \lambda {}^{(1)}\!\left(T_{a}^{\;\;b}\right)
  + 
  \frac{1}{2} \lambda^{2} {}^{(2)}\!\left(T_{a}^{\;\;b}\right)
  + O(\lambda^{3}).
\end{eqnarray}
The background energy momentum tensor $T_{a}^{\;\;b}$ is given
by the replacement $\bar{\varphi}\rightarrow\varphi$ in
Eq.~(\ref{eq:MFB-6.2-again}). 
Further, through the decompositions
(\ref{eq:linear-metric-decomp}),
(\ref{eq:H-ab-in-gauge-X-def-second-1}),
(\ref{eq:varphi-1-def}), and (\ref{eq:varphi-2-def}), the 
perturbations of the energy momentum tensor 
${}^{(1)}\!\left(T_{a}^{\;\;b}\right)$ and
${}^{(2)}\!\left(T_{a}^{\;\;b}\right)$ are also decomposed as
\begin{eqnarray}
  {}^{(1)}\!\left(T_{a}^{\;\;b}\right)
  &=:&
  {}^{(1)}\!{\cal T}_{a}^{\;\;b} + {\pounds}_{X}T_{a}^{\;\;b}
  \label{eq:first-order-energy-momentum-scalar-decomp}
  ,
  \\
  {}^{(2)}\!\left(T_{a}^{\;\;b}\right)
  &=:&
  {}^{(2)}\!{\cal T}_{a}^{\;\;b}
  + 2 {\pounds}_{X}{}^{(1)}\!\left(T_{a}^{\;\;b}\right)
  \nonumber\\
  &&
  + \left( {\pounds}_{Y} - {\pounds}_{X}^{2}\right) T_{a}^{\;\;b},
  \label{eq:second-order-energy-momentum-scalar-decomp}
\end{eqnarray}
where the gauge-invariant parts ${}^{(1)}\!{\cal T}_{a}^{\;\;b}$
and ${}^{(2)}\!{\cal T}_{a}^{\;\;b}$ of the first and the
second order are given by 
\begin{widetext}
\begin{eqnarray}
  {}^{(1)}\!{\cal T}_{a}^{\;\;b}
  &:=& 
  \nabla_{a}\varphi \nabla^{b}\varphi_{1} 
  - \nabla_{a}\varphi {\cal H}^{bc} \nabla_{c}\varphi 
  + \nabla_{a}\varphi_{1} \nabla^{b} \varphi 
  - \delta_{a}^{\;\;b}
  \left(
      \nabla_{c}\varphi\nabla^{c}\varphi_{1}
    - \frac{1}{2} \nabla_{c}\varphi {\cal H}^{dc} \nabla_{d} \varphi 
    + \varphi_{1} \frac{\partial V}{\partial\varphi}
  \right)
  \label{eq:first-order-energy-momentum-scalar-gauge-inv}
  , \\
  {}^{(2)}\!{\cal T}_{a}^{\;\;b}
  &:=&
  \nabla_{a}\varphi \nabla^{b}\varphi_{2} 
  + \nabla_{a}\varphi_{2} \nabla^{b}\varphi  
  - \nabla_{a}\varphi g^{bd} {\cal L}_{dc} \nabla^{c}\varphi
  - 2 \nabla_{a}\varphi {\cal H}^{bc} \nabla_{c}\varphi_{1} 
  + 2 \nabla_{a}\varphi {\cal H}^{bd}{\cal H}_{dc} \nabla^{c}\varphi
  + 2 \nabla_{a}\varphi_{1} \nabla^{b}\varphi_{1}
  \nonumber\\
  && 
  - 2 \nabla_{a}\varphi_{1} {\cal H}^{bc} \nabla_{c}\varphi 
  - \delta_{a}^{\;\;b}
  \left(
      \nabla_{c}\varphi\nabla^{c}\varphi_{2}
    - \frac{1}{2} \nabla^{c}\varphi {\cal L}_{dc}\nabla^{d}\varphi 
    + \nabla^{c}\varphi {\cal H}^{de}{\cal H}_{ec} \nabla_{d}\varphi 
    - 2 \nabla_{c}\varphi {\cal H}^{dc} \nabla_{d}\varphi_{1}
  \right.
  \nonumber\\
  && \quad\quad\quad\quad\quad\quad\quad\quad\quad\quad\quad\quad
  \left.
    + \nabla_{c}\varphi_{1}\nabla^{c}\varphi_{1} 
    + \varphi_{2}\frac{\partial V}{\partial\varphi}
    + \varphi_{1}^{2}\frac{\partial^{2}V}{\partial\varphi^{2}}
  \right)
  \label{eq:second-order-energy-momentum-scalar-gauge-inv}
  .
\end{eqnarray}
\end{widetext}
We note that
Eq.~(\ref{eq:first-order-energy-momentum-scalar-decomp}) and  
(\ref{eq:second-order-energy-momentum-scalar-decomp}) have the
same form as (\ref{eq:matter-gauge-inv-decomp-1.0}) and
(\ref{eq:matter-gauge-inv-decomp-2.0}), respectively.

%*******************************************************************

Next, we consider the perturbation of the Klein-Gordon equation 
\begin{eqnarray}
  \bar{C}_{(K)} := \bar{\nabla}^{a}\bar{\nabla}_{a}\bar{\varphi}
  - \frac{\partial V}{\partial\bar{\varphi}}(\bar{\varphi}) = 0.
  \label{eq:Klein-Gordon-equation}
\end{eqnarray}
Through the perturbative expansions
(\ref{eq:scalar-field-expansion-second-order}) and
(\ref{eq:metric-expansion}), the Klein-Gordon equation
(\ref{eq:Klein-Gordon-equation}) is expanded as
\begin{eqnarray}
  \bar{C}_{(K)} =: C_{(K)} + \lambda \stackrel{(1)}{C_{(K)}} + 
  \frac{1}{2} \lambda^{2} \stackrel{(2)}{C_{(K)}} +
  O(\lambda^{3}). 
\end{eqnarray}
$C_{(K)}$ is the background Klein-Gordon equation
\begin{eqnarray}
  \label{eq:Klein-Gordon-eq-background}
  C_{(K)}
  &:=&
  \nabla_{a}\nabla^{a}\varphi
  - \frac{\partial V}{\partial\bar{\varphi}}(\varphi)
  = 0
  .
\end{eqnarray}
The first- and the second-order perturbations
$\stackrel{(1)}{C_{(K)}}$ and $\stackrel{(2)}{C_{(K)}}$ are also
decomposed into the gauge-invariant and the gauge-variant parts as 
\begin{widetext}
\begin{eqnarray}
  \label{eq:Klein-Gordon-eq-first-decomp}
  &&
  \stackrel{(1)}{C_{(K)}}
  =:
  \stackrel{(1)}{{\cal C}_{(K)}}
  + {\pounds}_{X}C_{(K)},
  \quad
  \stackrel{(2)}{C_{(K)}}
  =:
  \stackrel{(2)}{{\cal C}_{(K)}}
  + 2 {\pounds}_{X}\stackrel{(1)}{C_{(K)}}
  + \left( {\pounds}_{Y} - {\pounds}_{X}^{2} \right) C_{(K)}
  ,
\end{eqnarray}
where
\begin{eqnarray}
  \stackrel{(1)}{{\cal C}_{(K)}}
  &:=&
  \nabla^{a}\nabla_{a}\varphi_{1}
  - H_{a}^{\;\;ac}[{\cal H}]\nabla_{c}\varphi
  - {\cal H}^{ab} \nabla_{a}\nabla_{b}\varphi
  - \varphi_{1} \frac{\partial^{2}V}{\partial\bar{\varphi}^{2}}(\varphi)
  \label{eq:Klein-Gordon-eq-first-gauge-inv-def}
  ,
  \\
  \stackrel{(2)}{{\cal C}_{(K)}}
  &:=&
      \nabla^{a}\nabla_{a}\varphi_{2} 
  -   H_{a}^{\;\;ac}[{\cal L}] \nabla_{c}\varphi
  + 2 H_{a}^{\;\;ad}[{\cal H}] {\cal H}_{cd} \nabla^{c}\varphi
  - 2 H_{a}^{\;\;ac}[{\cal H}] \nabla_{c}\varphi_{1}
  + 2 {\cal H}^{ab} H_{ab}^{\;\;\;\;c}[{\cal H}] \nabla_{c}\varphi
  \nonumber\\
  &&
  -   {\cal L}^{ab} \nabla_{a}\nabla_{b}\varphi
  + 2 {\cal H}^{a}_{\;\;d} {\cal H}^{db} \nabla_{a}\nabla_{b}\varphi
  - 2 {\cal H}^{ab} \nabla_{a}\nabla_{b}\varphi_{1}
  -   \varphi_{2} \frac{\partial^{2}V}{\partial\bar{\varphi}^{2}}(\varphi)
  -   (\varphi_{1})^{2}\frac{\partial^{3}V}{\partial\bar{\varphi}^{3}}(\varphi)
  \label{eq:Klein-Gordon-eq-second-gauge-inv-def}
  .
\end{eqnarray}
\end{widetext}
Here, we note that Eqs.~(\ref{eq:Klein-Gordon-eq-first-decomp})
have the same form as
Eqs.~(\ref{eq:matter-gauge-inv-decomp-1.0}) and
(\ref{eq:matter-gauge-inv-decomp-2.0}).

%*******************************************************************

By virtue of the order by order evaluations of the Klein-Gordon
equation, the first- and the second-order perturbation of the
Klein-Gordon equation are necessarily given in gauge-invariant
form as
\begin{eqnarray}
  \label{eq:Klein-Gordon-eq-first-second-gauge-inv}
  \stackrel{(1)}{{\cal C}_{(K)}} = 0, \quad 
  \stackrel{(2)}{{\cal C}_{(K)}} = 0.
\end{eqnarray}

%*******************************************************************

We should note that, in Ref.~\cite{kouchan-second-cosmo-matter},
we summarized the formulae of the energy momentum tensors for an
perfect fluid, an imperfect fluid, and a scalar field.
Further, we also summarized the equations of motion of these
three matter fields: i.e., the energy continuity equation and
the Euler equation for a perfect fluid; the energy continuity
equation and the Navier-Stokes equation for an imperfect fluid;
the Klein-Gordon equation for a scalar field.
All these formulae also have the same form as the decomposition 
formulae (\ref{eq:matter-gauge-inv-decomp-1.0}) and
(\ref{eq:matter-gauge-inv-decomp-2.0}).
In this sense, we may say that the decomposition formulae
(\ref{eq:matter-gauge-inv-decomp-1.0}) and
(\ref{eq:matter-gauge-inv-decomp-2.0}) are universal.

%*******************************************************************

%%%%%%%%%%%%%%%%%%%%%%%%%%%%%%%%%%%%%%%%%%%%%%%%%%%%%%%%%%%%%%%%%%%%%%
\subsection{Perturbations of the Einstein equation}
\label{sec:Perturbation-of-the-Einstein-equation}
%%%%%%%%%%%%%%%%%%%%%%%%%%%%%%%%%%%%%%%%%%%%%%%%%%%%%%%%%%%%%%%%%%%%%%

%****************************************************************

Finally, we impose the perturbed Einstein equation of each order, 
\begin{equation}
  {}^{(1)}G_{a}^{\;\;b} = 8\pi G \;\; {}^{(1)}T_{a}^{\;\;b},
  \quad
  {}^{(2)}G_{a}^{\;\;b} = 8\pi G \;\; {}^{(2)}T_{a}^{\;\;b}.
\end{equation}
Then, the perturbative Einstein equation is given by 
\begin{eqnarray}
  \label{eq:linear-order-Einstein-equation}
  {}^{(1)}\!{\cal G}_{a}^{\;\;b}\left[{\cal H}\right]
  &=&
  8\pi G {}^{(1)}{\cal T}_{a}^{\;\;b}
\end{eqnarray}
at linear order and
\begin{eqnarray}
  \label{eq:second-order-Einstein-equation}
  {}^{(1)}\!{\cal G}_{a}^{\;\;b}\left[{\cal L}\right]
  + {}^{(2)}\!{\cal G}_{a}^{\;\;b}\left[{\cal H}, {\cal H}\right]
  &=&
  8\pi G \;\; {}^{(2)}{\cal T}_{a}^{\;\;b} 
\end{eqnarray}
at second order.
These explicitly show that, order by order, the Einstein
equations are necessarily given in terms of gauge invariant
variables only.

%*********************************************************************

Together with Eqs.~(\ref{eq:Klein-Gordon-eq-first-second-gauge-inv}),
we have seen that the first- and the second-order perturbations
of the Einstein equations and the Klein-Gordon equation are
necessarily given in gauge-invariant form.
This implies that we do not have to consider the gauge degree of 
freedom, at least in the level where we concentrate only on the
equations of the system.

%*********************************************************************

We have reviewed the general outline of the second-order gauge
invariant perturbation theory.
We also note that the ingredients of this section are
independent of the explicit form of the background metric
$g_{ab}$, except for the decomposition assumption
(\ref{eq:linear-metric-decomp}) for the linear-order metric
perturbations and are valid not only in cosmological
perturbation case but also the other generic situations if
Eq.~(\ref{eq:linear-metric-decomp}) is correct.
Within this general framework, we develop a second-order
cosmological perturbation theory in terms of the gauge invariant
variables.

%*******************************************************************

%%%%%%%%%%%%%%%%%%%%%%%%%%%%%%%%%%%%%%%%%%%%%%%%%%%%%%%%%%%%%%%%%%%%%
%%%%%%%%%%%%%%%%%%%%%%%%%%%%%%%%%%%%%%%%%%%%%%%%%%%%%%%%%%%%%%%%%%%%%
\section{Cosmological background spacetime and equations}
\label{sec:Cosmological-Background-spacetime-equations}
%%%%%%%%%%%%%%%%%%%%%%%%%%%%%%%%%%%%%%%%%%%%%%%%%%%%%%%%%%%%%%%%%%%%%
%%%%%%%%%%%%%%%%%%%%%%%%%%%%%%%%%%%%%%%%%%%%%%%%%%%%%%%%%%%%%%%%%%%%%

The background spacetime ${\cal M}_{0}$ considered in
cosmological perturbation theory is a homogeneous, isotropic
universe that is foliated by the three-dimensional hypersurface
$\Sigma(\eta)$, which is parametrized by $\eta$.
Each hypersurface of $\Sigma(\eta)$ is a maximally symmetric
three-space\cite{Weinberg1972}, and the spacetime metric of this
universe is given by
\begin{eqnarray}
  g_{ab} = a^{2}(\eta)\left(
    - (d\eta)_{a}(d\eta)_{b}
    + \gamma_{ij}(dx^{i})_{a}(dx^{j})_{b}
  \right),
  \label{eq:background-metric}
\end{eqnarray}
where $a=a(\eta)$ is the scale factor, $\gamma_{ij}$ is the
metric on the maximally symmetric 3-space with curvature
constant $K$, and the indices $i,j,k,...$ for the spatial
components run from 1 to 3.

%*******************************************************************

To study the Einstein equation for this background spacetime,
we introduce the energy-momentum tensor for a scalar field,
which is given by 
\begin{eqnarray}
  T_{a}^{\;\;b}
  &=&
  \nabla_{a}\varphi\nabla^{b}\varphi -
  \frac{1}{2}\delta_{a}^{\;\;b}\left(\nabla_{c}\varphi\nabla^{c}\varphi +
    2V(\varphi)\right)
  \label{eq:energy-momentum-single-scalar}
  \\
  &=&
  -
  \left(
      \frac{1}{2a^{2}} (\partial_{\eta}\varphi)^{2}
    + V(\varphi)
  \right)
  (d\eta)_{a} \left(\frac{\partial}{\partial\eta}\right)^{b}
  \nonumber\\
  && \quad
  +
  \left(
    \frac{1}{2a^{2}} (\partial_{\eta}\varphi)^{2}
    - V(\varphi)
  \right)
  \gamma_{a}^{\;\;b},
  \label{eq:energy-momentum-single-scalar-homogeneous}
\end{eqnarray}
where we assumed that the scalar field $\varphi$ is
homogeneous
\begin{eqnarray}
  \label{eq:background-varphi-is-homogeneous}
  \varphi=\varphi(\eta)
\end{eqnarray}
and $\gamma_{a}^{\;\;b}$ are defined as 
\begin{eqnarray}
  \gamma_{ab} := \gamma_{ij}(dx^{i})_{a}(dx^{j})_{b}, \;\;
  \gamma_{a}^{\;\;b}:=\gamma_{i}^{\;\;j}(dx^{i})_{a}(\partial/\partial x^{j})^{b}.
  \label{eq:gammaab-gammaab-def}
\end{eqnarray}

%*******************************************************************

The background Einstein equations 
$G_{a}^{\;\;b}=8\pi GT_{a}^{\;\;b}$ for this background
spacetime filled with the single scalar field are given by
\begin{eqnarray}
  \label{eq:background-Einstein-equations-scalar-1}
  &&\!\!\!\!\!\!\!\!\!\!\!\!\!\!\!\! 
  {\cal H}^{2} + K = \frac{8 \pi G}{3} a^{2} \left(
    \frac{1}{2a^{2}} (\partial_{\eta}\varphi)^{2} + V(\varphi)
  \right)
  ,\\
  \label{eq:background-Einstein-equations-scalar-2}
  &&\!\!\!\!\!\!\!\!\!\!\!\!\!\!\!\! 
  2 \partial_{\eta}{\cal H} + {\cal H}^{2} + K = 8 \pi G 
  \left(-\frac{1}{2} (\partial_{\eta}\varphi)^{2} + a^{2} V(\varphi)\right)
  .
\end{eqnarray}
We also note that the equations
(\ref{eq:background-Einstein-equations-scalar-1}) and
(\ref{eq:background-Einstein-equations-scalar-2}) lead to 
\begin{eqnarray}
  \label{eq:background-Einstein-equations-scalar-3}
  {\cal H}^{2} + K - \partial_{\eta}{\cal H}
  = 4\pi G (\partial_{\eta}\varphi)^{2}.
\end{eqnarray}
Equation (\ref{eq:background-Einstein-equations-scalar-3})
is also useful when we derive the perturbative Einstein
equations.

%*******************************************************************

Next, we consider the background Klein-Gordon equation which is
the equation of motion $\nabla_{a}T_{b}^{\;\;a}=0$ for the
scalar field
\begin{eqnarray}
  \label{eq:background-Klein-Gordon-equation}
  \partial_{\eta}^{2}\varphi + 2 {\cal H} \partial_{\eta}\varphi
  +   a^{2} \frac{\partial V}{\partial\varphi}
  = 0.
\end{eqnarray}
The Klein-Gordon equation
(\ref{eq:background-Klein-Gordon-equation}) is also derived from 
the Einstein equations
(\ref{eq:background-Einstein-equations-scalar-1}) and 
(\ref{eq:background-Einstein-equations-scalar-2}).
This is a well known fact and is just due to the Bianchi
identity of the background spacetime.
However, these types of relation are useful to check whether the
derived system of equations is consistent.

%*******************************************************************

%%%%%%%%%%%%%%%%%%%%%%%%%%%%%%%%%%%%%%%%%%%%%%%%%%%%%%%%%%%%%%%%%%%%%
%%%%%%%%%%%%%%%%%%%%%%%%%%%%%%%%%%%%%%%%%%%%%%%%%%%%%%%%%%%%%%%%%%%%%
%%%%%%%%%%%%%%%%%%%%%%%%%%%%%%%%%%%%%%%%%%%%%%%%%%%%%%%%%%%%%%%%%%%%%
\section{Equations for the first-order cosmological perturbations}
\label{sec:Equations-for-the-first-order-cosmological-perturbations}
%%%%%%%%%%%%%%%%%%%%%%%%%%%%%%%%%%%%%%%%%%%%%%%%%%%%%%%%%%%%%%%%%%%%%
%%%%%%%%%%%%%%%%%%%%%%%%%%%%%%%%%%%%%%%%%%%%%%%%%%%%%%%%%%%%%%%%%%%%%
%%%%%%%%%%%%%%%%%%%%%%%%%%%%%%%%%%%%%%%%%%%%%%%%%%%%%%%%%%%%%%%%%%%%%

%*******************************************************************

On the cosmological background spacetime in the last section, we
develop the perturbation theory in the gauge-invariant manner.
In this section, we summarize the first-order perturbation of
the Einstein equation and the Klein-Gordon equations.
In Sec.~\ref{sec:Gauge-invariant-metric-perturbations}, we show
that the assumption on the decomposition
(\ref{eq:linear-metric-decomp}) of the linear-order metric
perturbation is correct.
In Sec.~\ref{sec:First-order-Einstein-equations}, we summarize
the first-order perturbation of the Einstein equation.
Finally, in Sec.~\ref{sec:First-order-Klein-Gordon-equations},
we show the first-order perturbation of the Klein-Gordon
equation.

%*******************************************************************

%%%%%%%%%%%%%%%%%%%%%%%%%%%%%%%%%%%%%%%%%%%%%%%%%%%%%%%%%%%%%%%%%%%%%
\subsection{Gauge-invariant metric perturbations}
\label{sec:Gauge-invariant-metric-perturbations}
%%%%%%%%%%%%%%%%%%%%%%%%%%%%%%%%%%%%%%%%%%%%%%%%%%%%%%%%%%%%%%%%%%%%%

%*******************************************************************

Here, we consider the first-order metric perturbation $h_{ab}$
and show the assumption on the decomposition
(\ref{eq:linear-metric-decomp}) is correct in the background
metric Eq.~(\ref{eq:background-metric}).
To accomplish the decomposition (\ref{eq:linear-metric-decomp}), 
first, we assume the existence of the Green functions
$\Delta^{-1}:=(D^{i}D_{i})^{-1}$, 
$(\Delta + 2 K)^{-1}$, and $(\Delta + 3 K)^{-1}$, where $D_{i}$
is the covariant derivative associated with the metric
$\gamma_{ij}$ and $K$ is the curvature constant of the
maximally symmetric three space.
Next, we consider the decomposition of the linear-order metric 
perturbation $h_{ab}$ as
\begin{eqnarray}
  h_{ab}
  &=&
  h_{\eta\eta}(d\eta)_{a}(d\eta)_{b}
  \nonumber\\
  &&
  + 2 \left(
    D_{i}h_{(VL)} + h_{(V)i}
  \right)(d\eta)_{(a}(dx^{i})_{b)}
  \\
  &&
  + a^{2} \left\{
    h_{(L)} \gamma_{ij}
    + \left(D_{i}D_{j} - \frac{1}{3}\gamma_{ij}\Delta\right)h_{(TL)}
  \right.
  \nonumber\\
  && \quad\quad\quad
  \left.
    + 2 D_{(i}h_{(TV)j)} + {h_{(TT)ij}}
  \right\} (dx^{i})_{a}(dx^{j})_{b},
  \nonumber
\end{eqnarray}
where $h_{(V)i}$, $h_{(TV)j}$, and ${h_{(TT)ij}}$ satisfy the
properties 
\begin{eqnarray}
  &&
  D^{i}h_{(V)i} = 0, \quad  
  D^{i} h_{(TV)i} = 0, 
  \nonumber\\
  &&
  h_{(TT)ij} = h_{(TT)ji}, \quad
  {h_{(T)}}^{i}_{\;\;i} := \gamma^{ij}{h_{(T)}}_{ij} = 0,
  \label{eq:bare-parturbative-property}
  \\
  &&
  D^{i} h_{(TT)ij} = 0.
  \nonumber
\end{eqnarray}
The gauge-transformation rules for the variables $h_{\eta\eta}$,
$h_{(VL)}$, $h_{(V)i}$, $h_{(L)}$, $h_{(TL)}$, $h_{(TV)j}$ and
${h_{(TT)ij}}$ are derived from
Eq.~(\ref{eq:first-order-gauge-trans-of-metric}).
Inspecting these gauge-transformation rules, we define the
gauge-variant part $X_{a}$ in
Eq.~(\ref{eq:linear-metric-decomp}):
\begin{eqnarray}
  X_{a}
  &:=&
  \left(
    h_{(VL)} - \frac{1}{2} a^{2}\partial_{\eta}h_{(TL)} 
  \right) (d\eta)_{a}
  \nonumber\\
  &&
  +
  a^{2} \left(
      h_{(TV)i}
    + \frac{1}{2} D_{i}h_{(TL)}
  \right)
  (dx^{i})_{a}.
\end{eqnarray}
We can easily check this vector field $X_{a}$ satisfies
Eq.~(\ref{eq:linear-metric-decomp-gauge-trans}).
Subtracting gauge variant-part ${\pounds}_{X}g_{ab}$ from
$h_{ab}$, we have the gauge-invariant part ${\cal H}_{ab}$ in
Eq.~(\ref{eq:linear-metric-decomp}):
\begin{eqnarray}
  {\cal H}_{ab}
  &=& a^{2} \left\{
    - 2 \stackrel{(1)}{\Phi} (d\eta)_{a}(d\eta)_{b}
    + 2 \stackrel{(1)}{\nu}_{i} (d\eta)_{(a}(dx^{i})_{b)}
  \right.
  \nonumber\\
  && \quad
  \left.
    + 
    \left( - 2 \stackrel{(1)}{\Psi} \gamma_{ij} 
      + \stackrel{(1)}{\chi}_{ij} \right)
    (dx^{i})_{a}(dx^{j})_{b}
  \right\},
  \label{eq:components-calHab}
\end{eqnarray}
where the properties $D^{i}\stackrel{(1)}{\nu}_{i}$ $:=$
$\gamma^{ij}D_{i}\stackrel{(1)}{\nu}_{j}$ $=$
$\stackrel{(1)}{\chi_{i}^{\;\;i}}$ $:=$
$\gamma^{ij}\stackrel{(1)}{\chi_{ij}}$ $=$ 
$D^{i}\stackrel{(1)}{\chi}_{ij} = 0$ are satisfied as
consequences of Eqs.~(\ref{eq:bare-parturbative-property}).

%*******************************************************************

Thus, we may say that our assumption for the decomposition 
(\ref{eq:linear-metric-decomp}) in linear-order metric
perturbation is correct in the case of cosmological
perturbations.
However, we have to note that to accomplish
Eq.~(\ref{eq:linear-metric-decomp}), we assumed the existence of
the Green functions $\Delta^{-1}$, $(\Delta + 2 K)^{-1}$, and
$(\Delta + 3 K)^{-1}$.
As shown in Ref.~\cite{kouchan-cosmo-second}, this assumption is 
necessary to guarantee the one to one correspondence between the
variables $\{h_{\eta\eta},h_{i\eta},h_{ij}\}$ and
$\{h_{\eta\eta},h_{(VL)},h_{(V)i},h_{(L)},h_{(TL)},h_{(TV)j},{h_{(TT)ij}}\}$,
but excludes some perturbative modes of the metric perturbations
which belong to the kernel of the operator $\Delta$,
$(\Delta+2K)$, and $(\Delta+3 K)$ from our consideration.
For example, homogeneous modes belong to the kernel of the
operator $\Delta$ and are excluded from our consideration. 
If we have to treat these modes, the separate treatments are
necessary.
In this paper, we ignore these modes, for simplicity.

%*******************************************************************

We also note the fact that the definition
(\ref{eq:linear-metric-decomp}) of the gauge-invariant variables
is not unique.
This comes from the fact that we can always construct new
gauge-invariant quantities by the combination of the
gauge-invariant variables.
For example,  using the gauge-invariant variables
$\stackrel{(1)}{\Phi}$ and $\stackrel{(1)}{\nu_{i}}$ of the
first-order metric perturbation, we can define a vector field
$Z_{a}$ by $Z_{a} := - a \stackrel{(1)}{\Phi} (d\eta)_{a} + a
\stackrel{(1)}{\nu_{i}} (dx^{i})_{a}$ which is gauge-invariant.
Then, we can rewrite the decomposition formula
(\ref{eq:linear-metric-decomp}) for the linear-order metric
perturbation as
\begin{eqnarray}
  h_{ab}
  &=&
  {\cal H}_{ab} - {\pounds}_{Z}g_{ab}
  + {\pounds}_{Z}g_{ab} + {\pounds}_{X}g_{ab},
  \nonumber\\
  &=:&
  {\cal K}_{ab} + {\pounds}_{X+Z}g_{ab},
  \label{eq:non-uniqueness-of-gauge-invariant-metric-perturbation}
\end{eqnarray}
where we have defined new gauge-invariant variable 
${\cal K}_{ab}$ by 
${\cal K}_{ab}:={\cal H}_{ab}-{\pounds}_{Z}g_{ab}$.
Clearly, ${\cal K}_{ab}$ is gauge-invariant and the vector field
$X^{a}+Z^{a}$ satisfies
Eq.~(\ref{eq:linear-metric-decomp-gauge-trans}).
In spite of this non-uniqueness, we specify the components of the
tensor ${\cal H}_{ab}$ as Eq.~(\ref{eq:components-calHab}),
which is the gauge-invariant part of the linear-order metric
perturbation associated with the longitudinal gauge.

%*******************************************************************

The non-uniqueness of the definitions of gauge-invariant
variables is related to the ``gauge-fixing'' for the linear-order
metric perturbations.
Due to this non-uniqueness, we can consider the gauge-fixing in
the first-order metric perturbation from two different points of
view.
The first point of view is that the gauge-fixing is to specify
the gauge-variant part $X^{a}$.
For example, the longitudinal gauge is realized by the gauge
fixing $X^{a}=0$.
Due to this gauge fixing $X^{a}=0$, we can regard the fact that
perturbative variables in the longitudinal gauge are the
completely gauge fixed variables.
On the other hand, we may also regard that the gauge fixing is
the specification of the gauge-invariant vector field $Z^{a}$ in
Eq.~(\ref{eq:non-uniqueness-of-gauge-invariant-metric-perturbation}).
In this point of view, we do not specify the vector field
$X^{a}$.
Instead, we have to specify the gauge-invariant vector $Z^{a}$
or equivalently to specify the gauge-invariant metric
perturbation ${\cal K}_{ab}$ without specifying $X^{a}$ so that
the first-order metric perturbation $h_{ab}$ coincides with the
gauge-invariant variables ${\cal K}_{ab}$ when we fix the gauge
$X^{a}$ so that $X^{a}+Z^{a}=0$.
These two different point of view of ``gauge fixing'' is
equivalent with each other due to the non-uniqueness of the
definition
(\ref{eq:non-uniqueness-of-gauge-invariant-metric-perturbation})
of the gauge-invariant variables.

%*******************************************************************

%%%%%%%%%%%%%%%%%%%%%%%%%%%%%%%%%%%%%%%%%%%%%%%%%%%%%%%%%%%%%%%%%%%%%
\subsection{First-order Einstein equations}
\label{sec:First-order-Einstein-equations}
%%%%%%%%%%%%%%%%%%%%%%%%%%%%%%%%%%%%%%%%%%%%%%%%%%%%%%%%%%%%%%%%%%%%%

%*******************************************************************

Here, we derive the linear-order Einstein equation
(\ref{eq:linear-order-Einstein-equation}).
To derive the components of the gauge invariant part of the
linearized Einstein tensor 
${}^{(1)}{\cal G}_{a}^{\;\;b}\left[{\cal H}\right]$, which is
defined by Eqs.~(\ref{eq:cal-G-def-linear}), we first derive the
components of the tensor 
$H_{ab}^{\;\;\;\;c}\left[{\cal H}\right]$, which is defined in
Eq.~(\ref{eq:Habc-def-1}) with $A_{ab} = {\cal H}_{ab}$ and its
component (\ref{eq:components-calHab}).
These components are summarized in Ref.~\cite{kouchan-cosmo-second}.

%*******************************************************************

From Eq.~(\ref{eq:cal-G-def-linear}), the component of 
${}^{(1)}{\cal G}_{a}^{\;\;b}\left[{\cal H}\right]$ are
summarized as 
\begin{widetext}
\begin{eqnarray}
  {}^{(1)}{\cal G}_{\eta}^{\;\;\eta}\left[{\cal H}\right]
  &=&
  - \frac{1}{a^{2}} \left\{
    \left(
      - 6 {\cal H} \partial_{\eta}
      + 2 \Delta
      + 6 K
    \right) \stackrel{(1)}{\Psi}
    - 6 {\cal H}^{2} \stackrel{(1)}{\Phi}
  \right\}
  \label{eq:kouchan-10.120}
  , \\
  {}^{(1)}{\cal G}_{i}^{\;\;\eta}\left[{\cal H}\right]
  &=&
  - \frac{1}{a^{2}}
  \left(
    2 \partial_{\eta} D_{i} \stackrel{(1)}{\Psi}
    + 2 {\cal H} D_{i} \stackrel{(1)}{\Phi} 
    - \frac{1}{2} \left(
      \Delta
      + 2 K
    \right)
    \stackrel{(1)\;\;}{\nu_{i}}  
  \right)
  \label{eq:kouchan-10.121}
  , \\
  {}^{(1)}{\cal G}_{\eta}^{\;\;i}\left[{\cal H}\right] 
  &=&
  \frac{1}{a^{2}} \left\{
    2 \partial_{\eta} D^{i} \stackrel{(1)}{\Psi}
    + 2 {\cal H} D^{i} \stackrel{(1)}{\Phi} 
    + \frac{1}{2} \left(
      - \Delta
      + 2 K
      + 4 {\cal H}^{2}
      - 4 \partial_{\eta}{\cal H}
    \right)
    \stackrel{(1)\;\;}{\nu^{i}}
  \right\}
  \label{eq:kouchan-10.122}
  , \\
  {}^{(1)}{\cal G}_{i}^{\;\;j}\left[{\cal H}\right]
  &=& 
  \frac{1}{a^{2}} \left[
    D_{i} D^{j} \left(\stackrel{(1)}{\Psi} - \stackrel{(1)}{\Phi}\right)
%  \right.
%  \nonumber\\
%  &&
%  \left.
    + 
    \left\{
      \left(
        -   \Delta
        + 2 \partial_{\eta}^{2} 
        + 4 {\cal H} \partial_{\eta}
        - 2 K
      \right)
      \stackrel{(1)}{\Psi}
      + \left(
          2 {\cal H} \partial_{\eta}
        + 4 \partial_{\eta}{\cal H}
        + 2 {\cal H}^{2}
        + \Delta
      \right)
      \stackrel{(1)}{\Phi}
    \right\}
    \gamma_{i}^{\;\;j}
  \right.
  \nonumber\\
  && \quad\quad
  \left.
    - \frac{1}{2a^{2}} \partial_{\eta} \left\{
      a^{2} \left( 
        D_{i} \stackrel{(1)\;\;}{\nu^{j}} + D^{j} \stackrel{(1)\;\;}{\nu_{i}}
      \right)
    \right\}
%  \right.
%  \nonumber\\
%  &&
%  \left.
    + \frac{1}{2} \left(
      \partial_{\eta}^{2}
      + 2 {\cal H} \partial_{\eta}
      + 2 K
      - \Delta
    \right) \stackrel{(1)\;\;\;\;}{\chi_{i}^{\;\;j}}
  \right]
  .
  \label{eq:kouchan-10.123}
\end{eqnarray}
\end{widetext}
Straightforward calculations show that these components of the
first-order gauge invariant perturbation
${}^{(1)}{\cal G}_{a}^{\;\;b}\left[{\cal H}\right]$ of the
Einstein tensor satisfies the identity
(\ref{eq:linear-order-divergence-of-calGab}).
Although this confirmation is also possible without
specification of the tensor ${\cal H}_{ab}$, the confirmation of
Eq.~(\ref{eq:linear-order-divergence-of-calGab}) through the
explicit components
(\ref{eq:kouchan-10.120})--(\ref{eq:kouchan-10.123}) implies
that we have derived the components of 
${}^{(1)}{\cal G}_{a}^{\;\;b}\left[{\cal H}\right]$ consistently.

%*******************************************************************

Next, we summarize the first-order perturbation of the energy
momentum tensor for a scalar field. 
Since, at the background level, we assume that the scalar field
$\varphi$ is homogeneous as
Eq.~(\ref{eq:background-varphi-is-homogeneous}), the components
of the gauge-invariant part of the first-order energy-momentum
tensor ${}^{(1)}\!{\cal T}_{a}^{\;\;b}$ are
given by 
\begin{widetext}
\begin{eqnarray}
  {}^{(1)}\!{\cal T}_{\eta}^{\;\;\eta}
  &=& 
  - \frac{1}{a^{2}} \left(
      \partial_{\eta}\varphi_{1}\partial_{\eta}\varphi
    - \stackrel{(1)}{\Phi} (\partial_{\eta}\varphi)^{2}
    + a^{2}\frac{dV}{d\varphi}\varphi_{1}
  \right)
  \label{eq:kouchan-10.161}
  ,
  \quad
  {}^{(1)}\!{\cal T}_{i}^{\;\;\eta}
  = 
  - \frac{1}{a^{2}}
  D_{i}\varphi_{1}\partial_{\eta}\varphi,
%  \label{eq:kouchan-10.162}
  \\
  {}^{(1)}\!{\cal T}_{\eta}^{\;\;i}
  &=& 
  \frac{1}{a^{2}} \partial_{\eta}\varphi 
  \left(
    D^{i}\varphi_{1}
    + (\partial_{\eta}\varphi) \stackrel{(1)\;\;}{\nu^{i}}
  \right)
  ,
  \label{eq:kouchan-10.163}
  \quad
  {}^{(1)}\!{\cal T}_{i}^{\;\;j}
  = 
  \frac{1}{a^{2}} \gamma_{i}^{\;\;j}\left(
      \partial_{\eta}\varphi_{1} \partial_{\eta}\varphi
    - \stackrel{(1)}{\Phi} (\partial_{\eta}\varphi)^{2}
    - a^{2} \frac{dV}{d\varphi} \varphi_{1}
  \right)
  .
%  \label{eq:kouchan-10.165}
\end{eqnarray}
\end{widetext}
The second equation in (\ref{eq:kouchan-10.163}) shows that
there is no anisotropic stress in the energy-momentum tensor of
the single scalar field.
Then, we obtain
\begin{eqnarray}
  \label{eq:absence-of-anisotropic-stress-Einstein-i-j-traceless-scalar}
  \stackrel{(1)}{\Phi} = \stackrel{(1)}{\Psi}.
\end{eqnarray}
From
Eqs.~(\ref{eq:kouchan-10.120})--(\ref{eq:kouchan-10.163}) and
(\ref{eq:absence-of-anisotropic-stress-Einstein-i-j-traceless-scalar}),
the components of scalar parts of the linearized Einstein equation
(\ref{eq:linear-order-Einstein-equation}) are given
as\cite{Mukhanov-Feldman-Brandenberger-1992}
\begin{eqnarray}
  &&
  \left(
        \Delta
    - 3 {\cal H} \partial_{\eta}
    + 4 K
    - \partial_{\eta}{\cal H}
    - 2 {\cal H}^{2}
  \right) \stackrel{(1)}{\Phi}
  \nonumber\\
  && \quad\quad\quad\quad
  = 
  4 \pi G \left(
    \partial_{\eta}\varphi_{1} \partial_{\eta}\varphi
    + a^{2}\frac{dV}{d\varphi}\varphi_{1}
  \right)
  \label{eq:kouchan-18.185}
  , \\
  &&
  \partial_{\eta}\stackrel{(1)}{\Phi} + {\cal H} \stackrel{(1)}{\Phi}
  =
  4 \pi G \varphi_{1} \partial_{\eta}\varphi
  \label{eq:kouchan-18.186}
  , \\
  &&
  \left(
        \partial_{\eta}^{2} 
    + 3 {\cal H} \partial_{\eta}
    +   \partial_{\eta}{\cal H}
    + 2 {\cal H}^{2}
  \right)
  \stackrel{(1)}{\Phi}
  \nonumber\\
  && \quad\quad\quad\quad =
  4 \pi G
  \left(
    \partial_{\eta}\varphi_{1} \partial_{\eta}\varphi
    - a^{2} \frac{dV}{d\varphi} \varphi_{1}
  \right)
  \label{eq:kouchan-18.187}
  .
\end{eqnarray}
In the derivation of
Eqs.~(\ref{eq:kouchan-18.185})--(\ref{eq:kouchan-18.187}), we
have used Eq.~(\ref{eq:background-Einstein-equations-scalar-3}).
We also note that only two of these equations are independent.
Further, the vector part of the component 
${}^{(1)}\!{\cal G}_{i}^{\;\;\eta}\left[{\cal H}\right]=8\pi G{}^{(1)}\!{\cal T}_{i}^{\;\;\eta}$ 
shows that
\begin{eqnarray}
  \label{eq:no-first-order-vector-mode-scalar-field-case}
  \stackrel{(1)}{\nu}_{i} = 0.
\end{eqnarray}
The equation for the tensor mode $\stackrel{(1)\;\;}{\chi_{ij}}$
is given by 
\begin{eqnarray}
  \label{eq:linearized-Einstein-i-j-traceless-tensor}
  \left(
        \partial_{\eta}^{2}
    + 2 {\cal H} \partial_{\eta}
    + 2 K
    - \Delta
  \right) \stackrel{(1)\;\;\;\;}{\chi_{i}^{\;\;j}}
  =
  0
  .
\end{eqnarray}

%*******************************************************************

Combining Eqs.~(\ref{eq:kouchan-18.185}) and
(\ref{eq:kouchan-18.187}), we eliminate the potential term of
the scalar field and thereby obtain
\begin{eqnarray}
  \left(
        \partial_{\eta}^{2} 
    +   \Delta
    + 4 K
  \right) \stackrel{(1)}{\Phi}
  = 
  8 \pi G \partial_{\eta}\varphi_{1} \partial_{\eta}\varphi.
  \label{eq:scalar-linearized-Einstein-scalar-master-eq-pre}
\end{eqnarray}
Further, using Eq.~(\ref{eq:kouchan-18.186}) to express
$\partial_{\eta}\varphi_{1}$ in terms of
$\partial_{\eta}\stackrel{(1)}{\Phi}$ and
$\stackrel{(1)}{\Phi}$, we also eliminate
$\partial_{\eta}\varphi_{1}$ in
Eq.~(\ref{eq:scalar-linearized-Einstein-scalar-master-eq-pre}).
Hence, we have
\begin{widetext}
\begin{eqnarray}
  \label{eq:scalar-linearized-Einstein-scalar-master-eq}
  \left\{
    \partial_{\eta}^{2}
    + 2 \left(
      {\cal H}
      - \frac{2\partial_{\eta}^{2}\varphi}{\partial_{\eta}\varphi}
    \right) \partial_{\eta}
    - \Delta
    - 4 K
    + 2 
    \left( \partial_{\eta}{\cal H}
      - \frac{{\cal H}\partial_{\eta}^{2}\varphi}{\partial_{\eta}\varphi}
    \right)
  \right\} \stackrel{(1)}{\Phi}
  = 
  0.
\end{eqnarray}
\end{widetext}
This is the master equation for the scalar mode perturbation of
the cosmological perturbation in universe filled with a single
scalar field.
It is also known that
Eq.~(\ref{eq:scalar-linearized-Einstein-scalar-master-eq})
reduces to a simple equation through a change of variables
\cite{Mukhanov-Feldman-Brandenberger-1992}.

%*******************************************************************

%%%%%%%%%%%%%%%%%%%%%%%%%%%%%%%%%%%%%%%%%%%%%%%%%%%%%%%%%%%%%%%%%%%%%
\subsection{First-order Klein-Gordon equations}
\label{sec:First-order-Klein-Gordon-equations}
%%%%%%%%%%%%%%%%%%%%%%%%%%%%%%%%%%%%%%%%%%%%%%%%%%%%%%%%%%%%%%%%%%%%%

%*******************************************************************

Next, we consider the first-order perturbation of the
Klein-Gordon equation
(\ref{eq:Klein-Gordon-eq-first-gauge-inv-def}). 
By the straightforward calculations using
Eqs.~(\ref{eq:background-metric}), (\ref{eq:components-calHab}),
(\ref{eq:background-varphi-is-homogeneous}),
(\ref{eq:background-Klein-Gordon-equation}), and the components
$H_{a}^{\;\;ac}$ summarized in Ref.~\cite{kouchan-cosmo-second},
the gauge-invariant part $\stackrel{(1)}{{\cal C}_{(K)}}$ of the
first-order Klein-Gordon equation defined by
Eq.~(\ref{eq:Klein-Gordon-eq-first-gauge-inv-def}) is given by 
\begin{eqnarray}
  - a^{2} \stackrel{(1)}{{\cal C}_{(K)}}
  &=&
      \partial_{\eta}^{2}\varphi_{1}
  + 2 {\cal H} \partial_{\eta}\varphi_{1}
  -   \Delta\varphi_{1}
  \nonumber\\
  &&
  - \left(
        \partial_{\eta}\stackrel{(1)}{\Phi}
    + 3 \partial_{\eta}\stackrel{(1)}{\Psi}
  \right) \partial_{\eta}\varphi
  \nonumber\\
  &&
  + 2 a^{2} \stackrel{(1)}{\Phi} \frac{\partial V}{\partial\bar{\varphi}}(\varphi)
  +   a^{2}\varphi_{1} \frac{\partial^{2}V}{\partial\bar{\varphi}^{2}}(\varphi)
  \nonumber\\
  &=& 0
  \label{eq:kouchan-17.806-first-explicit}
  .
\end{eqnarray}

%*******************************************************************

Through the background Einstein equations
(\ref{eq:background-Einstein-equations-scalar-1}),
(\ref{eq:background-Einstein-equations-scalar-2}), and the
first-order perturbations (\ref{eq:kouchan-18.186}) and
(\ref{eq:scalar-linearized-Einstein-scalar-master-eq}) of the
Einstein equation, we can easily derive the first-order
perturbation of the Klein-Gordon equation
(\ref{eq:kouchan-17.806-first-explicit})\cite{kouchan-second-cosmo-consistency}.
Hence, the first-order perturbation of the Klein-Gordon equation
is not independent of the background and the first-order
perturbation of the Einstein equation.
Therefore, from the viewpoint of the Cauchy problem, any
information obtained from the first-order perturbation of the
Klein-Gordon equation should also be obtained from the set of
the background and the first-order the Einstein equation, in
principle.

%*******************************************************************

%%%%%%%%%%%%%%%%%%%%%%%%%%%%%%%%%%%%%%%%%%%%%%%%%%%%%%%%%%%%%%%%%%%%%
%%%%%%%%%%%%%%%%%%%%%%%%%%%%%%%%%%%%%%%%%%%%%%%%%%%%%%%%%%%%%%%%%%%%%
%%%%%%%%%%%%%%%%%%%%%%%%%%%%%%%%%%%%%%%%%%%%%%%%%%%%%%%%%%%%%%%%%%%%%
\section{Equations for the second-order cosmological perturbations}
\label{sec:Equations-for-the-second-order-cosmological-perturbations}
%%%%%%%%%%%%%%%%%%%%%%%%%%%%%%%%%%%%%%%%%%%%%%%%%%%%%%%%%%%%%%%%%%%%%
%%%%%%%%%%%%%%%%%%%%%%%%%%%%%%%%%%%%%%%%%%%%%%%%%%%%%%%%%%%%%%%%%%%%%
%%%%%%%%%%%%%%%%%%%%%%%%%%%%%%%%%%%%%%%%%%%%%%%%%%%%%%%%%%%%%%%%%%%%%

%*******************************************************************

Now, we develop the second-order perturbation theory on the 
cosmological background spacetime in
Sec.~\ref{sec:Cosmological-Background-spacetime-equations}
within the general framework of the gauge-invariant perturbation
theory reviewed in
Sec.~\ref{sec:General-framework-of-GR-GI-perturbation-theory}.
Since we have already confirm the important step of our general
framework, i.e., the assumption for the decomposition
(\ref{eq:linear-metric-decomp}) of the linear-order metric
perturbation is correct.
Hence, the general framework reviewed in
Sec.~\ref{sec:General-framework-of-GR-GI-perturbation-theory}
is applicable.
Applying this framework, we define the second-order gauge
invariant variables of the metric perturbation in
Sec.~\ref{sec:Second-order-gauge-invariant-metric-variables}.
In Sec.~\ref{sec:Second-order-gauge-invariant-energy-momentum},
we summarize the explicit components of the gauge invariant
parts of the second-order perturbation of the Einstein tensor.
In
Sec.~\ref{sec:Second-order-gauge-invariant-energy-momentum-Klein-Gordon},
we summarize the explicit components of the second-order
perturbation of the energy-momentum tensor and the Klein-Gordon
equations.
Then, in Sec.~\ref{sec:Secnd-order-cosmological-Einstein-equations},
we derive the second-order Einstein equations in terms of
gauge-invariant variables.
The resulting equations have the source terms which constitute
of the quadratic terms of the linear-order perturbations.
Although these source terms have complicated forms, we give
identities which comes from the consistency of all the
second-order perturbations of the Einstein equation and the
Klein-Gordon equation in
Sec.~\ref{sec:Consistency-of-the-second-order-perturbations}.

%*******************************************************************

%%%%%%%%%%%%%%%%%%%%%%%%%%%%%%%%%%%%%%%%%%%%%%%%%%%%%%%%%%%%%%%%%%%%%
\subsection{Gauge-invariant metric perturbations}
\label{sec:Second-order-gauge-invariant-metric-variables}
%%%%%%%%%%%%%%%%%%%%%%%%%%%%%%%%%%%%%%%%%%%%%%%%%%%%%%%%%%%%%%%%%%%%%

%*******************************************************************

First, we consider the components of the gauge invariant
variables for the metric perturbation of second order.
The variable $\hat{L}_{ab}$ defined by Eq.~(\ref{eq:Lhatab-def})
is transformed as Eq.~(\ref{eq:kouchan-4.67}) under the gauge
transformation and we may regard the generator $\sigma_{a}$
defined by Eq.~(\ref{eq:sigma-def}) as an arbitrary vector field
on ${\cal M}_{0}$ from the fact that the generator $\xi_{2}^{a}$ in
Eq.~(\ref{eq:sigma-def}) is arbitrary.
We can apply the procedure to find gauge invariant variables for
the first-order metric perturbations
(\ref{eq:components-calHab}) in
Sec.~\ref{sec:Gauge-invariant-metric-perturbations}.
Then, we can accomplish the decomposition
(\ref{eq:Lhatab-decomposition}).
Following to the same argument as in the linear case, we may
choose the components of the gauge invariant variables 
${\cal L}_{ab}$ in Eq.~(\ref{eq:H-ab-in-gauge-X-def-second-1}) as
\begin{eqnarray}
  {\cal L}_{ab}
  &=& 
  - 2 a^{2} \stackrel{(2)}{\Phi} (d\eta)_{a}(d\eta)_{b}
  + 2 a^{2} \stackrel{(2)\;\;}{\nu_{i}} (d\eta)_{(a}(dx^{i})_{b)}
  \nonumber\\
  && \quad
  + a^{2} 
  \left( - 2 \stackrel{(2)}{\Psi} \gamma_{ij} 
    + \stackrel{(2)\;\;\;\;}{{\chi}_{ij}} \right)
  (dx^{i})_{a}(dx^{j})_{b},
  \label{eq:second-order-gauge-inv-metrc-pert-components}
\end{eqnarray}
where $\stackrel{(2)}{\nu}_{i}$ and
$\stackrel{(2)\;\;\;\;}{\chi_{ij}}$ satisfy the equations 
\begin{eqnarray}
  && D^{i}\stackrel{(2)\;\;}{\nu_{i}} = 0, \quad
  \stackrel{(2)\;\;\;\;}{\chi^{i}_{\;\;i}} = 0, \quad
   D^{i}\stackrel{(2)\;\;\;\;}{\chi_{ij}} = 0.
\end{eqnarray}
The gauge invariant variables $\stackrel{(2)}{\Phi}$ and
$\stackrel{(2)}{\Psi}$ are the scalar mode perturbations of
second order, and $\stackrel{(2)\;\;}{\nu_{i}}$ and
$\stackrel{(2)\;\;\;\;}{\chi_{ij}}$ are the second-order vector
and tensor modes of the metric perturbations, respectively.

%*******************************************************************

Here, we also note the fact that the decomposition
(\ref{eq:H-ab-in-gauge-X-def-second-1}) is not unique.
This situation is similar to the case of the linear-order, but
more complicated.
In the definition of the gauge invariant variables of the
second-order metric perturbation, we may replace 
\begin{eqnarray}
  \label{eq:non-uniqueness-of-gauge-invariant-2nd-order-metric-perturbation-1}
  X^{a} = X^{'a} - Z^{'a},
\end{eqnarray}
where $Z^{'a}$ is gauge invariant and $X^{'a}$ is transformed as 
\begin{eqnarray}
  \label{eq:non-uniqueness-of-gauge-invariant-2nd-order-metric-perturbation-2}
  {}_{{\cal Y}}\!X^{'a} - {}_{{\cal X}}\!X^{'a} = \xi^{a}_{1}
\end{eqnarray}
under the gauge transformation ${\cal X}_{\lambda}$
$\rightarrow$ ${\cal Y}_{\lambda}$. 
This $Z^{'a}$ may be different from the vector $Z^{a}$ in
Eq.~(\ref{eq:non-uniqueness-of-gauge-invariant-metric-perturbation}). 
By the replacement
(\ref{eq:non-uniqueness-of-gauge-invariant-2nd-order-metric-perturbation-1}), 
the second-order metric perturbation
(\ref{eq:H-ab-in-gauge-X-def-second-1}) is given in the form
\begin{eqnarray}
  l_{ab}
  &=:&
  {\cal J}_{ab}
  + 2 {\pounds}_{X'} h_{ab}
  + \left(
      {\pounds}_{Y'}
    - {\pounds}_{X'}^{2}
  \right)
  g_{ab},
  \label{eq:non-uniqueness-of-gauge-invariant-2nd-order-metric-perturbation-4}
\end{eqnarray}
where we defined
\begin{eqnarray}
  {\cal J}_{ab}
  &:=&
  {\cal L}_{ab}
  -   {\pounds}_{W}g_{ab}
  - 2 {\pounds}_{Z'}{\cal K}_{ab}
  \nonumber\\
  &&
  - 2 {\pounds}_{Z'}{\pounds}_{Z}g_{ab}
  +   {\pounds}_{Z'}^{2}g_{ab}
  \label{eq:calJ-def}
  , \\
  Y^{'a} &:=& Y^{a}+W^{a}+[X',Z']^{a}
  .
  \label{eq:Y-prime-transform}
\end{eqnarray}
Here, the vector field $W^{a}$ in
Eq.~(\ref{eq:Y-prime-transform}) constitute of some components 
of gauge invariant second-order metric perturbation 
${\cal L}_{ab}$ like $Z^{a}$ in
Eq.~(\ref{eq:non-uniqueness-of-gauge-invariant-metric-perturbation}). 
The tensor field ${\cal J}_{ab}$ is manifestly gauge invariant.
The gauge transformation rule of the new gauge-variant part
$Y^{'a}$ of the second-order metric perturbation is given by
\begin{eqnarray}
  {}_{{\cal Y}}\!Y^{'a}
  -
  {}_{{\cal X}}\!Y^{'a}
  &=&
  \xi_{(2)}^{a} + [\xi_{(1)},X']^{a}
  .
\end{eqnarray}
Although
Eq.~(\ref{eq:non-uniqueness-of-gauge-invariant-2nd-order-metric-perturbation-4}) 
is similar to Eq.~(\ref{eq:H-ab-in-gauge-X-def-second-1}),
the tensor fields ${\cal L}_{ab}$ and ${\cal J}_{ab}$ are
different from each other.
Thus, the definition of the gauge invariant variables for the
second-order metric perturbation is not unique in a more
complicated way than the linear order.
This non-uniqueness of gauge-invariant variables for the metric
perturbations propagates to the definition
(\ref{eq:matter-gauge-inv-def-1.0}) and
(\ref{eq:matter-gauge-inv-def-2.0}) of the gauge invariant
variables for matter fields.

%*******************************************************************

In spite of the existence of infinitely many definitions of the
gauge invariant variables, in this  paper, we consider the
components of ${\cal L}_{ab}$ given by 
Eq.~(\ref{eq:second-order-gauge-inv-metrc-pert-components}).
Eq.~(\ref{eq:second-order-gauge-inv-metrc-pert-components})
corresponds to the second-order extension of the longitudinal
gauge, which is called Poisson gauge $X^{a}=Y^{a}=0$.

%*******************************************************************

%%%%%%%%%%%%%%%%%%%%%%%%%%%%%%%%%%%%%%%%%%%%%%%%%%%%%%%%%%%%%%%%%%%%%
\subsection{Einstein tensor}
\label{sec:Second-order-gauge-invariant-energy-momentum}
%%%%%%%%%%%%%%%%%%%%%%%%%%%%%%%%%%%%%%%%%%%%%%%%%%%%%%%%%%%%%%%%%%%%%

%*******************************************************************

Here, we evaluate the second-order perturbation of the Einstein
tensor (\ref{eq:second-Einstein-2,0-0,2}) with the cosmological
background (\ref{eq:background-metric}). 
We evaluate the term 
${}^{(1)}\!{\cal G}_{a}^{\;\;b}\left[{\cal L}\right]$ and
${}^{(2)}\!{\cal G}_{a}^{\;\;b}\left[{\cal H}, {\cal H}\right]$ 
in the Einstein equation
(\ref{eq:second-order-Einstein-equation}).

%*******************************************************************

First, we evaluate the term
${}^{(1)}\!{\cal G}_{a}^{\;\;b}\left[{\cal L}\right]$ in the
Einstein equation (\ref{eq:second-order-Einstein-equation}).
Because the components
(\ref{eq:second-order-gauge-inv-metrc-pert-components}) of
${\cal L}_{ab}$ are obtained through the replacements
\begin{eqnarray}
  \label{eq:replacement-from-calHab-to-calLab}
  \stackrel{(1)}{\Phi} \rightarrow \stackrel{(2)}{\Phi}, \quad
  \stackrel{(1)\;\;}{\nu_{i}} \rightarrow \stackrel{(2)\;\;}{\nu}_{i}, \quad
  \stackrel{(1)}{\Psi} \rightarrow \stackrel{(2)}{\Psi}, \quad
  \stackrel{(1)\;\;\;\;}{\chi_{ij}} \rightarrow
  \stackrel{(2)\;\;\;\;}{\chi_{ij}} 
\end{eqnarray}
in the components (\ref{eq:components-calHab}) of 
${\cal H}_{ab}$, we easily obtain the components of 
${}^{(1)}{\cal G}_{a}^{\;\;b}\left[{\cal L}\right]$ through
the replacements (\ref{eq:replacement-from-calHab-to-calLab}) in
Eqs.~(\ref{eq:kouchan-10.120})--(\ref{eq:kouchan-10.123}).

%*******************************************************************

From Eq.~(\ref{eq:components-calHab}), we can derive the
components of 
${}^{(2)}\!{\cal G}_{a}^{\;\;b}={}^{(2)}\!{\cal G}_{a}^{\;\;b}[{\cal H},{\cal H}]$
defined by
Eqs.~(\ref{eq:cal-G-def-second})--(\ref{eq:Habc-def-2}) in a
straightforward manner.
Here, we use the results
(\ref{eq:absence-of-anisotropic-stress-Einstein-i-j-traceless-scalar})
and (\ref{eq:no-first-order-vector-mode-scalar-field-case}) of
the first-order Einstein equations, for simplicity.
Then the explicit components
${}^{(2)}\!{\cal G}_{a}^{\;\;b}={}^{(2)}\!{\cal G}_{a}^{\;\;b}[{\cal H},{\cal H}]$
are summarized as
\begin{widetext}
\begin{eqnarray}
  {}^{(2)}\!{\cal G}_{\eta}^{\;\;\eta}
  &=&
  \frac{2}{a^{2}}
  \left[
    -  3 D_{k}\stackrel{(1)}{\Phi} D^{k}\stackrel{(1)}{\Phi}
    -  8 \stackrel{(1)}{\Phi} \Delta\stackrel{(1)}{\Phi}
    -  3 \left(\partial_{\eta}\stackrel{(1)}{\Phi}\right)^{2}
    - 12 \left({\cal H}^{2} + K\right) \left(\stackrel{(1)}{\Phi}\right)^{2}
    +             D_{l}D_{k}\stackrel{(1)}{\Phi} \stackrel{(1)}{\chi^{lk}}
  \right.
  \nonumber\\
  && \quad\quad
  \left.
    + \frac{1}{8} \partial_{\eta}\stackrel{(1)}{\chi^{kl}}
    \left(
      \partial_{\eta}
      + 8 {\cal H}
    \right) \stackrel{(1)}{\chi_{kl}}
    + \frac{1}{2} D_{k}\stackrel{(1)}{\chi_{lm}} D^{[l}\stackrel{(1)}{\chi^{k]m}}
    - \frac{1}{8} D_{k}\stackrel{(1)}{\chi_{lm}} D^{k}\stackrel{(1)}{\chi^{ml}}
    - \frac{1}{2} \stackrel{(1)}{\chi^{lm}} \left(
      \Delta - K
    \right) \stackrel{(1)}{\chi_{lm}}
  \right]
  ,
  \label{eq:generic-2-calG-eta-eta}
  \\
%\end{eqnarray}
%\begin{eqnarray}
  {}^{(2)}\!{\cal G}_{\eta}^{\;\;i}
  &=&
  \frac{2}{a^{2}}
  \left[
               8  \stackrel{(1)}{\Phi} \partial_{\eta}D^{i}\stackrel{(1)}{\Phi}
    -             D_{j}\stackrel{(1)}{\Phi} \partial_{\eta}\stackrel{(1)}{\chi^{ji}}
    -             \left(
      \partial_{\eta}D_{j}\stackrel{(1)}{\Phi}
      + 2 {\cal H} D_{j}\stackrel{(1)}{\Phi}
    \right) \stackrel{(1)}{\chi^{ij}}
    + \frac{1}{4} \partial_{\eta}\stackrel{(1)}{\chi_{jk}} D^{i}\stackrel{(1)}{\chi^{kj}}
    + \stackrel{(1)}{\chi_{kl}} \partial_{\eta}D^{[i}\stackrel{(1)}{\chi^{k]l}}
  \right]
  ,
  \label{eq:generic-2-calG-eta-i}
  \\
%\end{eqnarray}
%\begin{eqnarray}
  {}^{(2)}\!{\cal G}_{i}^{\;\;\eta}
  &=&
  \frac{2}{a^{2}}
  \left[
      8 {\cal H} \stackrel{(1)}{\Phi} D_{i}\stackrel{(1)}{\Phi}
    - 2 D_{i}\stackrel{(1)}{\Phi} \partial_{\eta}\stackrel{(1)}{\Phi}
    +   D^{j}\stackrel{(1)}{\Phi} \partial_{\eta}\stackrel{(1)}{\chi_{ij}}
    -   \partial_{\eta}D^{j}\stackrel{(1)}{\Phi} \stackrel{(1)}{\chi_{ij}}
    - \frac{1}{4} \partial_{\eta}\stackrel{(1)}{\chi^{kj}} D_{i}\stackrel{(1)}{\chi_{kj}}
    +   \stackrel{(1)}{\chi^{kj}} \partial_{\eta}D_{[j}\stackrel{(1)}{\chi_{i]k}}
  \right]
  ,
  \label{eq:generic-2-calG-i-eta}
  \\
%\end{eqnarray}
%\begin{eqnarray}
  {}^{(2)}\!{\cal G}_{i}^{\;\;j}
  &=&
  \frac{2}{a^{2}}
  \left[
    \left\{
      -          3  D_{k}\stackrel{(1)}{\Phi} D^{k}\stackrel{(1)}{\Phi}
      - 4 \stackrel{(1)}{\Phi} \left(
        \Delta + K
      \right) \stackrel{(1)}{\Phi}
      - \partial_{\eta}\stackrel{(1)}{\Phi} \partial_{\eta}\stackrel{(1)}{\Phi}
      -          8  {\cal H} \stackrel{(1)}{\Phi} \partial_{\eta}\stackrel{(1)}{\Phi}
      - 4 \left(
          2 \partial_{\eta}{\cal H}
        +   {\cal H}^{2}
      \right) \left(\stackrel{(1)}{\Phi}\right)^{2}
    \right\} \gamma_{i}^{\;\;j}
  \right.
  \nonumber\\
  && \quad\quad
  \left.
    +          2  D_{i}\stackrel{(1)}{\Phi} D^{j}\stackrel{(1)}{\Phi}
    +          4  \stackrel{(1)}{\Phi} D_{i}D^{j}\stackrel{(1)}{\Phi}
    + \stackrel{(1)}{\chi_{i}^{\;\;j}} \left(
          \partial_{\eta}^{2}
      + 2 {\cal H} \partial_{\eta}
    \right) \stackrel{(1)}{\Phi}
    + D_{k}\stackrel{(1)}{\Phi}
    \left(
      D_{i}\stackrel{(1)}{\chi^{jk}} + D^{j}\stackrel{(1)}{\chi_{ik}}
    \right)
  \right.
  \nonumber\\
  && \quad\quad
  \left.
    - 2 D^{k}\stackrel{(1)}{\Phi} D_{k}\stackrel{(1)}{\chi_{i}^{\;\;j}}
    - 2 \stackrel{(1)}{\Phi} \left(
      \Delta - 2 K
    \right) \stackrel{(1)}{\chi_{i}^{\;\;j}}
    -             \Delta \stackrel{(1)}{\Phi} \stackrel{(1)}{\chi_{i}^{\;\;j}}
    +             D_{k}D_{i}\stackrel{(1)}{\Phi} \stackrel{(1)}{\chi^{jk}}
    +             D^{m}D^{j}\stackrel{(1)}{\Phi} \stackrel{(1)}{\chi_{im}}
    -             D_{l}D_{k}\stackrel{(1)}{\Phi} \stackrel{(1)}{\chi^{lk}} \gamma_{i}^{\;\;j}
  \right.
  \nonumber\\
  && \quad\quad
  \left.
    - \frac{1}{2} \partial_{\eta}\stackrel{(1)}{\chi_{ik}} \partial_{\eta}\stackrel{(1)}{\chi^{kj}}
    + D_{k}\stackrel{(1)}{\chi_{il}} D^{[k}\stackrel{(1)}{\chi^{l]j}}
    + \frac{1}{4} D^{j}\stackrel{(1)}{\chi_{lk}} D_{i}\stackrel{(1)}{\chi^{lk}}
    + \frac{1}{2} \stackrel{(1)}{\chi_{lm}} D_{i}D^{j}\stackrel{(1)}{\chi^{ml}}
    - \frac{1}{2} \stackrel{(1)}{\chi_{lm}} D^{l}D_{i}\stackrel{(1)}{\chi^{mj}}
  \right.
  \nonumber\\
  && \quad\quad
  \left.
    - \frac{1}{2} \stackrel{(1)}{\chi^{lm}} D_{l}D^{j}\stackrel{(1)}{\chi_{mi}}
    + \frac{1}{2} \stackrel{(1)}{\chi^{lm}} D_{m}D_{l}\stackrel{(1)}{\chi_{i}^{\;\;j}}
    - \frac{1}{2} \stackrel{(1)}{\chi^{jk}}  \left(
          \partial_{\eta}^{2}
      + 2 {\cal H} \partial_{\eta}
      -   \Delta
      + 2 K
    \right) \stackrel{(1)}{\chi_{ik}}
  \right.
  \nonumber\\
  && \quad\quad
  \left.
    + \frac{1}{2} \left\{
        \frac{3}{4} \partial_{\eta}\stackrel{(1)}{\chi_{lk}} \partial_{\eta}\stackrel{(1)}{\chi^{kl}}
      +   \stackrel{(1)}{\chi_{kl}} \left(
            \partial_{\eta}^{2}
        + 2 {\cal H} \partial_{\eta}
        -   \Delta
        +   K
      \right) \stackrel{(1)}{\chi^{lk}}
      - \frac{1}{4} D_{k}\stackrel{(1)}{\chi_{lm}} D^{k}\stackrel{(1)}{\chi^{ml}}
    \right.
  \right.
  \nonumber\\
  && \quad\quad\quad\quad\quad
  \left.
    \left.
      +             D_{k}\stackrel{(1)}{\chi_{lm}} D^{[l}\stackrel{(1)}{\chi^{k]m}}
    \right\} \gamma_{i}^{\;\;j}
  \right]
  .
  \label{eq:generic-2-calG-i-j}
\end{eqnarray}
\end{widetext}
We have checked the identity (\ref{eq:second-div-of-calGab-1,1})
through
Eqs.~(\ref{eq:generic-2-calG-eta-eta})--(\ref{eq:generic-2-calG-i-j}), 
Then, we may say that the expressions
(\ref{eq:generic-2-calG-eta-eta})--(\ref{eq:generic-2-calG-i-j})
are self-consistent.

%*******************************************************************

%%%%%%%%%%%%%%%%%%%%%%%%%%%%%%%%%%%%%%%%%%%%%%%%%%%%%%%%%%%%%%%%%%%%%
\subsection{Energy-momentum tensor and Klein-Gordon equation}
\label{sec:Second-order-gauge-invariant-energy-momentum-Klein-Gordon}
%%%%%%%%%%%%%%%%%%%%%%%%%%%%%%%%%%%%%%%%%%%%%%%%%%%%%%%%%%%%%%%%%%%%%

%*******************************************************************

Here, we summarize the explicit components of the gauge-invariant
parts (\ref{eq:second-order-energy-momentum-scalar-gauge-inv})
of the second-order perturbation of energy momentum tensor for a
single scalar field in terms of gauge-invariant variables.
Through Eqs.~(\ref{eq:background-varphi-is-homogeneous}),
(\ref{eq:components-calHab}),
(\ref{eq:second-order-gauge-inv-metrc-pert-components}), the
components of 
Eq.~(\ref{eq:second-order-energy-momentum-scalar-gauge-inv}) are
derived by the straightforward calculations.
In this paper, we just summarize the components of 
${}^{(2)}\!{\cal T}_{a}^{b}$ in the situation where the
first-order Einstein equations
(\ref{eq:absence-of-anisotropic-stress-Einstein-i-j-traceless-scalar})
and
(\ref{eq:no-first-order-vector-mode-scalar-field-case}) are
satisfied:
\begin{widetext}
\begin{eqnarray}
  a^{2} {}^{(2)}\!{\cal T}_{\eta}^{\;\;\eta}
  &=&
  -   \partial_{\eta}\varphi \partial_{\eta}\varphi_{2}
  +   (\partial_{\eta}\varphi)^{2} \stackrel{(2)}{\Phi}
  -   a^{2} \varphi_{2}\frac{\partial V}{\partial\varphi}
  + 4 \partial_{\eta}\varphi \stackrel{(1)}{\Phi} \partial_{\eta}\varphi_{1}
  - 4 (\partial_{\eta}\varphi)^{2} \left(\stackrel{(1)}{\Phi}\right)^{2}
  -   (\partial_{\eta}\varphi_{1})^{2}
  \nonumber\\
  && 
  -   D_{i}\varphi_{1} D^{i}\varphi_{1} 
  -   a^{2} (\varphi_{1})^{2} \frac{\partial^{2}V}{\partial\varphi^{2}}
  \label{eq:kouchan-19.209}
  ,
  \\
  a^{2} {}^{(2)}\!{\cal T}_{i}^{\;\;\eta}
  &=&
  - \partial_{\eta}\varphi D_{i}\varphi_{2}
  + 4 \partial_{\eta}\varphi D_{i}\varphi_{1} \stackrel{(1)}{\Phi}
  - 2 D_{i}\varphi_{1} \partial_{\eta}\varphi_{1}
  \label{eq:kouchan-19.211}
  ,
  \\
  a^{2} {}^{(2)}\!{\cal T}_{\eta}^{\;\;i}
  &=&
  \partial_{\eta}\varphi D^{i}\varphi_{2}
  + 2 \partial_{\eta}\varphi_{1} D^{i}\varphi_{1}
  + 4 \partial_{\eta}\varphi \stackrel{(1)}{\Phi} D^{i}\varphi_{1}
  - 2 \partial_{\eta}\varphi \stackrel{(1)}{\chi^{il}} D_{l}\varphi_{1}
  \label{eq:kouchan-19.210}
  ,
  \\
  a^{2} {}^{(2)}\!{\cal T}_{i}^{\;\;j}
  &=&
  D_{i}\varphi_{1} D^{j}\varphi_{1}
%  \nonumber\\
%  &&
  + \frac{1}{2} \gamma_{i}^{\;\;j}
  \left\{
    \partial_{\eta}\varphi \partial_{\eta}\varphi_{2}
    - 4 \partial_{\eta}\varphi \stackrel{(1)}{\Phi} \partial_{\eta}\varphi_{1} 
    + 4 (\partial_{\eta}\varphi)^{2} \left(\stackrel{(1)}{\Phi}\right)^{2}
    -   (\partial_{\eta}\varphi)^{2} \stackrel{(2)}{\Phi}
    +   (\partial_{\eta}\varphi_{1})^{2}
  \right.
  \nonumber\\
  && \quad\quad\quad\quad\quad\quad\quad\quad\quad\quad
  \left.
    -   D_{l}\varphi_{1} D^{l}\varphi_{1} 
    -   a^{2} \varphi_{2} \frac{\partial V}{\partial\varphi}
    -   a^{2} (\varphi_{1})^{2} \frac{\partial^{2}V}{\partial\varphi^{2}}
  \right\}
  \label{eq:kouchan-19.212}
  .
\end{eqnarray}
\end{widetext}
More generic formulae for the components of 
${}^{(2)}\!{\cal T}_{a}^{b}$ are given in
Ref.~\cite{kouchan-second-cosmo-matter}.

%*******************************************************************

Next, we show the gauge-invariant second-order the Klein-Gordon
equation.
We only consider the simple situation where
Eqs.~(\ref{eq:absence-of-anisotropic-stress-Einstein-i-j-traceless-scalar}) 
and (\ref{eq:no-first-order-vector-mode-scalar-field-case}) are
satisfied.
The formulae for more generic situation is given in
Ref.~\cite{kouchan-second-cosmo-matter}. 
Through Eqs.~(\ref{eq:components-calHab}),
(\ref{eq:second-order-gauge-inv-metrc-pert-components}),
(\ref{eq:background-varphi-is-homogeneous}), the second-order
perturbation of the Klein-Gordon equation
(\ref{eq:Klein-Gordon-eq-second-gauge-inv-def}) is given by
\begin{eqnarray}
  - a^{2} \stackrel{(2)}{{\cal C}_{(K)}}
  &=&
       \partial_{\eta}^{2}\varphi_{2} 
  +  2 {\cal H} \partial_{\eta}\varphi_{2} 
  -    \Delta\varphi_{2} 
  \nonumber\\
  && 
  -    \left(
    \partial_{\eta}\stackrel{(2)}{\Phi}
    +  3 \partial_{\eta}\stackrel{(2)}{\Psi}
  \right) \partial_{\eta}\varphi
  \nonumber\\
  && 
  +  2 a^{2} \stackrel{(2)}{\Phi} \frac{\partial V}{\partial\bar{\varphi}}(\varphi)
  +    a^{2}\varphi_{2}\frac{\partial^{2}V}{\partial\bar{\varphi}^{2}}(\varphi)
  \nonumber\\
  && 
  - \Xi_{(K)}
  \nonumber\\
  &=& 0
  ,
  \label{eq:Klein-Gordon-eq-second-gauge-inv-explicit}
\end{eqnarray}
where we defined
\begin{eqnarray}
  \Xi_{(K)}
  &:=&
     8 \partial_{\eta}\stackrel{(1)}{\Phi} \partial_{\eta}\varphi_{1}
  +  8 \stackrel{(1)}{\Phi} \Delta\varphi_{1}
  -  4 a^{2} \stackrel{(1)}{\Phi} \varphi_{1} \frac{\partial^{2}V}{\partial\bar{\varphi}^{2}}(\varphi)
  \nonumber\\
  &&
  -    a^{2} (\varphi_{1})^{2}\frac{\partial^{3}V}{\partial\bar{\varphi}^{3}}(\varphi)
  +  8 \stackrel{(1)}{\Phi} \partial_{\eta}\stackrel{(1)}{\Phi} \partial_{\eta}\varphi
  \nonumber\\
  && 
  -  2 \stackrel{(1)}{\chi^{ij}} D_{j}D_{i}\varphi_{1}
  +   \partial_{\eta}\varphi \stackrel{(1)}{\chi^{ij}} \partial_{\eta}\stackrel{(1)}{\chi_{ij}}
  .
  \label{eq:Klein-Gordon-eq-second-gauge-inv-explicit-source}
\end{eqnarray}

%*******************************************************************

In Eq.~(\ref{eq:Klein-Gordon-eq-second-gauge-inv-explicit}),
$\Xi_{(K)}$ is the source term which is the collection of the
quadratic terms of the linear-order perturbations in the
second-order perturbation of the Klein-Gordon equation.
If we ignore this source term,
Eq.~(\ref{eq:Klein-Gordon-eq-second-gauge-inv-explicit})
coincide with the first-order perturbation of the Klein-Gordon
equation.
From this source term
(\ref{eq:Klein-Gordon-eq-second-gauge-inv-explicit-source}) of
the Klein-Gordon equation, we can see that the mode-mode
coupling due to the non-linear effects appear in the
second-order Klein-Gordon equation.

%*******************************************************************

We cannot discuss solutions to
Eq.~(\ref{eq:Klein-Gordon-eq-second-gauge-inv-explicit}) only
through this equation, since this includes metric perturbations.
To determine the behavior of the metric perturbations, we have
to treat the Einstein equations simultaneously.
The second-order Einstein equation is shown in
Sec.~\ref{sec:Secnd-order-cosmological-Einstein-equations}.

%*******************************************************************

%%%%%%%%%%%%%%%%%%%%%%%%%%%%%%%%%%%%%%%%%%%%%%%%%%%%%%%%%%%%%%%%%%%%%
\subsection{Einstein equations}
\label{sec:Secnd-order-cosmological-Einstein-equations}
%%%%%%%%%%%%%%%%%%%%%%%%%%%%%%%%%%%%%%%%%%%%%%%%%%%%%%%%%%%%%%%%%%%%%

%*******************************************************************

Here, we show the all components of the second-order Einstein
equation (\ref{eq:second-order-Einstein-equation}).
All components of Eq.~(\ref{eq:second-order-Einstein-equation})
are summarized as
\begin{eqnarray}
  &&
  \left(
    - 3 {\cal H} \partial_{\eta}
    +   \Delta
    + 3 K
  \right) \stackrel{(2)}{\Psi}
%  \nonumber\\
%  && \quad
  +
  \left(
    -   \partial_{\eta}{\cal H}
    - 2 {\cal H}^{2}
    +   K
  \right)
  \stackrel{(2)}{\Phi} 
  \nonumber\\
  && \quad
  - 4 \pi G 
  \left(
        \partial_{\eta}\varphi \partial_{\eta}\varphi_{2} 
    +   a^{2} \varphi_{2} \frac{\partial V}{\partial\varphi}
  \right)
  =
  \Gamma_{0}
  ,
  \\
  \label{eq:kouchan-18.218}
%\end{eqnarray}
%\begin{eqnarray}
  &&
    2 \partial_{\eta} D_{i} \stackrel{(2)}{\Psi}
  + 2 {\cal H} D_{i} \stackrel{(2)}{\Phi}
  - \frac{1}{2} \left(
    \Delta
    + 2 K
  \right)
  \stackrel{(2)}{\nu_{i}}
  \nonumber\\
  &&
  -
  8\pi G D_{i}\varphi_{2} \partial_{\eta}\varphi  
  = 
  \Gamma_{i}
  ,
  \label{eq:kouchan-18.199}
\end{eqnarray}
\begin{widetext}
\begin{eqnarray}
  && 
  D_{i} D_{j} \left( \stackrel{(2)}{\Psi} - \stackrel{(2)}{\Phi} \right)
%  \nonumber\\
%  && 
  + 
  \left\{
    \left(
      -   \Delta
      + 2 \partial_{\eta}^{2} 
      + 4 {\cal H} \partial_{\eta}
      - 2 K
    \right)
    \stackrel{(2)}{\Psi}
%  \right.
%  \nonumber\\
%  && \quad\quad
%  \left.
    + \left(
        2 {\cal H} \partial_{\eta}
      + 2 \partial_{\eta}{\cal H}
      + 4 {\cal H}^{2}
      + \Delta
      + 2 K
    \right)
    \stackrel{(2)}{\Phi}
  \right\}
  \gamma_{ij}
  \nonumber\\
  &&
  - \frac{1}{a^{2}} \partial_{\eta} \left(
    a^{2} D_{(i} \stackrel{(2)}{\nu_{j)}}
  \right)
%  \nonumber\\
%  &&
  + \frac{1}{2} \left(
    \partial_{\eta}^{2}
    + 2 {\cal H} \partial_{\eta}
    + 2 K
    - \Delta
  \right) \stackrel{(2)}{\chi}_{ij}
%  \nonumber\\
%  &&
  - 8 \pi G \left(
    \partial_{\eta}\varphi\partial_{\eta}\varphi_{2}
    - a^{2} \varphi_{2}\frac{\partial V}{\partial\varphi}(\varphi)
  \right) \gamma_{ij} = \Gamma_{ij}
  \label{eq:kouchan-18.207}
  ,
\end{eqnarray}
where $\Gamma_{0}$, $\Gamma_{i}$ $\Gamma_{ij}$ are the
collection of the quadratic term of the first-order
perturbations as follows:
\begin{eqnarray}
  \Gamma_{0}
  &:=&
    4 \pi G \left(
        (\partial_{\eta}\varphi_{1})^{2}
    +   D_{i}\varphi_{1} D^{i}\varphi_{1} 
    +   a^{2} (\varphi_{1})^{2} \frac{\partial^{2}V}{\partial\varphi^{2}}
  \right)
%  \nonumber\\
%  && \quad\quad
  -          4  \partial_{\eta}{\cal H} \left(\stackrel{(1)}{\Phi}\right)^{2}
  -          2  \stackrel{(1)}{\Phi} \partial_{\eta}^{2}\stackrel{(1)}{\Phi}
  -          3  D_{k}\stackrel{(1)}{\Phi} D^{k}\stackrel{(1)}{\Phi}
  -         10  \stackrel{(1)}{\Phi} \Delta\stackrel{(1)}{\Phi}
  \nonumber\\
  &&
  -          3  \left(\partial_{\eta}\stackrel{(1)}{\Phi}\right)^{2}
  -         16  K \left(\stackrel{(1)}{\Phi}\right)^{2}
  -          8  {\cal H}^{2} \left(\stackrel{(1)}{\Phi}\right)^{2}
%  \nonumber\\
%  && 
  +             D_{l}D_{k}\stackrel{(1)}{\Phi} \stackrel{(1)}{\chi^{lk}}
%  \nonumber\\
%  && 
  + \frac{1}{8} \partial_{\eta}\stackrel{(1)}{\chi_{lk}} \partial_{\eta}\stackrel{(1)}{\chi^{kl}}
  +             {\cal H} \stackrel{(1)}{\chi_{kl}} \partial_{\eta}\stackrel{(1)}{\chi^{lk}}
  \nonumber\\
  && 
  - \frac{3}{8} D_{k}\stackrel{(1)}{\chi_{lm}} D^{k}\stackrel{(1)}{\chi^{ml}}
  + \frac{1}{4} D_{k}\stackrel{(1)}{\chi_{lm}} D^{l}\stackrel{(1)}{\chi^{mk}}
%  \nonumber\\
%  && \quad\quad
  - \frac{1}{2} \stackrel{(1)}{\chi^{lm}} \Delta\stackrel{(1)}{\chi_{lm}}
  + \frac{1}{2} K \stackrel{(1)}{\chi_{lm}} \stackrel{(1)}{\chi^{lm}}
  \label{eq:kouchan-19.337}
  ; \\
  \Gamma_{i}
  &:=&
            16  \pi G \partial_{\eta}\varphi_{1} D_{i}\varphi_{1}
  -          4  \partial_{\eta}\stackrel{(1)}{\Phi} D_{i}\stackrel{(1)}{\Phi}
  +          8  {\cal H} \stackrel{(1)}{\Phi} D_{i}\stackrel{(1)}{\Phi}
  -          8  \stackrel{(1)}{\Phi} \partial_{\eta}D_{i}\stackrel{(1)}{\Phi}
%  \nonumber\\
%  &&
  +          2  D^{j}\stackrel{(1)}{\Phi} \partial_{\eta}\stackrel{(1)}{\chi_{ji}}
  -          2  \partial_{\eta}D^{j}\stackrel{(1)}{\Phi} \stackrel{(1)}{\chi_{ij}}
  \nonumber\\
  &&
  - \frac{1}{2} \partial_{\eta}\stackrel{(1)}{\chi_{jk}} D_{i}\stackrel{(1)}{\chi^{kj}}
  -             \stackrel{(1)}{\chi_{kl}} \partial_{\eta}D_{i}\stackrel{(1)}{\chi^{lk}}
  +             \stackrel{(1)}{\chi^{kl}} \partial_{\eta}D_{k}\stackrel{(1)}{\chi_{il}}
  \label{eq:kouchan-19.338}
  ; \\
  \Gamma_{ij}
  &:=&
    16 \pi G D_{i}\varphi_{1} D_{j}\varphi_{1}
  +  8 \pi G \left\{
      (\partial_{\eta}\varphi_{1})^{2}
    - D_{l}\varphi_{1} D^{l}\varphi_{1}
    - a^{2} (\varphi_{1})^{2} \frac{\partial^{2}V}{\partial\varphi^{2}}
  \right\} \gamma_{ij}
%  \nonumber\\
%  && \quad\quad
  -          4  D_{i}\stackrel{(1)}{\Phi} D_{j}\stackrel{(1)}{\Phi}
  -          8  \stackrel{(1)}{\Phi} D_{i}D_{j}\stackrel{(1)}{\Phi}
  \nonumber\\
  &&
  + \left(
               6  D_{k}\stackrel{(1)}{\Phi} D^{k}\stackrel{(1)}{\Phi}
    +          4  \stackrel{(1)}{\Phi} \Delta\stackrel{(1)}{\Phi}
    +          2  \left(\partial_{\eta}\stackrel{(1)}{\Phi}\right)^{2}
    +          8  \partial_{\eta}{\cal H} \left(\stackrel{(1)}{\Phi}\right)^{2}
%  \right.
%  \nonumber\\
%  && \quad\quad\quad\quad
%  \left.
    +         16  {\cal H}^{2} \left(\stackrel{(1)}{\Phi}\right)^{2}
    +         16  {\cal H} \stackrel{(1)}{\Phi} \partial_{\eta}\stackrel{(1)}{\Phi}
    -          4  \stackrel{(1)}{\Phi} \partial_{\eta}^{2}\stackrel{(1)}{\Phi}
  \right) \gamma_{ij}
  \nonumber\\
  &&
  -          4  {\cal H} \partial_{\eta}\stackrel{(1)}{\Phi} \stackrel{(1)}{\chi_{ij}}
  -          2  \partial_{\eta}^{2}\stackrel{(1)}{\Phi} \stackrel{(1)}{\chi_{ij}}
  -          4  D^{k}\stackrel{(1)}{\Phi} D_{(i}\stackrel{(1)}{\chi_{j)k}}
  +          4  D^{k}\stackrel{(1)}{\Phi} D_{k}\stackrel{(1)}{\chi_{ij}}
  -          8  K \stackrel{(1)}{\Phi} \stackrel{(1)}{\chi_{ij}}
%  \nonumber\\
%  && \quad\quad
  +          4  \stackrel{(1)}{\Phi} \Delta\stackrel{(1)}{\chi_{ij}}
  -          4  D^{k}D_{(i}\stackrel{(1)}{\Phi} \stackrel{(1)}{\chi_{j)k}}
  \nonumber\\
  &&
  +          2  \Delta \stackrel{(1)}{\Phi} \stackrel{(1)}{\chi_{ij}}
  +          2  D_{l}D_{k}\stackrel{(1)}{\Phi} \stackrel{(1)}{\chi^{lk}} \gamma_{ij}
  +             \partial_{\eta}\stackrel{(1)}{\chi_{ik}} \partial_{\eta}\stackrel{(1)}{\chi_{j}^{\;\;k}}
  -             D^{k}\stackrel{(1)}{\chi_{il}} D_{k}\stackrel{(1)}{\chi_{j}^{\;\;l}}
  +             D^{k}\stackrel{(1)}{\chi_{il}} D^{l}\stackrel{(1)}{\chi_{jk}}
  - \frac{1}{2} D_{i}\stackrel{(1)}{\chi^{lk}} D_{j}\stackrel{(1)}{\chi_{lk}}
  \nonumber\\
  &&
  -             \stackrel{(1)}{\chi_{lm}} D_{i}D_{j}\stackrel{(1)}{\chi^{ml}}
  +          2  \stackrel{(1)}{\chi^{lm}} D_{l}D_{(i}\stackrel{(1)}{\chi_{j)m}}
  -             \stackrel{(1)}{\chi^{lm}} D_{m}D_{l}\stackrel{(1)}{\chi_{ij}}
  \nonumber\\
  &&
  - \frac{1}{4} \left(
      3 \partial_{\eta}\stackrel{(1)}{\chi_{lk}} \partial_{\eta}\stackrel{(1)}{\chi^{kl}}
    - 3 D_{k}\stackrel{(1)}{\chi_{lm}} D^{k}\stackrel{(1)}{\chi^{ml}}
    + 2 D_{k}\stackrel{(1)}{\chi_{lm}} D^{l}\stackrel{(1)}{\chi^{mk}}
%  \right.
%  \nonumber\\
%  && \quad\quad\quad\quad\quad
%  \left.
    - 4 K \stackrel{(1)}{\chi_{lm}} \stackrel{(1)}{\chi^{lm}}
  \right) \gamma_{ij}
  \label{eq:kouchan-19.339}
  .
\end{eqnarray}
Here, we used Eqs.~(\ref{eq:background-Einstein-equations-scalar-3}),
(\ref{eq:absence-of-anisotropic-stress-Einstein-i-j-traceless-scalar}),
(\ref{eq:kouchan-18.186}),
(\ref{eq:no-first-order-vector-mode-scalar-field-case})
and (\ref{eq:scalar-linearized-Einstein-scalar-master-eq-pre}).

%**************************************************************

The tensor part of Eq.~(\ref{eq:kouchan-18.207}) is given by
\begin{eqnarray}
  \left(
    \partial_{\eta}^{2} + 2 {\cal H} \partial_{\eta} + 2 K  - \Delta
  \right)
  \stackrel{(2)\;\;\;\;}{\chi_{ij}}
  &=&
  2 \Gamma_{ij}
  - \frac{2}{3} \gamma_{ij} \Gamma_{k}^{\;\;k}
  - 3
  \left(
    D_{i}D_{j} - \frac{1}{3} \gamma_{ij} \Delta
  \right) 
  \left( \Delta + 3 K \right)^{-1}
  \left(
    \Delta^{-1} D^{k}D_{l}\Gamma_{k}^{\;\;l}
    - \frac{1}{3} \Gamma_{k}^{\;\;k}
  \right)
  \nonumber\\
  &&
  + 4
  \left\{ 
      D_{(i} (\Delta+2K)^{-1} D_{j)}\Delta^{-1}D^{l}D_{k}\Gamma_{l}^{\;\;k}
    - D_{(i}(\Delta+2K)^{-1}D^{k}\Gamma_{j)k}
  \right\}
  .
  \label{eq:kouchan-18.215}
\end{eqnarray}
This tensor mode is also called the second-order gravitational
waves.

%**************************************************************

Further, the vector part of Eqs.~(\ref{eq:kouchan-18.199}) and
(\ref{eq:kouchan-18.207}) yields the initial value constraint and
the evolution equation of the vector mode
$\stackrel{(2)\;\;}{\nu_{j}}$: 
\begin{eqnarray}
  &&
  \stackrel{(2)}{\nu_{i}}
  = 
  \frac{2}{\Delta + 2 K}
  \left\{
    D_{i} \Delta^{-1} D^{k} \Gamma_{k}
    - \Gamma_{i}
  \right\}
%  \label{eq:kouchan-18.199-3}
  ,
  \quad
  \partial_{\eta}
  \left(
    a^{2} \stackrel{(2)}{\nu_{i}}
  \right)
  =
  \frac{2 a^{2}}{\Delta + 2 K}
  \left\{
    D_{i}\Delta^{-1} D^{k}D_{l}\Gamma_{k}^{\;\;l}
    - D_{k}\Gamma_{i}^{\;\;k}
  \right\}
  .
  \label{eq:kouchan-18.214}
\end{eqnarray}

%**************************************************************

Finally, scalar part of
Eqs.~(\ref{eq:kouchan-18.218})--(\ref{eq:kouchan-18.207}) are
summarized as 
\begin{eqnarray}
  &&
    2 \partial_{\eta} \stackrel{(2)}{\Psi}
  + 2 {\cal H} \stackrel{(2)}{\Phi}
  -
  8\pi G \varphi_{2} \partial_{\eta}\varphi  
  = 
  \Delta^{-1} D^{k} \Gamma_{k}
  \label{eq:kouchan-18.199-2}
  , \\
  &&
  \stackrel{(2)}{\Psi} - \stackrel{(2)}{\Phi}
  = 
  \frac{3}{2}
  (\Delta + 3 K)^{-1}
  \left\{
    \Delta^{-1} D^{i}D_{j}\Gamma_{i}^{\;\;j} - \frac{1}{3} \Gamma_{k}^{\;\;k}
  \right\}
  \label{eq:kouchan-18.213}
  , \\
  &&
  \left(
    -   \partial_{\eta}^{2} 
    - 5 {\cal H} \partial_{\eta}
    + \frac{4}{3} \Delta
    + 4 K
  \right) \stackrel{(2)}{\Psi}
  -
  \left(
      2 \partial_{\eta}{\cal H}
    +   {\cal H} \partial_{\eta}
    + 4 {\cal H}^{2}
    +   \frac{1}{3} \Delta
  \right)
  \stackrel{(2)}{\Phi} 
%  \nonumber\\
%  && \quad\quad\quad\quad\quad\quad\quad\quad\quad\quad\quad\quad\quad\quad\quad\quad\quad
  - 8 \pi G a^{2} \varphi_{2} \frac{\partial V}{\partial\varphi}
  =
  \Gamma_{0} - \frac{1}{6} \Gamma_{k}^{\;\;k}
  ,
  \label{eq:kouchan-18.228-2}
  \\
  && 
  \left\{
    \partial_{\eta}^{2}
    + 2 \left(
      {\cal H}
      - \frac{\partial_{\eta}^{2}\varphi}{\partial_{\eta}\varphi}
    \right)
    \partial_{\eta}
    -             \Delta
    -          4  K
    + 2 \left(
      \partial_{\eta}{\cal H}
      - \frac{\partial_{\eta}^{2}\varphi}{\partial_{\eta}\varphi} {\cal H}
    \right)
  \right\}
  \stackrel{(2)}{\Phi}
  \nonumber\\
  && 
  \quad\quad\quad\quad
  =
  - \Gamma_{0}
  - \frac{1}{2} \Gamma_{k}^{\;\;k}
  +
  \Delta^{-1} D^{i}D_{j}\Gamma_{i}^{\;\;j}
  + \left(
    \partial_{\eta}
    - \frac{\partial_{\eta}^{2}\varphi}{\partial_{\eta}\varphi}
  \right)
  \Delta^{-1}D^{k}\Gamma_{k}
  \nonumber\\
  && 
  \quad\quad\quad\quad\quad\quad
  -
  \frac{3}{2}
  \left\{
    \partial_{\eta}^{2}
    - \left(
      \frac{2\partial_{\eta}^{2}\varphi}{\partial_{\eta}\varphi} - {\cal H}
    \right)
    \partial_{\eta}
  \right\}
  (\Delta + 3 K)^{-1}
  \left\{
    \Delta^{-1} D^{i}D_{j}\Gamma_{i}^{\;\;j} - \frac{1}{3} \Gamma_{k}^{\;\;k}
  \right\}.
  \label{eq:kouchan-18.233}
\end{eqnarray}
\end{widetext}
where $\Gamma_{i}^{\;\;j} := \gamma^{kj}\Gamma_{ik}$ and
$\Gamma_{k}^{\;\;k} = \gamma^{ij}\Gamma_{ij}$.
Eq.~(\ref{eq:kouchan-18.233}) is the second-order extension of
Eq.~(\ref{eq:scalar-linearized-Einstein-scalar-master-eq}),
which is the master equation of scalar mode of the second-order
cosmological perturbation in a universe filled with a single
scalar field.

%**************************************************************

Thus, we have a set of ten equations for the second-order
perturbations of a universe filled with a single scalar field,
Eqs.~(\ref{eq:kouchan-18.215})--(\ref{eq:kouchan-18.233}).
To solve this system of equations of the second-order Einstein
equation, first of all, we have to solve the linear-order system.
This is accomplished by solving
Eq.~(\ref{eq:scalar-linearized-Einstein-scalar-master-eq}) to
obtain the potential $\stackrel{(1)}{\Phi}$, $\varphi_{1}$ is
given through (\ref{eq:kouchan-18.186}), and the tensor mode
$\stackrel{(1)}{\chi}_{ij}$ is given by solving
Eq.~(\ref{eq:linearized-Einstein-i-j-traceless-tensor}).  
Next, we evaluate the quadratic terms, $\Gamma_{0}$,
$\Gamma_{i}$ and $\Gamma_{ij}$ of the linear-order
perturbations, which are defined by
Eqs.~(\ref{eq:kouchan-19.337})--(\ref{eq:kouchan-19.339}). 
Then, using the information of
Eqs.~(\ref{eq:kouchan-19.337})--(\ref{eq:kouchan-19.339}), we
estimate the source term in Eq.~(\ref{eq:kouchan-18.233}). 
If we know the two independent solutions to the linear-order
master equation
(\ref{eq:scalar-linearized-Einstein-scalar-master-eq}), we can
solve Eq.~(\ref{eq:kouchan-18.233}) through the method using the
Green functions.  
After constructing the solution $\stackrel{(2)}{\Phi}$ to
Eq.~(\ref{eq:kouchan-18.233}), we can obtain the second-order
metric perturbation $\stackrel{(2)}{\Psi}$ through
Eq.~(\ref{eq:kouchan-18.213}).
Thus, we have obtained the second-order gauge invariant
perturbation $\varphi_{2}$ of the scalar field through
Eq.~(\ref{eq:kouchan-18.199-2}).
Thus, the all scalar modes $\stackrel{(2)}{\Phi}$,
$\stackrel{(2)}{\Psi}$, $\varphi_{2}$ are obtained.
Equation (\ref{eq:kouchan-18.228-2}) is then used to check the
consistency of the second-order perturbation of the Klein Gordon
equation (\ref{eq:Klein-Gordon-eq-second-gauge-inv-explicit}) as
in Sec.~\ref{sec:Consistency-of-the-second-order-perturbations}.

%*******************************************************************

For the vector-mode, $\stackrel{(1)\;\;}{\nu_{i}}$ of the
first-order identically vanishes due to the momentum constraint
(\ref{eq:no-first-order-vector-mode-scalar-field-case}) for the
linear-order metric perturbations.
On the other hand, in the second-order, we have
evolution equation (\ref{eq:kouchan-18.214}) of the vector mode
$\stackrel{(2)\;\;}{\nu_{i}}$ with the initial value constraint.
This evolution equation of the second-order vector mode
should be consistent with the initial value constraint, which is
confirmed in
Sec.~\ref{sec:Consistency-of-the-second-order-perturbations}.
Equations (\ref{eq:kouchan-18.214}) also imply that the
second-order vector-mode perturbation may be generated by the 
mode couplings of the linear order perturbations.
As the simple situations, the generation of the second-order
vector mode due to the scalar-scalar mode coupling is discussed
in Refs.~\cite{Mena:2007ve-Lu:2008ju-Lu:2008ju}.

%*******************************************************************

The second-order tensor mode is also generated by the
mode-coupling of the linear-order perturbations through the
source term in Eq.~(\ref{eq:kouchan-18.215}).
Note that Eq.~(\ref{eq:kouchan-18.215}) is almost same as
Eq.~(\ref{eq:linearized-Einstein-i-j-traceless-tensor}) for the
linear-order tensor mode, except for the existence of the source
term in Eq.~(\ref{eq:kouchan-18.215}).
If we know the solution to the linear-order Einstein equations
(\ref{eq:linearized-Einstein-i-j-traceless-tensor}) and
(\ref{eq:scalar-linearized-Einstein-scalar-master-eq}), we can
evaluate the source term in Eq.~(\ref{eq:kouchan-18.215}). 
Further, we can solve Eq.~(\ref{eq:kouchan-18.215}) through
the Green function method.
This leads the generation of the gravitational wave of the
second order.
Actually, in the simple situation where the first-order tensor
mode neglected, the generation of the second-order gravitational
waves discussed in some
literature\cite{Ananda:2006af-Osano:2006ew-etc}.

%*******************************************************************

%%%%%%%%%%%%%%%%%%%%%%%%%%%%%%%%%%%%%%%%%%%%%%%%%%%%%%%%%%%%%%%%%%%%%
\subsection{Consistency of equations for second-order perturbations}
\label{sec:Consistency-of-the-second-order-perturbations}
%%%%%%%%%%%%%%%%%%%%%%%%%%%%%%%%%%%%%%%%%%%%%%%%%%%%%%%%%%%%%%%%%%%%%

%****************************************************************

Now, we consider the consistency of the second-order
perturbations of the Einstein equations
(\ref{eq:kouchan-18.199-2})--(\ref{eq:kouchan-18.233}) for
the scalar modes, Eqs.~(\ref{eq:kouchan-18.214}) for vector
mode, and the Klein-Gordon equation
(\ref{eq:Klein-Gordon-eq-second-gauge-inv-explicit}).
The consistency check of these equations are necessary to
guarantee that the derived equations are correct, since the
second-order Einstein equations have complicated forms owing to
the quadratic terms of the linear-order perturbations that arise
from the nonlinear effects of the Einstein equations.

%****************************************************************

Since the first equation in Eqs.~(\ref{eq:kouchan-18.214}) is
the initial value constraint for the vector mode
$\stackrel{(2)}{\nu_{i}}$ and it should be consistent with the
evolution equation, i.e., the second equation of
Eqs.~(\ref{eq:kouchan-18.214}). these equations should be
consistent with each other from the general arguments of the
Einstein equation.
Explicitly, these equations are consistent with each other if
the equation
\begin{eqnarray}
  \partial_{\eta}\Gamma_{k}
  + 2 {\cal H} \Gamma_{k}
  - D^{l}\Gamma_{lk} = 0
  \label{eq:kouchan-19.358}
\end{eqnarray}
is satisfied.
Actually, through the first-order perturbative Einstein
equations (\ref{eq:kouchan-18.186}),
(\ref{eq:scalar-linearized-Einstein-scalar-master-eq}),
(\ref{eq:linearized-Einstein-i-j-traceless-tensor}),
we can confirm the equation (\ref{eq:kouchan-19.358}).
This is a trivial result from a general viewpoint,
because the Einstein equation is the first class constrained
system.
However, this trivial result implies that we have derived the
source terms $\Gamma_{i}$ and $\Gamma_{ij}$ of the second-order
Einstein equations consistently.

%**************************************************************

Next, we consider Eq.~(\ref{eq:kouchan-18.228-2}).
Through the second-order Einstein equations
(\ref{eq:kouchan-18.199-2}), (\ref{eq:kouchan-18.213}),
(\ref{eq:kouchan-18.233}), and the background Klein-Gordon
equation (\ref{eq:background-Klein-Gordon-equation}), 
we can confirm that Eq.~(\ref{eq:kouchan-18.228-2}) is
consistent with the set of the background, first-order and other
second-order Einstein equation if the equation
\begin{eqnarray}
  \left(
                  \partial_{\eta}
    +          2  {\cal H}
  \right) D^{k}\Gamma_{k}
  - D^{j}D^{i}\Gamma_{ij}
  =
  0
  \label{eq:kouchan-19.345}
\end{eqnarray}
is satisfied under the background and first-order Einstein
equations.
Actually, we have already seen that
Eq.~(\ref{eq:kouchan-19.358}) is satisfied under the background
and first-order Einstein equations.
Taking the divergence of Eq.~(\ref{eq:kouchan-19.358}), we can
immediately confirm Eq.~(\ref{eq:kouchan-19.345}).
Then, Eq.~(\ref{eq:kouchan-18.228-2}) gives no information.

%**************************************************************

Thus, we have seen that the derived Einstein equations of the
second order
(\ref{eq:kouchan-18.214})--(\ref{eq:kouchan-18.233}) are
consistent with each other through
Eq.~(\ref{eq:kouchan-19.358}).
This fact implies that the derived source terms $\Gamma_{i}$ and 
$\Gamma_{ij}$ of the second-order perturbations of the Einstein
equations, which are defined by Eqs.~(\ref{eq:kouchan-19.338})
and (\ref{eq:kouchan-19.339}), are correct source terms of the
second-order Einstein equations.
On the other hand, for $\Gamma_{0}$, we have to consider the
consistency between the perturbative Einstein equations and the
perturbative Klein-Gordon equation as seen below.

%**************************************************************

Now, we consider the consistency of the second-order
perturbation of the Klein-Gordon equation and the Einstein
equations.
The second-order perturbation of the Klein-Gordon equation is
given by
Eq.~(\ref{eq:Klein-Gordon-eq-second-gauge-inv-explicit}) with
the source term
(\ref{eq:Klein-Gordon-eq-second-gauge-inv-explicit-source}). 
Since the vector mode $\stackrel{(2)}{\nu_{i}}$ and tensor mode
$\stackrel{(2)}{\chi}_{ij}$ of the second-order do not appear in
the expressions
(\ref{eq:Klein-Gordon-eq-second-gauge-inv-explicit}) of the
second-order perturbation of the Klein-Gordon equation, we may
concentrate on the Einstein equations for scalar mode of the
second order, i.e.,
Eqs.~(\ref{eq:kouchan-18.199-2}), (\ref{eq:kouchan-18.213}), and
(\ref{eq:kouchan-18.233}) with the definitions 
(\ref{eq:kouchan-19.337})--(\ref{eq:kouchan-19.339}) of the
source terms.
As in the linear case, the second-order perturbation of the
Klein-Gordon equation should also be derived from the set of
equations consisting of the second-order perturbations of the
Einstein equations (\ref{eq:kouchan-18.199-2}),
(\ref{eq:kouchan-18.213}), (\ref{eq:kouchan-18.233}), the
first-order perturbations of the Einstein equations
(\ref{eq:absence-of-anisotropic-stress-Einstein-i-j-traceless-scalar}),
(\ref{eq:kouchan-18.186}),
(\ref{eq:scalar-linearized-Einstein-scalar-master-eq}), 
and the background Einstein equations
(\ref{eq:background-Einstein-equations-scalar-1}) and
(\ref{eq:background-Einstein-equations-scalar-2}).
Actually, from these equation, we can show that the second-order
perturbation of the Klein-Gordon equation is consistent with the
background and the second-order Einstein equations if the
equation 
\begin{eqnarray}
  &&
  2 \left(
    \partial_{\eta} + {\cal H}
  \right) \Gamma_{0}
  -    D^{k}\Gamma_{k}
  +    {\cal H} \Gamma_{k}^{\;\;k}
  \nonumber\\
  && \quad\quad
  + 8 \pi G \partial_{\eta}\varphi \Xi_{(K)}
  = 0
  \label{eq:kouchan-19.374}
\end{eqnarray}
is satisfied under the background and the first-order Einstein 
equations.
Further, we can also confirm Eq.~(\ref{eq:kouchan-19.374})
through the background Einstein equations
(\ref{eq:background-Einstein-equations-scalar-1}) and
(\ref{eq:background-Einstein-equations-scalar-2}), the
scalar part of the first-order perturbation of the momentum
constraint (\ref{eq:kouchan-18.186}), the evolution equations
(\ref{eq:scalar-linearized-Einstein-scalar-master-eq}) and 
(\ref{eq:linearized-Einstein-i-j-traceless-tensor}) for the
first order scalar and tensor modes in the Einstein equation.

%****************************************************************

As shown in Ref.~\cite{kouchan-second-cosmo-consistency},
the first-order perturbation of the Klein-Gordon equation is
derived from the background and the first-order perturbations 
of the Einstein equation.
In the case of the second-order perturbation, the Klein-Gordon
equation (\ref{eq:Klein-Gordon-eq-second-gauge-inv-explicit})
can be also derived from the background, the first-order, and
the second-order Einstein equations.
The second-order perturbations of the Einstein equation and the
Klein-Gordon equation include the source terms $\Gamma_{0}$,
$\Gamma_{i}$, $\Gamma_{ij}$, and $\Xi_{(K)}$ due to the
mode-coupling of the linear-order perturbations.
The second-order perturbation of the Klein-Gordon equation gives
the relation (\ref{eq:kouchan-19.374}) between the source terms
$\Gamma_{0}$, $\Gamma_{i}$, $\Gamma_{ij}$, $\Xi_{(K)}$ and we
have also confirmed that Eq.~(\ref{eq:kouchan-19.374}) is
satisfied due to the background, the first-order perturbation of
the Einstein equations, and the Klein-Gordon equation.
Thus, the second-order perturbation of the Klein-Gordon equation
is not independent of the set of the background, the
first-order, and the second-order Einstein equations if we
impose on the Einstein equation at any conformal time $\eta$. 
This also implies that the derived formulae of the source terms
$\Gamma_{0}$, $\Gamma_{i}$, $\Gamma_{ij}$, and $\Xi_{(K)}$ are
consistent with each other.
In this sense, we may say that the formulae
(\ref{eq:kouchan-19.337})--(\ref{eq:kouchan-19.339}) and
(\ref{eq:Klein-Gordon-eq-second-gauge-inv-explicit-source}) for
these source terms are correct.

%**************************************************************

%%%%%%%%%%%%%%%%%%%%%%%%%%%%%%%%%%%%%%%%%%%%%%%%%%%%%%%%%%%%%%%%%%%%%%
\section{Summary and discussions}
\label{sec:summary}
%%%%%%%%%%%%%%%%%%%%%%%%%%%%%%%%%%%%%%%%%%%%%%%%%%%%%%%%%%%%%%%%%%%%%%

%*******************************************************************

In this paper, we summarized the current status of the
formulation of the gauge-invariant second-order cosmological
perturbations.
Although the presentation in this paper is restricted to the
case of the universe filled by a single scalar field, the
essence of the general framework of the gauge-invariant
perturbation theory is transparent through this simple case.
The general framework of the general relativistic higher-order
gauge-invariant perturbation theory can be separated into three
parts.
First one is the general formulation to derive the
gauge-transformation rules (\ref{eq:Bruni-47-one}) and
(\ref{eq:Bruni-49-one}).
Second one is the construction of the gauge-invariant variables
for the perturbations on the generic background spacetime
inspecting gauge-transformation rules (\ref{eq:Bruni-47-one}) and
(\ref{eq:Bruni-49-one}) and the decomposition formula
(\ref{eq:matter-gauge-inv-decomp-1.0}) and
(\ref{eq:matter-gauge-inv-decomp-2.0}) for perturbations of any
tensor field.
Third one is the application of the above general framework of
the gauge-invariant perturbation theory to the cosmological
situations.

%*******************************************************************

To derive the gauge-transformation rules (\ref{eq:Bruni-47-one})
and (\ref{eq:Bruni-49-one}), we considered the general arguments
on the Taylor expansion of an arbitrary tensor field on a
manifold, the general class of the diffeomorphism which is wider
than the usual exponential map, and the general formulation of
the perturbation theory.
This general class of diffeomorphism is represented in terms of
the Taylor expansion (\ref{eq:Taylor-expansion-of-f}) of its
pull-back.
As commented in
Sec.~\ref{sec:Taylor-expansion-of-tensors-on-a-manifold}, this
general class of diffeomorphism does not form a one-parameter
group of diffeomorphism as shown through
Eq.~(\ref{eq:Phi-is-not-one-parameter-group-of-diffeomorphism}).
However, the properties
(\ref{eq:Phi-is-not-one-parameter-group-of-diffeomorphism}) do 
not directly mean that this general class of diffeomorphism does
not form a group. 
One of the key points of the properties of this diffeomorphism
is the non-commutativity of generators $\xi_{1}^{a}$ and
$\xi_{2}^{a}$ of each order. 
Although the expression of the $n$-th order Taylor expansion of
the pull-back of this general class is discussed in
Ref.~\cite{M.Bruni-S.Sonego-CQG1999}, when we consider the
situation of the $n$-th order perturbation, this
non-commutativity becomes important\cite{kouchan-gauge-inv}.
Therefore, to clarify the properties of this general class of
diffeomorphism, we have to take care of this non-commutativity
of generators. 
Thus, there is a room to clarify the properties of this general
class of diffeomorphism.

%*******************************************************************

Further, in Sec.~\ref{sec:Formulation-of-perturbation-theory},
we introduced a gauge choice ${\cal X}_{\lambda}$ as an
exponential map, for simplicity.
On the other hand, we have the concept of the general class of
diffeomorphism which is wider than the class of the exponential
map.
Therefore, we may introduce a gauge choice as one of the element
of this general class of diffeomorphism. 
However, the gauge-transformation rules (\ref{eq:Bruni-47-one})
and (\ref{eq:Bruni-49-one}) will not be changed even if we
generalize the definition of a each gauge choice as emphasized
in Sec.~\ref{sec:Formulation-of-perturbation-theory}. 
Although there is a room to sophisticate in logical arguments
to derive the gauge-transformation rules (\ref{eq:Bruni-47-one})
and (\ref{eq:Bruni-49-one}), these are harmless to the
development of the general framework of the gauge-invariant
perturbation theory shown in
Secs.~\ref{sec:Formulation-of-perturbation-theory},
\ref{sec:gauge-invariant-variables},
\ref{sec:Perturbation-of-the-field-equations}, and their
application to cosmological perturbations in
Sec.~\ref{sec:Cosmological-Background-spacetime-equations}.

%*******************************************************************

As emphasize in Sec.~\ref{sec:gauge-invariant-variables}, our
starting point to construct gauge invariant variables is the
assumption that {\it we already know the procedure for finding
  gauge invariant variables for the linear metric
  perturbations as Eq.~(\ref{eq:linear-metric-decomp})}.
This is highly nontrivial assumption on a generic background
spacetime.
The procedure to accomplish the decomposition
(\ref{eq:linear-metric-decomp}) completely depends on the
details of the background spacetime. 
In spite of this non-triviality, this assumption is almost
correct in some background
spacetime\cite{kouchan-papers}.
Further, once we accept this assumption, we can develop the
higher-order perturbation theory in an independent manner of the 
details of the background spacetime.
We also expect that this general framework of the
gauge-invariant perturbation theory is extensible to $n$-th
order perturbation theory, since our procedure to construct
gauge-invariant variables can be extended to the third-order
perturbation theory with two-parameter\cite{kouchan-gauge-inv}.
Due to this situation, in
Ref.~\cite{kouchan-second-cosmo-matter}, we propose the
conjecture which state that the above assumption for the
decomposition of the linear-order metric perturbation is correct
for any background spacetime.
We may also say that the most nontrivial part of our general
framework of higher-order gauge-invariant perturbation theory is
in the above assumption.
Further, as emphasized in
Sec.~\ref{sec:Gauge-invariant-metric-perturbations}, we assumed
the existence of some Green functions to accomplish the
decomposition (\ref{eq:linear-metric-decomp}) and this
assumption exclude some perturbative modes of the metric 
perturbations from our consideration, even in the case of
cosmological perturbations.
For example, homogeneous modes of perturbations are excluded in
our current arguments of the cosmological perturbation theory.
These homogeneous modes will be necessary to discuss the
comparison with the arguments based on the long-wavelength
approximation.
Therefore, we have to say that there is a room to clarify even
in the cosmological perturbation theory.

%*******************************************************************

Even if the assumption is correct on any background spacetime,
the other problem is in the interpretations of the
gauge-invariant variables.
We have commented on the non-uniqueness in the definitions of
the gauge-invariant variables through
Eqs.~(\ref{eq:non-uniqueness-of-gauge-invariant-metric-perturbation})
and
(\ref{eq:non-uniqueness-of-gauge-invariant-2nd-order-metric-perturbation-4}). 
This non-uniqueness in the definition of gauge-invariant
variables also leads some ambiguities in the interpretations of 
gauge-invariant variables.
On the other hand, as emphasize in
Sec.~\ref{sec:Formulation-of-perturbation-theory}, any
observations and experiments are carried out only on the
physical spacetime through the physical processes on the
physical spacetime.
For this reason, any direct observables in any observations and 
experiments should be independent of the gauge choice.
Further, the non-uniqueness in the definitions 
the gauge-invariant variables expressed by 
Eqs.~(\ref{eq:non-uniqueness-of-gauge-invariant-metric-perturbation})
and
(\ref{eq:non-uniqueness-of-gauge-invariant-2nd-order-metric-perturbation-4}) 
have the same form as the decomposition formulae
(\ref{eq:matter-gauge-inv-decomp-1.0}) and
(\ref{eq:matter-gauge-inv-decomp-2.0}).
Therefore, if the statement that {\it any direct observables in
  any observations and experiments is independent of the
  gauge choice} is really true, this also confirm that the
non-uniqueness of the definition of the gauge-invariant variables
also have nothing to do with the direct observables in
observations and experiments.
These will be confirmed by the clarification of the relations
between gauge-invariant variables and observables in experiments
and observations.
To accomplish this, we have to specify the concrete process of
experiments and observations and clarify the problem what are 
the direct observables in the experiments and observations and
derive the relations between the gauge-invariant variables and
observables in concrete observations and experiments.
If these arguments are completed, we will be able to show that
the gauge degree of freedom is just unphysical degree of freedom
and the non-uniqueness of the gauge-invariant variables have
nothing to do with the direct observables in the concrete
observation or experiment, and then, we will be able to clarify
the precise physical interpretation of the gauge-invariant
variables.

%*******************************************************************

For example, in the case of the CMB physics, we can easily see
that the first-order perturbation of the CMB temperature is
automatically gauge invariant from
Eq.~(\ref{eq:matter-gauge-inv-decomp-1.0}), because the
background temperature of CMB is homogeneous. 
On the other hand, the decomposition formula
(\ref{eq:matter-gauge-inv-decomp-2.0}) of the second order
yields that the theoretical prediction of the second-order
perturbation of the CMB temperature may depend on gauge choice,
since we do know the existence of the first-order fluctuations
as the temperature anisotropy in CMB.
However, as emphasized above, the direct observables in
observations should be gauge invariant and the gauge-variant
term in Eq.~(\ref{eq:matter-gauge-inv-decomp-2.0}) should be
disappear in the direct observables.
Therefore, we have to clarify the how gauge-invariant variables
are related to the observed temperature fluctuations and the
gauge-variant term disappear in the observable.

%*******************************************************************

Although there are some rooms to accomplish the complete
formulation of the second-order cosmological perturbation
theory, we derived all the components of the second-order
perturbation of the Einstein equation without ignoring any
types modes (scalar-, vector-, tensor-types) of perturbations in
the case of a scalar field system.
In our formulation, any gauge fixing is not necessary and we
can obtain all equations in the gauge-invariant form, which are
equivalent to the complete gauge fixing.
In other words, our formulation gives complete gauge-fixed
equations without any gauge fixing.
Therefore, equations obtained in a gauge-invariant manner cannot
be reduced without physical restrictions any more.
In this sense, the equations shown here are irreducible.
This is one of the advantages of the gauge-invariant perturbation
theory.

%*******************************************************************

The resulting Einstein equations of the second order show that
any type of mode-coupling appears as the quadratic terms of the
linear-order perturbations owing to the nonlinear effect of the
Einstein equations, in principle.
Perturbations in cosmological situations are classified into
three types: scalar, vector, and tensor.
In the second-order perturbations, we also have these three
types of perturbations as in the case of the first-order
perturbations.
Furthermore, in the equations for the second-order perturbations,
there are many quadratic terms of linear-order perturbations
owing to the nonlinear effects of the system.
Owing to these nonlinear effects, the above three types of
perturbations couple with each other.
In the scalar field system shown in this paper, the first-order
vector mode does not appear due to the momentum constraint of
the first-order perturbation of the Einstein equation.
Therefore, we have seen that three types of mode-coupling
appear in the second-order Einstein equations, i.e.,
scalar-scalar, scalar-tensor, and tensor-tensor type of mode
coupling.
In general, all types of mode-coupling may appear in the
second-order Einstein equations.
Actually, in Ref.~\cite{kouchan-second-cosmo-consistency}, we
also derived the all components of the Einstein equations for
a perfect fluid system and we can see all types of
mode-coupling, i.e., scalar-scalar, scalar-vector,
scalar-tensor, vector-vector, vector-tensor, tensor-tensor types
mode-coupling, appear in the second-order Einstein equation, in
general.
Of course, in the some realistic situations of cosmology, we may 
neglect some modes.
In this case, we may neglect some mode-coupling.
However, even in this case, we should keep in mind the fact that
all types of mode-couplings may appear in principle when we
discuss the realistic situations of cosmology.
We cannot deny the possibility that the mode-couplings of any
type produces observable effects when the quite high accuracy of
observations is accomplished.

%*******************************************************************

Even in the case of the single scalar field discussed in this
paper, the source terms of the second-order Einstein equation
show the mode-coupling of scalar-scalar, scalar-tensor, and the 
tensor-tensor types as mentioned above. 
Since the tensor mode of the linear order is also generated due
to quantum fluctuations during the inflationary phase, the 
mode-couplings of the scalar-tensor and tensor-tensor types may
appear in the inflation.
If these mode-couplings occur during the inflationary phase,
these effects will depend on the scalar-tensor ratio $r$.
If so, there is a possibility that the accurate observations of
the second-order effects in the fluctuations of the scalar type
in our universe also restrict the scalar-tensor ratio $r$ or
give some consistency relations between the other observations
such as the measurements of the B-mode of the polarization of
CMB.
This will be a new effect that gives some information on the
scalar-tensor ratio $r$.

%*******************************************************************

Furthermore, we have also checked the consistency between the
second-order perturbations of the equations of motion of matter
field and the Einstein equations. 
In the case of a scalar field, we checked the consistency
between the second-order perturbations of the Klein-Gordon
equation and the Einstein equations.
Due to this consistency check, we have obtained the
consistency relations between the source terms in these
equations $\Gamma_{0}$, $\Gamma_{i}$, $\Gamma_{ij}$, and
$\Xi_{(K)}$, which are given by Eqs.~(\ref{eq:kouchan-19.358})
and (\ref{eq:kouchan-19.374}).
We note that the relation (\ref{eq:kouchan-19.358}) comes
from the consistency in the Einstein equations of
the second order by itself, while the relation
(\ref{eq:kouchan-19.374}) comes from the consistency between the
second-order perturbation of the Klein-Gordon equation and the
Einstein equation.
We also showed that these relations between the
source terms are satisfied through the background and the
first-order perturbation of the Einstein equations in
Ref.~\cite{kouchan-second-cosmo-consistency}.
This implies that the set of all equations are self-consistent
and the derived source terms $\Gamma_{0}$, $\Gamma_{i}$,
$\Gamma_{ij}$, and $\Xi_{(K)}$ are correct.
We also note that these relations are independent of the
details of the potential of the scalar field.

%*******************************************************************

Thus, we have derived the self-consistent set of equations of
the second-order perturbation of the Einstein equations and the
evolution equations of matter fields in terms of gauge-invariant
variables.
As the current status of the second-order gauge-invariant
cosmological perturbation theory, we may say that the curvature
terms in the second-order Einstein tensor
(\ref{eq:second-order-Einstein-equation}), i.e., the
second-order perturbations of the Einstein tensor, are almost
completely derived although there remains the problem of
homogeneous modes as mentioned above.
After complete the problem of homogeneous modes, we have to
clarify the physical behaviors of the second-order cosmological
perturbation in the single scalar field system in the context of
the inflationary scenario.
This is the preliminary step to clarify the quantum behaviors of 
second-order perturbations in the inflationary universe.
Further, we also have to carry out the comparison with the
result by long-wavelength approximations.
If these issues are completed, we may say that we have
completely understood the properties of the second-order
perturbation of the Einstein tensor.
The next task is to clarify the nature of the second-order
perturbation of the energy-momentum tensor through the extension
to multi-fluid or multi-field systems.
Further, we also have to extend our arguments to the Einstein
Boltzmann system to discuss CMB physics, since we have to treat
photon and neutrinos through the Boltzmann distribution
functions. 
This issue is also discussed in some
literature\cite{Non-Gaussianity-in-CMB,Pitro-2007-2009}. 
If we accomplish these extension, we will be able to clarify the
Non-linear effects in CMB physics.

%*******************************************************************

Finally, readers might think that the ingredients of this paper
is too mathematical as Astronomy.
However, we have to emphasize that a high degree of the
theoretical sophistication leads unambiguous theoretical
predictions in many case.
As in the case of the linear-order cosmological perturbation
theory, the developments in observations are also supported by
the theoretical sophistication and the theoretical
sophistication are accomplished motivated by observations.
In this sense, now, we have an opportunity to develop the
general relativistic second-order perturbation theory to a high
degree of sophistication which is motivated by observations.
We also expect that this theoretical sophistication will be also
useful to discuss the theoretical predictions of Non-Gaussianity
in CMB and comparison with observations.
Therefore, I think that this opportunity is opened not only for
observational cosmologists but also for theoretical and
mathematical physicists.

%*******************************************************************

%%%%%%%%%%%%%%%%%%%%%%%%%%%%%%%%%%%%%%%%%%%%%%%%%%%%%%%%%%%%%%%%%%%%%%
\section*{Acknowledgments}
%%%%%%%%%%%%%%%%%%%%%%%%%%%%%%%%%%%%%%%%%%%%%%%%%%%%%%%%%%%%%%%%%%%%%%
The author thanks participants in the GCOE/YITP workshop
YITP-W-0901 on ``Non-linear cosmological perturbations'' which
was held at YITP in Kyoto, Japan in April, 2009, for valuable
discussions, in particular, Prof. M.~Bruni, Prof. R.~Maartens,
Prof. M.~Sasaki, Prof. T.~Tanaka, and Prof. K.~Tomita.
This review is an extension of the contribution to this workshop
by the author.

%****************************************************************

%%%%%%%%%%%%%%%%%%%%%%%%%%%%%%%%%%%%%%%%%%%%%%%%%%%%%%%%%%%%%%%%%%%%%%
%%%%%%%%%%%%%%%%%%%%%%%%%%%%%%%%%%%%%%%%%%%%%%%%%%%%%%%%%%%%%%%%%%%%%%
\appendix
%%%%%%%%%%%%%%%%%%%%%%%%%%%%%%%%%%%%%%%%%%%%%%%%%%%%%%%%%%%%%%%%%%%%%%
%%%%%%%%%%%%%%%%%%%%%%%%%%%%%%%%%%%%%%%%%%%%%%%%%%%%%%%%%%%%%%%%%%%%%%
%%%%%%%%%%%%%%%%%%%%%%%%%%%%%%%%%%%%%%%%%%%%%%%%%%%%%%%%%%%%%%%%%%%%%%
\section{Derivation of the generic representation of the Taylor
  expansion of tensors on a manifold}
\label{sec:derivation-of-Taylor-expansion}
%%%%%%%%%%%%%%%%%%%%%%%%%%%%%%%%%%%%%%%%%%%%%%%%%%%%%%%%%%%%%%%%%%%%%%
%%%%%%%%%%%%%%%%%%%%%%%%%%%%%%%%%%%%%%%%%%%%%%%%%%%%%%%%%%%%%%%%%%%%%%
%%%%%%%%%%%%%%%%%%%%%%%%%%%%%%%%%%%%%%%%%%%%%%%%%%%%%%%%%%%%%%%%%%%%%%

%*******************************************************************

In this section, we derive the representation of the
coefficients of the formal Taylor expansion
(\ref{eq:Taylor-expansion-of-f}) of the pull-back of a
diffeomorphism in terms of the suitable derivative operators.
The guide principle of our arguments is the following
theorem\cite{Bruni-Gualtieri-Sopuerta-2003,Kobayashi-Nomizu-I-1996}.

%*******************************************************************

\begin{theorem}
  \label{theorem:Bruni-Gualtieri-Sopuerta-2003-Appendix}
  Let ${\cal D}$ be a derivative operator acting on the set of
  all the tensor fields defined on a differentiable manifold
  ${\cal M}$ and satisfying the following conditions: (i) it is
  linear and satisfies the Leibniz rule; (ii) it is tensor-type
  preserving; (iii) it commutes with every contraction of a
  tensor field; and (iv) it commutes with the exterior
  differentiation $d$.
  Then, ${\cal D}$ is equivalent to the Lie derivative operator
  with respect to some vector field $\xi$, i.e., 
  ${\cal D}={\pounds}_{\xi}$. 
\end{theorem}

%*******************************************************************

The prove of the assertion of Theorem
\ref{theorem:Bruni-Gualtieri-Sopuerta-2003-Appendix} is given in
Ref.~\cite{Bruni-Gualtieri-Sopuerta-2003} as follows.
When acting on functions, the derivative operator ${\cal D}$
defines a vector field $\xi$ through the relation
\begin{eqnarray}
  \label{eq:Bruni-Gualtieri-Sopuerta-2003-A.1}
  {\cal D}f =: \xi(f) = {\pounds}_{\xi}f, 
  \quad \forall f\in{\cal F}(M).
\end{eqnarray}
The assertion of the Theorem for an arbitrary tensor field is
hold iff the assertions for an arbitrary scalar function and for
an arbitrary vector field $V$ are hold.
To do this, we consider the scalar function $V(f)$ and we obtain
\begin{eqnarray}
  {\cal D}(V(f)) = \xi(V(f)) 
\end{eqnarray}
through Eq.~(\ref{eq:Bruni-Gualtieri-Sopuerta-2003-A.1}).
Through the conditions (i)-(iv) of ${\cal D}$, ${\cal D}(V(f))$
is also given by
\begin{eqnarray}
  {\cal D}(V(f))
  &=& {\cal D}(df(V))
  = {\cal D}\left\{
    {\cal C}(df\otimes V)
  \right\}
  \nonumber\\
  &=&
  {\cal C}\left\{
    {\cal D}(df\otimes V)
  \right\}
  \nonumber\\
  &=&
  {\cal C}\left\{
    {\cal D}(df)\otimes V
    +
    df \otimes{\cal D}V
  \right\}
  \nonumber\\
  &=&
  {\cal C}\left\{
    d({\cal D}f)\otimes V
    +
    df \otimes{\cal D}V
  \right\}
  \nonumber\\
  &=&
  d({\cal D}f)(V) + df({\cal D}V)
  \nonumber\\
  &=&
  V({\cal D}f) + ({\cal D}V)(f)
\end{eqnarray}
Then we obtain 
\begin{eqnarray}
  ({\cal D}V)(f) &=& \xi(V(f)) - V(\xi(f))
  = \left[\xi, V\right](f)
  \nonumber\\
  &=& ({\pounds}_{\xi}V)(f)
\end{eqnarray}
for an arbitrary $f$, i.e.,
\begin{eqnarray}
  \label{eq:Bruni-Gualtieri-Sopuerta-2003-A.4}
  {\cal D}V = {\pounds}_{\xi}V.
\end{eqnarray}
Through Eqs.~(\ref{eq:Bruni-Gualtieri-Sopuerta-2003-A.1}) and
(\ref{eq:Bruni-Gualtieri-Sopuerta-2003-A.4}), we can recursively
show 
\begin{eqnarray}
  \label{eq:Bruni-Gualtieri-Sopuerta-2003-A.2}
  {\cal D}Q = {\pounds}_{\xi}Q
\end{eqnarray}
for an arbitrary tensor field $Q$\cite{Kobayashi-Nomizu-I-1996}.

%*******************************************************************

Now, we consider the derivation of the Taylor expansion
(\ref{eq:symbolic-Taylor-expansion-of-f}). 
As in the main text, we first consider the representation of the
Taylor expansion of $\Phi^{*}_{\lambda}f$ for an arbitrary
scalar function $f\in{\cal F}(M)$:
\begin{eqnarray}
  (\Phi^{*}_{\lambda}f)(p)
  &=&
  f(p)
  +
  \lambda 
  \left\{\frac{\partial}{\partial\lambda}(\Phi^{*}_{\lambda}f)\right\}_{\lambda=0}
  \nonumber\\
  && 
  +
  \frac{1}{2} \lambda^{2} 
  \left\{\frac{\partial^{2}}{\partial\lambda^{2}}(\Phi^{*}_{\lambda}f)\right\}_{\lambda=0}
  + O(\lambda^{3}),
  \label{eq:symbolic-Taylor-expansion-of-f-appendix}
\end{eqnarray}
where ${\cal F}(M)$ denotes the algebra of $C^{\infty}$
functions on ${\cal M}$.
Although the operator $\partial/\partial\lambda$ in the bracket
$\{*\}_{\lambda=0}$ of
Eq.~(\ref{eq:symbolic-Taylor-expansion-of-f-appendix}) are 
simply symbolic notation, we stipulate the properties
\begin{eqnarray}
  \left\{
    \frac{\partial^{2}}{\partial\lambda^{2}}(\Phi^{*}_{\lambda}f)
  \right\}_{\lambda=0}
  &=& 
  \left\{
    \frac{\partial}{\partial\lambda}\left(
      \frac{\partial}{\partial\lambda}(\Phi^{*}_{\lambda}f)
    \right)
  \right\}_{\lambda=0}
  \label{eq:properties-of-partial-over-partial-lambda-1}
  ,\\
  \left\{
    \frac{\partial}{\partial\lambda}(\Phi^{*}_{\lambda}f)^{2}
  \right\}_{\lambda=0}
  &=& 
  \left\{
    2 \Phi^{*}_{\lambda}f \frac{\partial}{\partial\lambda}(\Phi^{*}_{\lambda}f)
  \right\}_{\lambda=0}
  \label{eq:properties-of-partial-over-partial-lambda-2}
  .
\end{eqnarray}
for $\forall f\in{\cal F}({\cal M})$, where $n$ is an arbitrary 
finite integer. 
These properties imply that the operator
$\partial/\partial\lambda$ is in fact not simply symbolic
notation but indeed the usual partial differential operator on
$\RF$.
We note that the property
(\ref{eq:properties-of-partial-over-partial-lambda-2}) is the
Leibniz rule, which plays important roles when we derive the
representation of the Taylor expansion
(\ref{eq:symbolic-Taylor-expansion-of-f-appendix}) in terms of
suitable Lie derivatives.

%*******************************************************************

Together with the property
(\ref{eq:properties-of-partial-over-partial-lambda-2}), Theorem 
\ref{theorem:Bruni-Gualtieri-Sopuerta-2003-Appendix} yields 
that there exists a vector field $\xi_{1}$ so that 
\begin{eqnarray}
  \left\{
    \frac{\partial}{\partial\lambda}(\Phi^{*}_{\lambda}f)
  \right\}_{\lambda=0}
  &=:& 
  {\pounds}_{\xi_{1}} f
  .
  \label{eq:def-of-calL1}
\end{eqnarray}
Actually, the conditions (ii)-(iv) in Theorem
\ref{theorem:Bruni-Gualtieri-Sopuerta-2003-Appendix} are
satisfied from the fact that $\Phi_{\lambda}^{*}$ is the
pull-back of a diffeomorphism $\Phi_{\lambda}$ and (i) is
satisfied due to the property
(\ref{eq:properties-of-partial-over-partial-lambda-2}).

%*******************************************************************

Next, we consider the second-order term in
Eq.~(\ref{eq:symbolic-Taylor-expansion-of-f-appendix}).
Since we easily expect that the second-order term in
Eq.~(\ref{eq:symbolic-Taylor-expansion-of-f-appendix}) may
includes ${\cal L}_{\xi_{1}}^{2}$, we define the derivative
operator ${\cal L}_{2}$ by 
\begin{eqnarray}
  \left\{
    \frac{\partial^{2}}{\partial\lambda^{2}}(\Phi^{*}_{\lambda}f)
  \right\}_{\lambda=0}
  &=:& 
  \left({\cal L}_{(2)} + a {\pounds}_{\xi_{1}}^{2}\right) f
  \label{eq:def-of-calL2}
  ,
\end{eqnarray}
where $a$ is determined so that ${\cal L}_{2}$ satisfy the
conditions of Theorem
\ref{theorem:Bruni-Gualtieri-Sopuerta-2003-Appendix}. 
The conditions (ii)-(iv) in Theorem
\ref{theorem:Bruni-Gualtieri-Sopuerta-2003-Appendix} for 
${\cal L}_{2}$ are satisfied from the fact that
$\Phi_{\lambda}^{*}$ is the pull-back of a diffeomorphism
$\Phi_{\lambda}$.
Further, ${\cal L}_{2}$ is obviously linear but we have to check
${\cal L}_{2}$ satisfy the Leibniz rule, i.e.,  
\begin{eqnarray}
  \label{eq:Leibnitz-rule-of-calL2}
  {\cal L}_{2}\left(f^{2}\right) = 2 f {\cal L}_{2} f
\end{eqnarray}
for $\forall f \in {\cal F}({\cal M})$.
To do this, we use the properties
(\ref{eq:properties-of-partial-over-partial-lambda-1}) and
(\ref{eq:properties-of-partial-over-partial-lambda-2}), then we
can easily see that the Leibniz rule
(\ref{eq:Leibnitz-rule-of-calL2}) is satisfied iff $a=1$ and we
may regard ${\cal L}_{2}$ as the Lie derivative with respect to
some vector field.
Then, when and only when $a=1$, there exists a vector field
$\xi_{2}$ such that 
\begin{eqnarray}
  {\cal L}_{2}f = {\pounds}_{\xi_{2}}f
  \label{eq:def-of-calL2-is-Lie}
\end{eqnarray}
and 
\begin{eqnarray}
  \left\{
    \frac{\partial^{2}}{\partial\lambda^{2}}(\Phi^{*}_{\lambda}f)
  \right\}_{\lambda=0}
  &=:& 
  \left({\pounds}_{\xi_{2}} + {\pounds}_{\xi_{1}}^{2}\right) f
  \label{eq:second-order-term-of-Taylor-expansion-is-Lie}
  .
\end{eqnarray}
Thus, we have seen that the Taylor expansion
(\ref{eq:symbolic-Taylor-expansion-of-f-appendix}) for an
arbitrary scalar function $f$ is given by
Eq.~(\ref{eq:Taylor-expansion-of-f}).

%*******************************************************************

Although the formula (\ref{eq:Taylor-expansion-of-f}) of the
Taylor expansion is for an arbitrary scalar function, we can 
easily extend this formula to that for an arbitrary tensor field 
$Q$ as the assertion of Theorem
\ref{theorem:Bruni-Gualtieri-Sopuerta-2003-Appendix}.
The proof of the extension of the formula
(\ref{eq:Taylor-expansion-of-f}) to an arbitrary tensor field
$Q$ is completely parallel to the proof of the formula
(\ref{eq:Taylor-expansion-of-f}) for an arbitrary scalar function
if we stipulate the properties
\begin{eqnarray}
  \left\{
    \frac{\partial^{2}}{\partial\lambda^{2}}(\Phi^{*}_{\lambda}Q)
  \right\}_{\lambda=0}
  &=& 
  \left\{
    \frac{\partial}{\partial\lambda}\left(
      \frac{\partial}{\partial\lambda}(\Phi^{*}_{\lambda}Q)
    \right)
  \right\}_{\lambda=0}
  \label{eq:properties-of-partial-over-partial-lambda-1-tensor}
  ,\\
  \left\{
    \frac{\partial}{\partial\lambda}(\Phi^{*}_{\lambda}Q)^{2}
  \right\}_{\lambda=0}
  &=& 
  \left\{
    2 \Phi^{*}_{\lambda}Q \frac{\partial}{\partial\lambda}(\Phi^{*}_{\lambda}Q)
  \right\}_{\lambda=0}
  \label{eq:properties-of-partial-over-partial-lambda-2-tensor}
\end{eqnarray}
instead of
Eqs.~(\ref{eq:properties-of-partial-over-partial-lambda-1}) and
(\ref{eq:properties-of-partial-over-partial-lambda-2}). 
As the result, we obtain the representation of the Taylor
expansion for an arbitrary tensor field $Q$.

%*******************************************************************

%%%%%%%%%%%%%%%%%%%%%%%%%%%%%%%%%%%%%%%%%%%%%%%%%%%%%%%%%%%%%%%%%%%%%%
%%%%%%%%%%%%%%%%%%%%%%%%%%%%%%%%%%%%%%%%%%%%%%%%%%%%%%%%%%%%%%%%%%%%%%
%%%%%%%%%%%%%%%%%%%%%%%%%%%%%%%%%%%%%%%%%%%%%%%%%%%%%%%%%%%%%%%%%%%%%%
\section{Derivation of the perturbative Einstein tensors}
\label{sec:derivation-of-pert-Einstein-tensors}
%%%%%%%%%%%%%%%%%%%%%%%%%%%%%%%%%%%%%%%%%%%%%%%%%%%%%%%%%%%%%%%%%%%%%%
%%%%%%%%%%%%%%%%%%%%%%%%%%%%%%%%%%%%%%%%%%%%%%%%%%%%%%%%%%%%%%%%%%%%%%
%%%%%%%%%%%%%%%%%%%%%%%%%%%%%%%%%%%%%%%%%%%%%%%%%%%%%%%%%%%%%%%%%%%%%%

%*******************************************************************

Following the outline of the calculations explained in
Sec.~\ref{sec:Perturbation-of-the-Einstein-tensor}, we first
calculate the perturbative expansion of the inverse metric.
The perturbative expansion of the inverse metric can be easily
derived from Eq.~(\ref{eq:metric-expansion}) and the definition
of the inverse metric 
\begin{eqnarray}
  \label{eq:inverse-metric-def}
  \bar{g}^{ab}\bar{g}_{bc} = \delta^{a}_{c}.
\end{eqnarray}
We also expand the inverse metric $\bar{g}^{ab}$ in the form
\begin{eqnarray}
  \label{eq:inverse-metric-expansion}
  \bar{g}^{ab} = g^{ab} + \lambda {}^{(1)}\!\bar{g}^{ab} +
  \frac{1}{2} \lambda^{2} {}^{(2)}\!\bar{g}^{ab}.
\end{eqnarray}
Then, each term of the expansion of the inverse metric is given
by 
\begin{eqnarray}
  \label{eq:inverse-metric-each-order}
  {}^{(1)}\!\bar{g}^{ab} = - h^{ab}, \quad
  {}^{(2)}\!\bar{g}^{ab} = 2 h^{ac} h_{c}^{\;\;b} - l^{ab}.
\end{eqnarray}

%*********************************************************************

To derive the formulae for the perturbative expansion of the
Riemann curvature, we have to derive the formulae for the
perturbative expansion of the tensor $C^{c}_{\;\;ab}$ given by
Eq.~(\ref{eq:c-connection}).
The tensor $C^{c}_{\;\;ab}$ is also expanded in the same form as
Eq.~(\ref{eq:Bruni-39-one}).
The first-order perturbations of $C^{c}_{\;\;ab}$ have the
well-known form\cite{Wald-book}
\begin{eqnarray}
  {}^{(1)}\!C^{c}_{\;\;ab}
  =
  \nabla_{(a}h_{b)}^{\;\;c} - \frac{1}{2} \nabla^{c}h_{ab}
  =:
  H_{ab}^{\;\;\;\;c}\left[h\right],
  \label{eq:KN2005-3.12}
\end{eqnarray}
where $H_{ab}^{\;\;\;\;c}\left[A\right]$ is defined by
Eq.~(\ref{eq:Habc-def-1}) for an arbitrary tensor field $A_{ab}$
defined on the background spacetime ${\cal M}_{0}$.
In terms of the tensor field $H_{ab}^{\;\;\;\;c}$ defined by 
(\ref{eq:Habc-def-1}) the second-order perturbation
${}^{(2)}\!C^{c}_{\;\;ab}$ of the tensor field $C^{c}_{\;\;ab}$
is given by
\begin{eqnarray}
  {}^{(2)}\!C^{c}_{\;\;ab}
  =
  H_{ab}^{\;\;\;\;c}\left[l\right] - 2 h^{cd} H_{abd}\left[h\right].
  \label{eq:KN2005-3.13}
\end{eqnarray}
The Riemann curvature (\ref{eq:phys-riemann-back-riemann-rel})
on the physical spacetime ${\cal M}_{\lambda}$ is also expanded
in the form (\ref{eq:Bruni-39-one}):
\begin{eqnarray}
  \bar{R}_{abc}^{\;\;\;\;\;\;d}
  &=:&
  R_{abc}^{\;\;\;\;\;\;d}
  +
  \lambda {}^{(1)}\!R_{abc}^{\;\;\;\;\;\;d}
  +
  +
  \frac{1}{2} \lambda^{2} {}^{(2)}\!R_{abc}^{\;\;\;\;\;\;d}
  \nonumber\\
  &&
  + O(\lambda^{3}).
\end{eqnarray}
The first- and the second-order perturbation of the Riemann
curvature are given by 
\begin{eqnarray}
  {}^{(1)}\!R_{abc}^{\;\;\;\;\;\;d} 
  &=&
  - 2 \nabla_{[a} {}^{(1)}\!C^{d}_{\;\;b]c},
  \label{eq:KN2005-3.15}
  \\
  {}^{(2)}\!R_{abc}^{\;\;\;\;\;\;d} 
  &=&
  - 2 \nabla_{[a} {}^{(2)}\!C^{d}_{\;\;b]c}
  + 4 {}^{(1)}\!C^{e}_{\;\;c[a} {}^{(1)}\!C^{d}_{\;\;b]e}
  \label{eq:KN2005-3.16}
\end{eqnarray}
Substituting Eqs.~(\ref{eq:KN2005-3.12}) and
(\ref{eq:KN2005-3.13}) into Eqs.~(\ref{eq:KN2005-3.15}) and
(\ref{eq:KN2005-3.16}), we obtain the perturbative form of the
Riemann curvature in terms of the variables defined by
Eq.~(\ref{eq:Habc-def-1}) and (\ref{eq:Habc-def-2}):
\begin{eqnarray}
  {}^{(1)}\!R_{abc}^{\;\;\;\;\;\;d} 
  &=&
  - 2 \nabla_{[a} H_{b]c}^{\;\;\;\;\;d}\left[h\right],
  \label{eq:KN2005-3.15-2}
  \\
  {}^{(2)}\!R_{abc}^{\;\;\;\;\;\;d} 
  &=&
  - 2 \nabla_{[a} H_{b]c}^{\;\;\;\;\;d}\left[l\right]
  + 4 H_{[a}^{\;\;\;de}\left[h\right] H_{b]ce}\left[h\right]
  \nonumber\\
  && 
  + 4 h^{de} \nabla_{[a} H_{b]ce}\left[h\right].
  \label{eq:KN2005-3.16-2}
\end{eqnarray}

%*********************************************************************

To write down the perturbative curvatures
(\ref{eq:KN2005-3.15-2}) and (\ref{eq:KN2005-3.16-2}) in terms
of the gauge invariant and variant variables defined by
Eqs.~(\ref{eq:linear-metric-decomp}) and
(\ref{eq:H-ab-in-gauge-X-def-second-1}), we first derive an 
expression for the tensor field $H_{abc}[h]$ in terms of the
gauge invariant variables, and then, we derive a perturbative
expression for the Riemann curvature.

%*********************************************************************

First, we consider the linear-order perturbation
(\ref{eq:KN2005-3.15-2}) of the Riemann curvature. 
Using the decomposition (\ref{eq:linear-metric-decomp}) and the
identity $R_{[abc]}^{\;\;\;\;\;\;\;\;d}=0$, we can easily
derive the relation
\begin{eqnarray}
  \label{eq:KN2005-3.20}
  H_{abc}\left[h\right]
  =
  H_{abc}\left[{\cal H}\right]
  +
  \nabla_{a}\nabla_{b}X_{c}
  +
  R_{bca}^{\;\;\;\;\;\;d} X_{d}
  ,
\end{eqnarray}
where the variable $H_{abc}\left[{\cal H}\right]$ is defined by
Eqs.~(\ref{eq:Habc-def-1}) and (\ref{eq:Habc-def-2}) with
$A_{ab}={\cal H}_{ab}$.
Clearly, the variable $H_{ab}^{\;\;\;\;c}\left[{\cal H}\right]$
is gauge invariant.
Taking the derivative and using the Bianchi identity
$\nabla_{[a}R_{bc]de}=0$, we obtain 
\begin{eqnarray}
  {}^{(1)}\!R_{abc}^{\;\;\;\;\;\;d} 
  &=&
  - 2 \nabla_{[a} H_{b]c}^{\;\;\;\;\;d}\left[{\cal H}\right]
  + 
  {\pounds}_{X}R_{abc}^{\;\;\;\;\;\;d}
  \label{eq:KN2005-3.23}
  .
\end{eqnarray}
Similar but some cumbersome calculations yield
\begin{eqnarray}
  {}^{(2)}\!R_{abc}^{\;\;\;\;\;d} 
  \!\!\!\!
  &=&
  \!\!\!\!
  - 2 \nabla_{[a} H_{b]c}^{\;\;\;\;d}\left[{\cal L}\right]
  + 
  4 H_{[a}^{\;\;\;de}\left[{\cal H}\right] 
  H_{b]ce}\left[{\cal H}\right]
  \nonumber\\
  &&
  + 
  4 {\cal H}_{e}^{\;\;d}
  \nabla_{[a}H_{b]c}^{\;\;\;\;e}\left[{\cal H}\right]
  \nonumber\\
  &&
  + 2 {\pounds}_{X}{}^{(1)}\!R_{abc}^{\;\;\;\;\;d} 
  + \left(
    {\pounds}_{Y} - {\pounds}_{X}^{2}
  \right)R_{abc}^{\;\;\;\;\;d} 
  \label{eq:KN2005-3.32}
  .
\end{eqnarray}
Equations (\ref{eq:KN2005-3.23}) and (\ref{eq:KN2005-3.32}) have
the same for as the decomposition formulae
(\ref{eq:matter-gauge-inv-decomp-1.0}) and
(\ref{eq:matter-gauge-inv-decomp-2.0}), respectively.

%*********************************************************************

Contracting the indices $b$ and $d$ in
Eqs.~(\ref{eq:KN2005-3.23}) and (\ref{eq:KN2005-3.32}) of the
perturbative Riemann curvature, we can directly derive the
formulae for the perturbative expansion of the Ricci curvature:
expanding the Ricci curvature
\begin{eqnarray}
  \bar{R}_{ab}
  =:
  R_{ab}
  +
  \lambda {}^{(1)}\!R_{ab}
  +
  +
  \frac{1}{2} \lambda^{2} {}^{(2)}\!R_{ab}
  + O(\lambda^{3}),
\end{eqnarray}
we obtain the first-order Ricci curvature as
\begin{eqnarray}
   {}^{(1)}\!R_{ab}
   &=&
   - 2 \nabla_{[a}H_{c]b}^{\;\;\;\;\;c}\left[{\cal H}\right]
   +
   {\pounds}_{X}R_{ab}
  \label{eq:KN2005-3.45}
  .
\end{eqnarray}
and we also obtain the second-order Ricci curvature as 
\begin{eqnarray}
  {}^{(2)}\!R_{ab}
  &=&
  - 2 \nabla_{[a}H_{c]b}^{\;\;\;\;\;c}\left[{\cal L}\right]
  +
   4 H_{[a}^{\;\;\;cd}\left[{\cal H}\right] H_{c]bd}\left[{\cal H}\right]
  \nonumber\\
  &&
  +
  4 {\cal H}_{d}^{\;\;c} \nabla_{[a}H_{b]c}^{\;\;\;\;\;d}\left[{\cal H}\right]
  \nonumber\\
  &&
  +
  2 {\pounds}_{X}{}^{(1)}\!R_{ab}
  + \left(
    {\pounds}_{Y} - {\pounds}_{X}^{2}
  \right)R_{ab}
  \label{eq:KN2005-3.46}
  .
\end{eqnarray}

%*********************************************************************

The scalar curvature on the physical spacetime ${\cal M}$ is
given by $\bar{R} = \bar{g}^{ab}\bar{R}_{ab}$.
To obtain the perturbative form of the scalar curvature, we
expand the $\bar{R}$ in the form (\ref{eq:Bruni-39-one}), i.e., 
\begin{eqnarray}
  \bar{R} =: R + \lambda {}^{(1)}\!R
  + \frac{1}{2}\lambda^{2} {}^{(2)}\!R + O(\lambda^{3})
\end{eqnarray}
and $\bar{g}^{ab}\bar{R}_{ab}$ is expanded through the Leibniz
rule.
Then, the perturbative formula for the scalar curvature at each
order is derived from perturbative form of the inverse metric
(\ref{eq:inverse-metric-each-order}) and the Ricci curvature
(\ref{eq:KN2005-3.45}) and (\ref{eq:KN2005-3.46}).
Straightforward calculations lead to the expansion of the scalar
curvature as
\begin{eqnarray}
  {}^{(1)}\!R
  &=&
  - 2 \nabla_{[a} H_{b]}^{\;\;\;ab}\left[{\cal H}\right]
  - R_{ab} {\cal H}^{ab}
  + {\pounds}_{X}R
  \label{eq:KN2005-3.52}
  , \\
  {}^{(2)}\!R
  &=&
  - 2 \nabla_{[a}H_{b]}^{\;\;\;ab}\left[{\cal L}\right]
  + R^{ab} \left(
    2 {\cal H}_{ca} {\cal H}_{b}^{\;\;c}
    - {\cal L}_{ab}
  \right)
  \nonumber\\
  &&
  + 4 H_{[a}^{\;\;\;cd}\left[{\cal H}\right]
  H_{c]\;\;d}^{\;\;\;a}\left[{\cal H}\right]
  + 4 {\cal H}_{c}^{\;\;b} \nabla_{[a}H_{b]}^{\;\;\;ac}\left[{\cal H}\right]
  \nonumber\\
  &&
  + 4 {\cal H}^{ab} \nabla_{[a} H_{d]b}^{\;\;\;\;\;d}\left[{\cal H}\right]
  \nonumber\\
  &&
  + 2 {\pounds}_{X}{}^{(1)}\!R
  + \left(
    {\pounds}_{Y} - {\pounds}_{X}^{2}
  \right)R
  \label{eq:KN2005-3.54}
  .
\end{eqnarray}
We also note that the expansion formulae
(\ref{eq:KN2005-3.52}) and (\ref{eq:KN2005-3.54}) have the same
for as the decomposition formulae
(\ref{eq:matter-gauge-inv-decomp-1.0}) and
(\ref{eq:matter-gauge-inv-decomp-2.0}), respectively.

%*********************************************************************

Next, we consider the perturbative form of the Einstein tensor
$\bar{G}_{ab}:=\bar{R}_{ab}-\frac{1}{2}\bar{g}_{ab}\bar{R}$ and
we expand $\bar{G}_{ab}$ as in the form (\ref{eq:Bruni-39-one}):
\begin{eqnarray}
  \bar{G}_{ab} =: G_{ab} + \lambda {}^{(1)}\!\left(G_{ab}\right)
  + \frac{1}{2}\lambda^{2} {}^{(2)}\!\left(G_{ab}\right) + O(\lambda^{3})
  .
\end{eqnarray}
As in the case of the scalar curvature, straightforward
calculations lead
\begin{widetext}
\begin{eqnarray}
  {}^{(1)}\!\left(G_{ab}\right)
  &=&
  - 2 \nabla_{[a} H_{d]b}^{\;\;\;\;\;d}\left[{\cal H}\right]
  + g_{ab} \nabla_{[c} H_{d]}^{\;\;\;cd}\left[{\cal H}\right]
  - \frac{1}{2} R {\cal H}_{ab}
  + \frac{1}{2} g_{ab} R_{cd} {\cal H}^{cd}
  + {\pounds}_{X}G_{ab}
  \label{eq:KN2005-3.59}
  ,
  \\
  {}^{(2)}\!\left(G_{ab}\right)
  &=&
  - 2 \nabla_{[a} H_{c]b}^{\;\;\;\;\;c}\left[{\cal L}\right]
  + 4 H_{[a}^{\;\;\;cd}\left[{\cal H}\right] H_{c]bd}\left[{\cal H}\right]
  + 4 {\cal H}_{c}^{\;\;d} \nabla_{[a} H_{d]b}^{\;\;\;\;\;c}\left[{\cal H}\right]
  \nonumber\\
  &&
  - \frac{1}{2} g_{ab} \left(
    - 2 \nabla_{[c} H_{d]}^{\;\;\;cd}\left[{\cal L}\right]
    + 2 R_{de}{\cal H}_{c}^{\;\;d}{\cal H}^{ec}
    - R_{de}{\cal L}^{de}
    + 4 H_{[c}^{\;\;\;de}\left[{\cal H}\right]H_{d]\;\;e}^{\;\;\;c}\left[{\cal H}\right]
  \right.
  \nonumber\\
  && \quad\quad\quad\quad
  \left.
    + 4 {\cal H}_{e}^{\;\;d} \nabla_{[c} H_{d]}^{\;\;\;ce}\left[{\cal H}\right]
    + 4 {\cal H}^{ce} \nabla_{[c} H_{d]e}^{\;\;\;\;\;d}\left[{\cal H}\right]
  \right)
  + 2 {\cal H}_{ab} \nabla_{[c}H_{d]}^{\;\;\;cd}\left[{\cal H}\right]
  + {\cal H}_{ab} {\cal H}^{cd} R_{cd}
  - \frac{1}{2} R {\cal L}_{ab}
  \nonumber\\
  &&
  + 2 {\pounds}_{X}{}^{(1)}\!\left(G_{ab}\right)
  + \left({\pounds}_{Y} - {\pounds}_{X}^{2}\right)G_{ab}
  \label{eq:KN2005-3.61}
  .
\end{eqnarray}
\end{widetext}
We note again that Eqs.~(\ref{eq:KN2005-3.59}) and
(\ref{eq:KN2005-3.61}) have the same form as the decomposition
formulae (\ref{eq:matter-gauge-inv-decomp-1.0}) and
(\ref{eq:matter-gauge-inv-decomp-2.0}), respectively.

%*********************************************************************

The perturbative formulae for the perturbation of the Einstein
tensor 
\begin{eqnarray}
  \bar{G}_{a}^{\;\;b}=\bar{g}^{bc}\bar{G}_{ac}
\end{eqnarray}
is derived by the similar manner to the case of the
perturbations of the scalar curvature.
Through these formulae summarized above, straightforward
calculations leads 
Eqs.~(\ref{eq:linear-Einstein})--(\ref{eq:(2)Sigma-def-second}).
We have to note that to derive the formulae
(\ref{eq:cal-G-def-second}) with
Eq.~(\ref{eq:(2)Sigma-def-second}), we have to consider the
general relativistic gauge-invariant perturbation theory with
two infinitesimal parameters which is developed in
Refs.~\cite{kouchan-gauge-inv,kouchan-second}, as commented in
the main text.

%*********************************************************************

%%%%%%%%%%%%%%%%%%%%%%%%%%%%%%%%%%%%%%%%%%%%%%%%%%%%%%%%%%%%%
%%%%%%%%%%%%%%%%%%%%%%%%%%%%%%%%%%%%%%%%%%%%%%%%%%%%%%%%%%%%%

\end{document}